\documentclass[twocolumn]{aastex631}

\usepackage{amsmath,amstext}
\usepackage[T1]{fontenc}
\usepackage[figure,figure*]{hypcap}
\usepackage{comment}
\usepackage{graphicx}
\usepackage{natbib}
\usepackage{epsfig}
\usepackage{epstopdf}
\usepackage{enumerate}
\usepackage{tabularx}
\usepackage{multirow}
\usepackage{booktabs}
\usepackage{longtable} 
\usepackage{multirow, bigdelim}
\usepackage{url}
\usepackage{appendix}
\usepackage[ruled,vlined]{algorithm2e}

\SetKwInput{KwDefine}{Notation}
\SetKwInput{KwInput}{Input}
\SetKwInput{KwOutput}{Output}

\newcommand{\msun}{$M_{\odot}$\xspace}
\newcommand{\mrem}{M_{\rm rem}\xspace}
\newcommand{\rrem}{R_{\rm rem}\xspace}
\newcommand{\mpns}{M_{\rm PNS}\xspace}
\newcommand{\rpns}{R_{\rm PNS}\xspace}

\newcommand{\kpush}{$k_{\rm push}$\xspace}
\newcommand{\trise}{$t_{\rm rise}$\xspace}
\newcommand{\tpb}{t_{\mathrm{pb}}\xspace}

\newcommand{\code}[1]{\texttt{\detokenize{#1}}}

\newcommand{\nprog}{174} 
\newcommand{\nsims}{1,684} 
\newcommand{\minmass}{10.8} 
\newcommand{\maxmass}{40}  

\newcommand{\nns}{1,057}
\newcommand{\nbh}{477} 
\newcommand{\nweird}{130} 

\newcommand{\nnotrun}{20}

\newcommand{\muslope}{2.47 \times 10^{-5}}
\newcommand{\muslopeerr}{2.19 \times 10^{-7}}
\newcommand{\muintercept}{-0.015}
\newcommand{\muintercepterr}{2.13 \times 10^{-4}}
\newcommand{\mrerr}{4.26 \times 10^{-5}}

\graphicspath{{./}{figures/}{figures/xfigures/}{figures/app-single-times/}{figures/app-outliers/}}

\received{\today}
\revised{}
\accepted{}
\submitjournal{ApJ}

\shorttitle{Gravitational Waves from CCSNe}
\shortauthors{Wolfe et al.}

\begin{document}

\title{Gravitational Wave Eigenfrequencies from Neutrino-Driven Core-Collapse Supernovae}

\correspondingauthor{Fr\"ohlich, Wolfe}
\email{cfrohli@ncsu.edu, newolfe@mit.edu}

\author[0000-0003-2540-3845]{Noah E. Wolfe}
\affiliation{Department of Physics, North Carolina State University, Raleigh NC 27695, USA}
\affiliation{LIGO Laboratory, Massachusetts Institute of Technology, 185 Albany St, Cambridge, MA 02139, USA}
\affiliation{Department of Physics and Kavli Institute for Astrophysics and Space Research, Massachusetts Institute of Technology, 77 Massachusetts Ave, Cambridge, MA 02139, USA}

\author[0000-0003-0191-2477]{Carla Fr\"ohlich}
\affiliation{Department of Physics, North Carolina State University, Raleigh NC 27695, USA}

\author[0000-0001-6432-7860]{Jonah M. Miller}
\affiliation{
    CCS-2, Computational Physics and Methods, 
    Los Alamos National Laboratory, 
    Los Alamos NM 87544, USA
}
\affiliation{
    Center for Theoretical Astrophysics, 
    Los Alamos National Laboratory, 
    Los Alamos NM 87544, USA
}

\author[0000-0001-8709-5118]{Alejandro Torres-Forn\'e}
\affiliation{Departamento de Astronom\'ia y Astrof\'isica, Universitat de Val\`encia,
46100 Burjassot (Val\`encia), Spain}
\address{ Observatori Astronòmic, Universitat de València, E-46980, Paterna (València), Spain}

\author[0000-0003-4293-340X]{Pablo Cerd\'a-Dur\'an}
\affiliation{Departamento de Astronom\'ia y Astrof\'isica, Universitat de Val\`encia,
46100 Burjassot (Val\`encia), Spain}
\address{ Observatori Astronòmic, Universitat de València, E-46980, Paterna (València), Spain}

\begin{abstract}
Core-collapse supernovae are predicted to produce gravitational waves (GWs) that may be detectable by Advanced LIGO/Virgo. These GW signals carry information from the heart of these catacylsmic events, where matter reaches nuclear densities. 
Recent studies have shown that it may be possible to infer properties of the proto-neutron star (PNS) via gravitational waves generated by hydrodynamic perturbations of the PNS. 
However, we lack a comprehensive understanding of how these relationships may change with the properties of core-collapse supernovae.
In this work, we build a self-consistent suite of over 1000 exploding core-collapse supernovae from a grid of progenitor masses and metallicities combined with six different nuclear equations of state. 
Performing a linear perturbation analysis on each model, we compute the resonant gravitational-wave frequencies of the PNS, and we motivate a time-agnostic method for identifying characteristic frequencies of the dominant gravitational-wave emission. 
From this, we identify two characteristic frequencies, of the early- and late-time signal, that measure the surface gravity of the cold remnant neutron star, and simultaneously constrain the hot nuclear equation of state. 
However, we find that the details of the core-collapse supernova model, such as the treatment of gravity or the neutrino transport, and whether it explodes, noticeably change the magnitude and evolution of the PNS eigenfrequencies.
\end{abstract}



\section{Introduction} \label{sec:intro}

Core-collapse supernovae (CCSNe) are the deaths of massive stars $\gtrsim 8 \ \mathrm{M}_{\odot}$\footnote{ 
Stars of 8--10 \msun undergo collapse of the ONeMg core due to electron capture reactions, while stars $\gtrsim 10$ \msun undergo gravitational collapse of the Fe-core, though the exact limits are uncertain and depend also on the metallicity.}.
At the end of their stellar lifetime, massive stars have developed a dense core of iron-group nuclei that cannot undergo further fusion reactions to produce the excess energy required to support the star against gravitational collapse.
Eventually, the stellar core begins to contract and then collapse in on itself under gravity.
The weight of the infalling star compresses the iron core, inducing electron capture reactions onto the iron nuclei, converting protons into neutrons.
At the end of this neutronization, the iron core is predominantly composed of neutrons.
The neutron degeneracy pressure is sufficient to halt further compression, ``core bounce'' occurs, and an outward moving shock wave is formed, which soon stalls and becomes a standing accretion shock.
In the neutrino-driven supernova mechanism, if the combined influence of neutrino (re-)absorption and hydrodynamic instabilities behind the shock are sufficiently large, the shock is revived and propagates through the collapsing star, gravitationally unbinding the outer layers in a successful supernova explosion.
If the shock cannot be successfully revived, the infalling stellar material will continue to accrete through the shock and increase the mass of the central object, eventually forming a black hole.
By achieving core densities at or beyond the nuclear saturation density, $\gtrsim 10^{14}$ g/cm$^3$, at high temperatures $\mathcal{O}(10 - 100 \ \mathrm{MeV})$, core-collapse supernovae are compelling laboratories for extreme nuclear matter.

Gravitational waves (GWs) are a long-postulated astrophysical messenger from CCSNe that will yield new insights into fundamental nuclear physics \citep{ruffini_1971}. Generated by quadrupole mass oscillations, gravitational waves emitted during core collapse are expected to originate near or within the proto-neutron star (PNS), from rotation of the core \citep[][and references therein]{moenchmeyer_rotating-core-collapse-gw_1991, dimmelmeier_relativistic-ccsne-gw_2001, richers_2017},
convection within the PNS \citep{gw_pns-convection_burrows1996, gw_pns-convection_muller1997, gw_pns-convection_muller2004}, or hydrodynamic plumes outside of the PNS striking its surface and perturbing its dense nuclear material \citep{murphy_2009, warren_2020}. 
It is this latter mechanism for GW generation in core-collapse supernovae that is thought to be the excitation that dominates the GW signal amplitude \citep{marek_equation-of-state_2009, murphy_2009, muller_fpeak_2013, kuroda_new_2016, nakamura_multimessenger_2016, oconnor_exploring_2018, morozova_eigenfrequencies_2018, torres-forne_I_2018, powell_gravitational_2019, radice_characterizing_2019}.

Gravitational waves emitted by core collapse are expected to be generated without interference from other regions or physical processes in dying stars, making these signals ideal probes into the behavior of matter at the extreme nuclear densities of the PNS.
There are already real detection prospects for these signals; at design sensitivity, the twin Advanced LIGO detectors \citep{LIGO} plus Advanced Virgo detector \citep{Virgo} are expected to be sensitive to non-rotating core-collapse supernovae within ${\sim}10$ kpc (within the Milky Way), and in highly-rotating models of core collapse, as far as $50$ kpc \citep{szczepanczy_cWB-ccsne-detect_2021}\footnote{For LIGO O4/O5 design sensitivities, using the coherent WaveBurst \citep{klimenko_cWB_2016} pipeline \citep[for predictions based on other pipelines and interferometer sensitivity curves, see][]{arnaud_gw-bursts_1999, takiwaki_ccsne-network-analysis_2015, gw_ccsne_current_gossan2016}.}. 
Though this distance is well within the Milky Way and its satellite galaxies, we expect at least one such ``galactic'' supernova every century \citep{Adam2013, Rozwadowska2021}.
Future detectors, like Cosmic Explorer \citep{ce_horizon_study_evans2021}, Einstein Telescope \citep{Hild2011} and NEMO \citep{nemo}, will be more sensitive to the predicted frequency band of core-collapse gravitational wave emission. 
Thus, it is likely that we will observe at least one core-collapse event in gravitational-waves within the next few decades.

Many studies have sought to understand the hydrodynamic source of gravitational waves and detection prospects of these signals using two- and three-dimensional models of core collapse \citep{marek_equation-of-state_2009, murphy_2009, muller_fpeak_2013, yakunin_gravitational_2015, kuroda_new_2016, nakamura_multimessenger_2016, andresen_gravitational_2017, powell_gravitational_2019, radice_characterizing_2019, mezzacappa_gravitational-wave_2020, vartanyan_2020, andresen_2021, takiwaki_2021, bugli_2022, mezzacappa_gravitational-wave_2023, vartanyan_2023}. 
In pursuit of core-collapse gravitational-wave phenomenology, some recent work has found that the frequency structure of gravitational waves produced by the collapse of astrophysically rare progenitors may depend on the equation of state; for example, \citet{richers_2017} with rotating models of core-collapse and \citet{jakobus-eos-2023} with two high-mass (35~\msun{} and 85~\msun{}), zero-metallicity progenitors.
However, due to the extreme computational cost of multi-dimensional models, it is not yet possible to fully explore the observational consequences of core-collapse gravitational waves across a landscape of stellar progenitor properties, nuclear equation of state configurations, and other relevant (astro)physics.
In addition, multi-dimensional models often rely on approximations to general relativity that have a non-negligible impact on the spectrum of gravitational waves emitted during core collapse  \citep{dimmelmeier_relativistic-ccsne-gw_2001, fryer_ccsne-gw-review_2010s, muller_fpeak_2013, morozova_eigenfrequencies_2018, torres-forne_II_2019, powell_gravitational_2019, Sotani.2020}.

Nevertheless, multi-dimensional studies serve as vital guides to the morphology of core-collapse gravitational-wave signals.
In particular, multi-dimensional models consistently produce gravitational-wave signals with a stochastic amplitude, but well-structured time-frequency evolution\footnote{This is observationally-convenient as gravitational-wave detectors are more sensitive to the frequency evolution than the amplitude evolution \citep{2021arXiv211206861T}.}.
%
%
Asteroseismology studies in spherical symmetry which focus on the time-frequency evolution (but forgo GW amplitude information) have emerged as a complimentary approach for studying the GW signal from CCSNe.

Neutron star asteroseismology is well established for cold neutron stars settled in a steady state after their violent birth, particularly in the context of searches for continuous gravitational-wave signals \citep[see for example the review by][and references therein]{ns-continuous-gw-review-2019}. 
Recent studies have applied linear perturbation analyses for asteroseismology of the hot, \textit{proto}-neutron star during core collapse, using simplified models of its evolution \citep{sotani_simple_2016} as well as full metric and hydrodynamic data in the proto-neutron star domain generated by numerical simulations \citep{torres-forne_I_2018, morozova_eigenfrequencies_2018, torres-forne_II_2019}. 
These works has clarified the role of general relativity in such analyses \citep{morozova_eigenfrequencies_2018, torres-forne_II_2019, Sotani.2020} and attempted to analytically fit different frequency modes as a function of bulk PNS properties \citep{torres-forne_universal_2019, Sotani.2021,Mori2023}. 
However, it remains difficult to draw conclusions about the observational consequences of CCSN gravitational waves due to uncertainties introduced by the choice of mode classification schemes \citep{torres-forne_I_2018, torres-forne_II_2019}, by the choice of boundary conditions \citep{Sotani2019, Sotani.2021}, and by the setup of the physical assumptions made in different simulations of core collapse (such as neutrino or gravitational physics, among others).

Here, we seek to build a single, self-consistent phenomenology of core-collapse gravitational waves around neutron star structure in general and the nuclear equation of state in particular. 
To this end, we conduct an asteroseismological analysis of a large suite of supernova models. 
We use the spherically-symmetric \code{PUSH} code to perform \nsims{} simulations of stellar collapse and explosion for a grid of \nprog{} progenitors between \minmass{} and \maxmass{} \msun at three different metallicities (solar, sub-solar, and zero) and six different nuclear equations of state (EOS). 
The detailed information on the time evolution of the PNS from our core collapse simulations is used as input to the \code{GREAT} code to calculate the resonant frequencies of the PNS, following the method of \citet{torres-forne_II_2019} (TF19 hereafter).
We then apply modern statistical methods to these resonant eigenfrequencies to identify correlations between the early (late) time frequencies, the surface gravity of the remnant neutron star, and the nuclear equation of state.

In Section~\ref{sec:models} we describe our numerical setup and the nuclear equations of state used.
We also provide a summary of all core-collapse simulations performed. 
Then, in Section~\ref{sec:eigenfreqs}, we detail the linear perturbation analysis that yields eigenfrequencies for each model.
In Section~\ref{sec:results2}, we develop a method for characterizing the frequency structure of these eigenfrequencies, driven by correlations between the evolution of the eigenfrequencies and both the remnant neutron star mass and equation of state.
We further investigate these relationships in Section~\ref{sec:results3}, and find two characteristic frequencies of the dominant part of the gravitational-wave signal which can constrain the nuclear equation of state and measure the remnant surface gravity.
Finally, we summarize and discuss our results in Section \ref{sec:summary}.

\section{Numerical setup and models} 
\label{sec:models}
In this study, we analyze the gravitational wave eigenfrequencies of \nsims{} core-collapse supernova simulations spanning a range of progenitor star ZAMS masses (\minmass{} to \maxmass{} \msun) at three different metallicities ($Z/Z_{\odot}=1$, $10^{-4}$, $0$).

\begin{figure*}
    \centering
    \includegraphics[width=\linewidth]{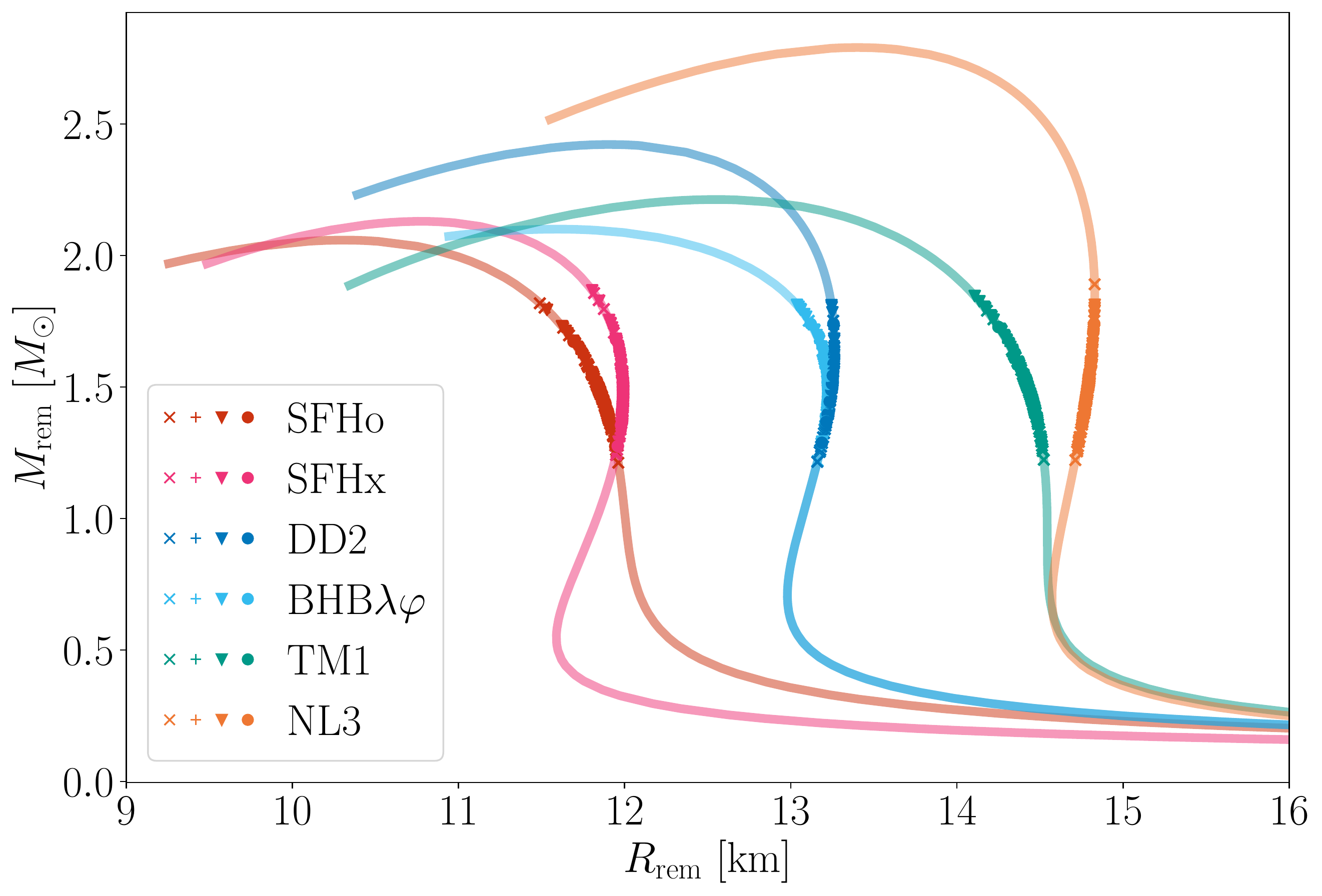}
    \caption{
        Mass-radius relationship for the six nuclear EOSs considered in this work. The remnant mass $\mrem{}$ is the gravitational birth mass of cold neutron stars and and $\rrem{}$ is the corresponding radius. The markers indicate the \nns{} exploding models from this work. Different markers correspond to different progenitor metallicities. The transparent lines are the complete mass-radius relations for each EOS.  
    }
    \label{fig:ns-coverage}
\end{figure*}

All core-collapse supernova simulations are performed with the code \code{Agile-IDSA}, which uses the spherically-symmetric, general relativistic, adaptive-mesh hydrodynamics code \code{Agile} \citep{Liebendoerfer.Agile}, the Isotropic Diffusion Source Approximation (IDSA) for the transport of electron-flavor neutrinos and antineutrinos \citep{Liebendoerfer.IDSA:2009}, and the Advanced Spectral Leakage (ASL) scheme for the transport of heavy-flavor (muon and tau) neutrinos and anti-neutrinos \citep{perego16.asl}. For matter at high density in nuclear statistical equilibrium, we employ a set of six finite-temperature nuclear equations of state for matter: DD2 \citep{dd2_1}, SFHo \citep{sfho}, SFHx \citep{sfho}, BHB$\lambda\phi$ \citep{bhblp}, TM1 \citep{dd2_1}, and NL3 \citep{dd2_1}.
These equations of state are largely differentiated by how they parameterize interactions between nucleons in dense nuclear matter, and the experimental and theoretical data they use to calibrate these models. 
The SFHo and SFHx equations of state share the same underlying model, with additional calibrations to known neutron star masses and radii. 
The TM1 and NL3 equations of state are similar parameterizations, where TM1 uses slightly more recent experimental data. 
Finally, DD2 and BHB$\lambda \varphi$ treat nuclear matter in the same manner, however BHB$\lambda \varphi$ also includes hyperons at high densities. 
All six nuclear EOS considered here allow for a maximum neutron star mass above 2~\msun. 
Combined, these equations of state cover a large range of possible scenarios for the behavior of matter at extreme densities, as can be seen in Figure~\ref{fig:ns-coverage}.

All simulations in this work rely on the PUSH method \citep{push1,push2} to trigger explosions in spherical symmetry. 
In the PUSH method, a fraction of the heavy-flavor neutrino energy is deposited behind the shock, mimicking in spherically-symmetric CCSN simulations the enhanced neutrino-heating observed in multi-dimensional simulations \citep{push1}. 
This energy deposition is formulated via a parametrized heating term $Q^+_{\mathrm{push}}(t,r)$  (energy per unit mass and time) for each radial position $r$ and each time $t$. There are two free parameters, \kpush and \trise which control the temporal behavior of the push-heating. Following \citet{push2}, we set $t_{\mathrm{rise}}= 400$~ms and \kpush is a parabolic function of compactness $\xi_M$.
We follow \citet{oconnor_compactness_2011} for the definition of compactness,
\begin{equation} \label{eq:compactness}
    \xi_M = \frac{M/M_{\odot}}{R(M)/1000 \,\mathrm{km}},
\end{equation}
where $R(M)$ is the radius which encloses the mass $M$.
As in \citet{push2}, we evaluate the compactness for an enclosed mass $M = 2.0$~\msun{} at the time of bounce.

The initial conditions for the simulations are \nprog{} progenitor models taken from the \code{KEPLER} stellar evolution code with zero-age main sequence (ZAMS) masses between \minmass{} \msun and \maxmass{} \msun at solar, sub-solar, and zero metallicity \citep{Woosley.Heger:2002,Woosley.Heger:2007}, as summarized in Table \ref{tab:progenitors}. For all progenitors, the pre-explosion material up to the helium shell (corresponding to a radial coordinate of $2 \times 10^9$~cm -- $3 \times 10^{10}$~cm) was mapped to 180 radial zones in \code{Agile-IDSA}. 
The adaptive grid algorithm in the \code{Agile} code places more radial zones near steep thermodynamic gradients, achieving the highest radial resolution at the PNS surface with a typical resolution among our models of 20 zones (out of 180 in total) in the region with densities of $10^{10} \,\mathrm{g/cm}^3 < \rho < 10^{12} \,\mathrm{g/cm}^3$. 

\begin{table}    
\begin{center}
	\caption{Progenitor models used in this study.
	} \label{tab:progenitors}
	\begin{tabular}{llllllc}
	\tableline \tableline 
Series & Metallicity    & $M_{\mathrm{min}}$  & $M_{\mathrm{max}}$      & $\Delta m$    & EOS \\
       &  ($Z_{\odot}$) & ($M_{\odot}$)  & ($M_{\odot}$) & ($M_{\odot}$) &   &  \\
	\tableline
   s   & 1 & $10.8$ & $28.2$ & $0.2$ & all \\
       &   & $29.0$ & $40.0$ & $1.0$ & all \\
   w   & 1 & $12.0$ & $33.0$ & $1.0$ & all \\
       &   & $35.0$ & $40.0$ & $5.0$ & all \\
   u   & $10^{-4}$ & $11.0$ & $40.0$ & $1.0$ & DD2 \\
   u   & $10^{-4}$ & $11.0$ & $40.0$ & $0.2$ & not DD2 \\
   z   & 0         & $11.0$ & $40.0$ & $1.0$ & all \\
	\tableline
	\end{tabular}
\end{center}
\tablecomments{All simulations were performed using six nuclear EOS (SFHo, SFHx, DD2, BHB$\lambda\varphi$, TM1, NL3), for a total of \nsims~ supernova models, of which \nns{} are exploding models and are analyzed in this work. 
Note that the simulations with the DD2 EOS are taken from \citet{push2} and \citet{push4}. In the case of the u-series, the simulations with the DD2 EOS have a larger mass spacing $\Delta m$ than for the other five EOSs. 
The progenitors of the s-, u-, and z-series are from \citet{Woosley.Heger:2002}; those of the w-series are from \citet{Woosley.Heger:2007}. When referencing specific models studied in this work, we will label them as \code{{metallicity series}{ZAMS mass}_{EOS}. }
}
\end{table}

All simulations were run for a total time of 7~s, except those using the DD2 EOS which are taken from \citet{push2} and were run for a total of 5~s. At the end of a simulation with \code{Agile-IDSA}, we post-process the simulation data to compute quantities such as the explosion energy ($E_{\mathrm{exp}}$) and the mass cut ($m_{\mathrm{cut}}$), as described in detail in Section 2.5.1 of \citet{push1}.
If the final explosion energy $E_{\mathrm{exp}}>0$, we categorize the model as an exploding model and otherwise as a non-exploding model.
We define the proto-neutron star as the region where the density $\rho \geq 10^{11}$ g/cm${}^3$ and we use $\mpns{}$ to denote the mass enclosed by this density. 
Finally, we use $\mrem{}$ to denote the gravitational birth mass of the final cold neutron star.

In this work, we evolved \nprog{} progenitors from the onset of core collapse through explosion (or failure to explode), using six nuclear equations of state for each progenitor for a total of \nsims{} models.
Of these, \nns{} successfully exploded while \nbh{} failed to explode. An additional \nweird{} simulations could not be confidently categorized as a successful or failed explosion, and hence were excluded from this analysis.
An additional \nnotrun{} models are not run yet. 
Hence, the following sections only include the \nns{} simulations which successfully exploded and formed a neutron star.

In Figure~\ref{fig:ns-coverage}, we show for all \nns{} simulations the gravitational birth mass of the resulting cold neutron star as a function of its radius, overplotted with the mass-radius relation for each equation of state included in this study.
Our models densely cover a range of neutron star masses at and above the theoretical minimum of ${\sim}1.4$~\msun{} but well below the largest observed neutron star masses. 
The mass distribution of NSs from our models is consistent with that of isolated neutron stars observed to date \citep{meskhi_eos-remnant-distribution_2022}. 
We note that the large (${\sim}15$~km) radii of our neutron stars with the TM1 and the NL3 equations of state are largely inconsistent with constraints on neutron star radii from GW observations of a binary neutron star merger \citep{lvk-ns-radii-2018} and from x-ray observations of pulsars with NICER \citep{raaijmakers_2021}. 
These models are still useful to include to expand the parameter space of nuclear physics considered in our study.

\section{Gravitational Wave Eigenfrequencies} 
\label{sec:eigenfreqs}

\subsection{PNS eigenfrequencies from perturbation analysis}

Gravitational waves are, at lowest-order, quadrupolar phenomena, generated by aspherical motions of matter.
In the weak-field limit, for gravitational radiation from an astrophysical source, the amplitude (`strain') and frequencies of gravitational waves are determined by a wave equation in 3+1 spacetime. 
A spherically-symmetric matter distribution lacks a quadrupole moment, hence our supernova models produce no gravitational radiation.
Instead, we use a linear perturbation analysis to calculate the eigenfrequencies of the proto-neutron star.

Two- and three-dimensional simulations of core collapse often produce a loud, monotonically increasing feature in the time-frequency evolution of the gravitational waves that dominates the signal (e.g. see Figure~7 of \citet{powell_gravitational_2019} or Figure~5 of \citet{radice_characterizing_2019}). 
Several works have applied a linear perturbation analysis of the PNS in 2D core-collapse simulations \citep[][e.g.]{morozova_eigenfrequencies_2018,torres-forne_I_2018,torres-forne_II_2019} and find that some modes trace this `loud' feature of the GW spectrogram. 
Our approach in this work is the same as in TF19. 
Thus, we expect certain eigenfrequencies to be the frequencies of gravitational waves generated by perturbations of the compact proto-neutron star.

First, we briefly summarize the key elements of the fully-general relativistic linear perturbation analysis of a spherically-symmetric, self-gravitating fluid in equilibrium from TF19.
This method assumes a static, equilibrium solution to the hydrodynamic background equations. 
Then, the background solution is linearly perturbed, allowing for perturbations to the lapse and the conformal factor, i.e.\  to all metric terms.
Eulerian perturbations are denoted with $\delta$ and Lagrangian perturbations with $\Delta$.
So, the linear Eulerian perturbation of a hydrodynamic variable $y$ is $y \rightarrow y + \delta y$, and this can be related to the Lagrangian perturbation of that variable as
\begin{equation}
    \Delta y = \delta y + \xi^i \frac{\partial y}{\partial x^i},
\end{equation}
where the latin index $i$ denotes a spatial coordinate and $\xi^i$ is the magnitude of the Lagrangian displacement of a fluid element along the spatial coordinate $x^i$. 
Perturbing the background hydrodynamic equations by an Eulerian perturbation thus introduces the Lagrangian displacements $\xi^i$ throughout the system of equations, and so we may solve for these displacements. 
The Eulerian displacement of a hydrodynamic variable can be decomposed using spherical harmonics\footnote{The spherical harmonics form a complete orthonormal basis for real-valued functions on the surface of a sphere of radius $r$.} to separately expose the (time-dependent) angular and radial dependence, as 
\begin{equation} \label{eqn:sph-harm-decompose}
    \delta y = \delta \hat{y} (r) \, Y_{lm} (\theta, \varphi) \, e^{-i \sigma t}.
\end{equation}
In this equation, $i = \sqrt{-1}$ is the unit imaginary number, $t$ denotes time, $\delta \hat{y}(r)$ is the scalar magnitude of the perturbation $\delta y$ and the radial dependence of the perturbation, $Y_{lm}(\theta, \varphi)$, is the standard spherical harmonic function of degree $l$ and order $m$, and $\sigma$ is an eigenvalue for which a perturbation satisfies the linearized general-relativistic hydrodynamics equations and appropriate outer boundary condition.

In this form, we can identify the Lagrangian displacements for each spatial coordinate $(r, \theta, \varphi)$ as
\begin{align}
    \xi^r &= \eta_r(r) Y_{lm}(\theta, \varphi) e^{-i \sigma t}, \\
    \xi^{\theta}  &= \eta_{\perp}(r) \frac{1}{r^2} \frac{ \partial Y_{lm} }{\partial \theta} e^{-i \sigma t}, \\
    \xi^{\varphi} &= \eta_{\perp}(r) \frac{1}{r^2 \sin^2 \theta} \frac{ \partial Y_{lm} }{ \partial \varphi } e^{-i \sigma t},
\end{align}
where $\eta_r(r)$ and $\eta_{\perp}(r)$ are the magnitude of the radial and non-radial perturbation associated with a particular $l, m$.
In this work we only consider perturbations of degree $l = 2$ (the leading order), as perturbations of higher degrees are expected to contribute negligibly to the signal amplitude \citep{torres-forne_I_2018}\footnote{We need not specify $m$ as the magnitudes $\eta$ are independent of $m$ in spherically-symmetric backgrounds.}.

The eigenfrequencies are calculated by integrating the linearized equations of general relativistic hydrodynamics from the center of the PNS to its surface, defined as the density contour at $\rho = 10^{11}$ g/cm${}^3$, at different values of the radial frequency $\sigma$. 
We enforce an inner boundary condition of $\eta_r = 0$ at the origin.
At the PNS surface, we assume that the pressure is in equilibrium, i.e.\ $\Delta P = 0$, yielding the following outer boundary condition
\begin{equation}
    q \sigma^2 \eta_{\perp} + \rho h \left( \frac{\delta \hat{\psi}}{\psi} - \frac{\delta \hat{Q}}{Q} \right) + \eta_r \frac{\partial P}{\partial r} = 0,
\end{equation}
where $h$ is the relativistic enthalpy and $\sigma$ is as defined for Equation~\ref{eqn:sph-harm-decompose}. Finally, $q = \rho h \alpha^{-2} \, \psi^4$ and $Q = \alpha \psi$ follows the notation of TF19. 
Enforcing $\Delta P = 0$ at the surface of the PNS is equivalent to treating the PNS as being in vacuum. 
An alternative choice of outer boundary condition would be to enforce $\eta_r = 0$ at the shock radius, as in \cite{torres-forne_I_2018} and in TF19, which implies no displacement beyond the shock. 
While this is a more correct physical assumption than $\Delta P = 0$ at the PNS surface, $\eta_r(r)$ suffers noticeable numerical noise due to the relatively low-resolution of radial zones near the shock radius in the Lagrangian treatment of \code{Agile}. Thus, additional manual tuning is required to identify modes, which is not feasible for \nns{} models. However, we have checked that, for the modes considered in this work, the mode frequency does not change by more than ${\sim}5\%$ if either of these boundary conditions is used.

The inner boundary condition is enforced via the shooting method, and then the system of equations is integrated for valid inner boundary conditions, yielding $\eta_r$, $\eta_{\perp}$, $\delta \hat{Q}$, and $\delta \hat{\psi}$.
Upon completion of this integration, the outer boundary condition is evaluated for a series of test values of $\sigma$, with the terms $\rho$, $h$, $\psi$, $Q$ and $\partial P / \partial r$ taken from the input hydrodynamic background.
Those values of $\sigma$ that satisfy the outer boundary condition yield eigenfrequencies $\sigma / 2 \pi$.

\subsection{Application to General Relativistic, Spherically Symmetric Models}

In this work, we post-process our core-collapse simulations with the method of TF19 to calculate the gravitational-wave eigenfrequencies of the PNS in each model. 
We utilize v1.2 of the publically-available code \code{GREAT} from the same authors to perform this analysis.
However, a key difference between their work and this work lies in the treatment of gravity in the simulations of core collapse.
In TF19, the method was applied to two-dimensional simulations mostly using an approximate relativistic treatment of gravity (``CoCoNuT''  models being the exception, see also the caption to Figure~\ref{fig:universal-relation-comparison}).
Our models from \code{Agile} are spherically symmetric using a fully general-relativistic (GR) treatment of gravity. 
These mathematically and physically different treatments of gravity require different approaches to correctly calculate the gravitational wave frequency structure.

The primary difference between these treatments of gravity is the chosen gauge. In geometric units of $c = G = 1$, the metric used in TF19 is
\begin{equation}
    ds^2 = -\alpha^2 dt^2 + \psi^4(t, r, \theta, \varphi) \left( dr^2 + r^2 d\Omega^2 \right),
\end{equation}
where $\alpha$ is the lapse function, $d\Omega^2 = d\theta^2 + \sin^2(\theta) d\varphi^2$ is the solid angle element, and $\psi(t, r, \theta, \varphi)$ is the conformal factor\footnote{Generally, this decomposition of the metric into a timelike component and flat spatial metric multiplied by a factor $\psi$ is known as a conformally-flat metric \citep{alcubierre}. Note, however, that a conformally flat metric is not actually spatially flat if $\psi$ is not constant in space.}. 
Meanwhile in \code{Agile}, the metric is
\begin{equation}
    ds^2 = -\alpha^2 dt^2 + (4 \pi r^2 \rho(t,r))^{-2} dr^2 + r^2 d\Omega^2,
\end{equation}
where $\rho(t,r)$ is the density of material in the zone at radius $r$ at time $t$.
There is no immediately obvious reason --- in particular for the PNS --- that the spatial part of the fully-general relativistic \code{Agile} metric should be flat (i.e.\ that $\psi$ should be constant in space after an appropriate coordinate transformation), and thus we need to calculate $\psi$ from the output data from \code{Agile}. 
We achieve this by finding a conformal radius $\tilde{r}$ that yields a conformally-flat decomposition of the \code{Agile} metric, and use this to calculate $\psi$. By providing $\tilde{r}$ in place of the areal radius as well as $\psi$ as inputs to \code{GREAT}, we do not have to make any modifications to \code{GREAT} to account for the different treatment of gravity in the core-collapse simulations. The details of this transformation can be found in Appendix \ref{app:spacetime}.

\subsection{Fundamental Mode Identification}

\begin{figure}
    \centering
    \includegraphics[width=\linewidth]{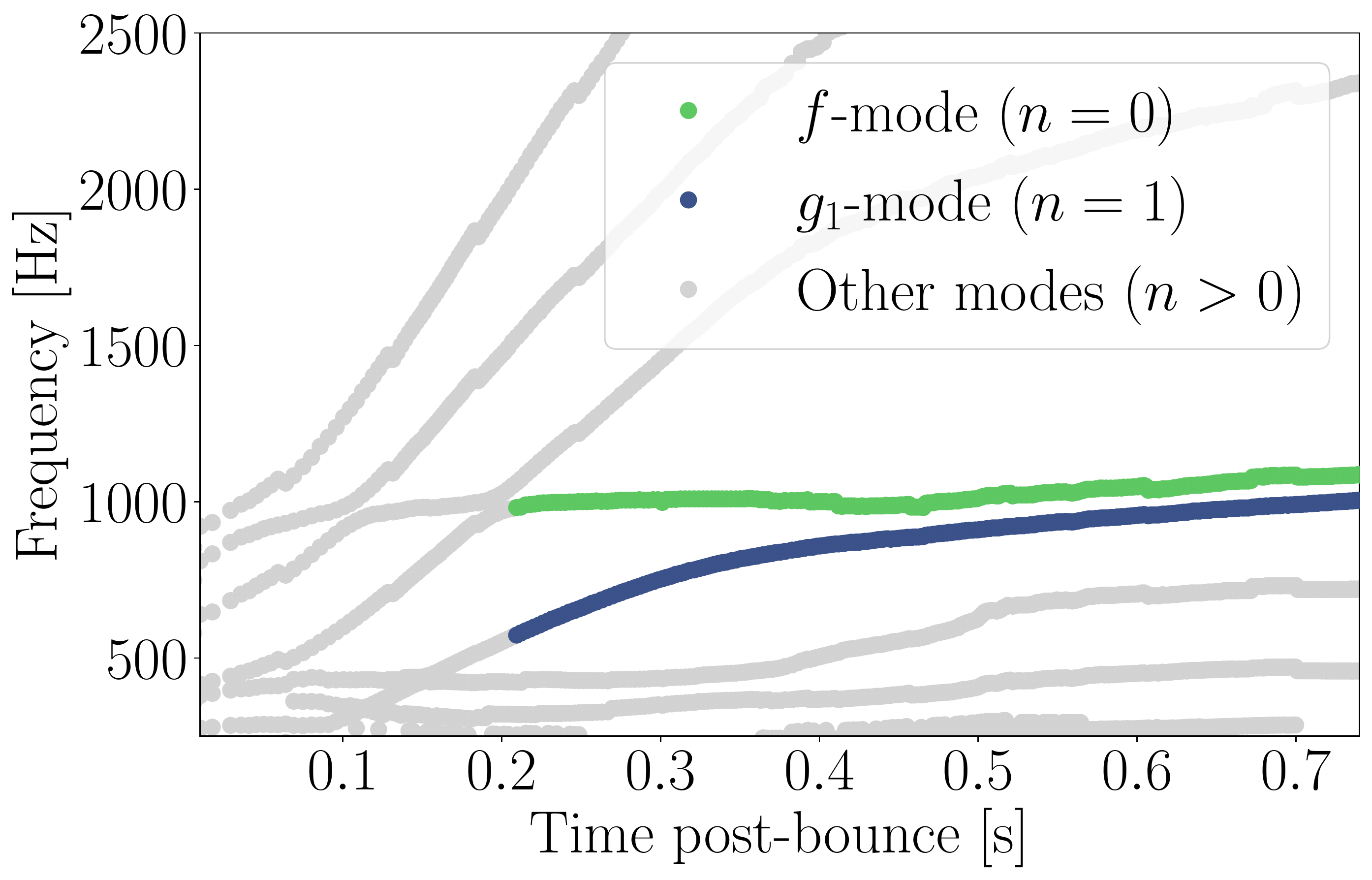}
    \caption{
        Eigenfrequencies for the \code{s19.0_SFHo} model as a function of the time post bounce. 
        The fundamental mode (green), $g_1$-mode (blue), and other eigenmodes (grey) are identified using the Cowling scheme, determined by the number of zero-crossings $n$ found in the $\eta_r(r)$ function associated with each frequency.
    \label{fig:eigenfrequencies}
    }
\end{figure}

When provided hydrodynamic background data from one of our models as input, the \code{GREAT} code yields a set of eigenfrequencies at each timestep, and a discrete function $\eta_r(r)$ associated with each eigenfrequency.
To classify these eigenfrequencies, we employ the Cowling classification scheme \citep{cowling_1941}. 
In this scheme, the mode number associated with a particular eigenfrequency is the number of zero-crossings, $n$, in the $\eta_r(r)$ function associated with that eigenfrequency. 
The fundamental ($f$) mode is the nodeless mode ($n = 0$). The $g$-modes ($p$-modes) are identified by decreasing (increasing) frequency for increasing $n$, among eigenfrequencies below (above) the $f$-mode.
Generally, buoyancy (gravity) is the restoring force for $g$-modes and pressure is the restoring force for $p$-modes.
We note, however, that there is an on-going discussion in the literature about the detailed nature of these modes (e.g. TF19 and \citet{torres-forne_universal_2019}).
In Figure~\ref{fig:eigenfrequencies}, we show the eigenfrequencies calculated with \code{GREAT} for an example model, \code{s19.0_SFHo}, with the $f$-mode (green) and the $g_1$-mode (blue) identified.
The other modes are displayed in grey. 

The goal of this classification is to find the dominant mode of the gravitational-wave emission.
Although at late times the $f$-mode of the PNS appears to be the dominant mode, at early times the $g_1$-mode dominates gravitational-wave emission (see discussion in Section~\ref{sec:avoided-crossing}).
The change of character of the dominant mode occurs due to an avoided crossing taking place typically at about $0.4$~s after bounce (see Figure~\ref{fig:eigenfrequencies}).
Alternative classifications (e.g matching classification, see TF19) based on similarity of eigenfunctions, provide much more consistent identification of the dominant mode\footnote{For comparison with the work of \cite{torres-forne_universal_2019} using the matching classification, the $f$-mode (at late times) of this work corresponds to the $^2g_2$ mode of theirs.}.
However, they require a degree of manual tuning that is not feasible to conduct for the \nns{} eigenfrequency analyses of this work.
The consequences of our choice are discussed in the next sections. 
\accepted{accepted this paragraph and footnote 5}

\section{Characteristic GW Frequencies}
\label{sec:results2}

\subsection{Time-frequency Evolution} \label{sec:results-patterns}

In this section, we investigate the evolution of the fundamental mode in the first 0.7 seconds post-bounce.
With the fundamental mode only emerging at ${\sim}0.2$ seconds post-bounce in most of our models, this yields a window of ${\sim}0.5$ seconds that lines up with the approximate window in which we expect it to be possible to reconstruct the time-frequency evolution of the core-collapse signal \citep{powell_ccsne-inference_2022, Bizouard2021, Bruel2023}.

\begin{figure*}
    \centering
    \includegraphics[width=\linewidth]{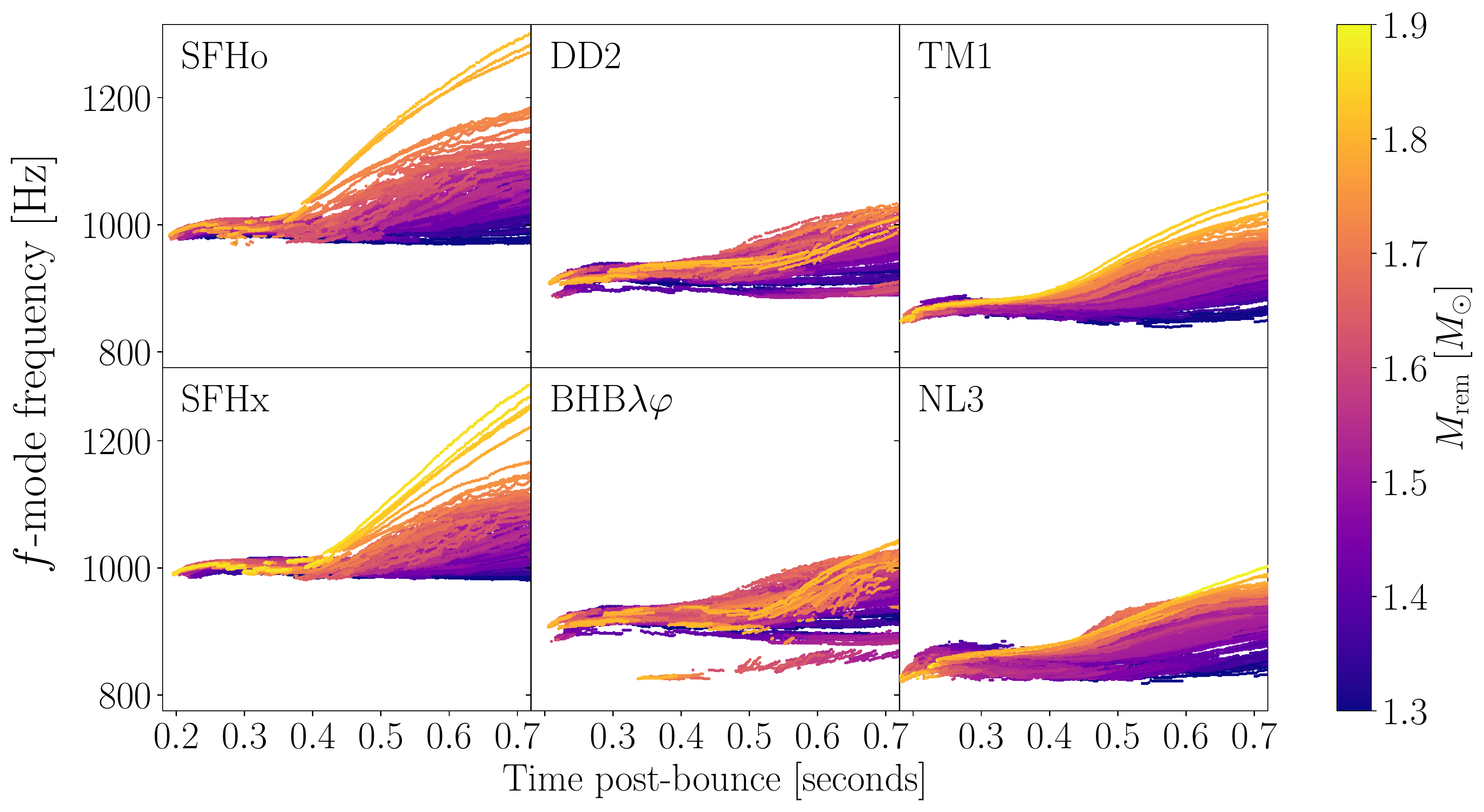}
    \caption{
        The time-frequency evolution of the $f$-mode for each model in the time interval from 0.2 to 0.7 seconds post-bounce, organized by nuclear equation of state and colored by the cold neutron star remnant mass.
    }
    \label{fig:fmode-mrem-lines}
\end{figure*}

Following the procedure for eigenfrequency calculation and mode classification described in Section \ref{sec:eigenfreqs}, we identified the $f$-mode as a function of the post-bounce time $\tpb$ for all \nns{} models that successfully exploded, yielding a discrete function $f(\tpb)$ for each model. 
In Figure \ref{fig:fmode-mrem-lines} we show for all models of this work the $f$-mode frequency as a function of the time post bounce, per equation of state and colored by the (gravitational) remnant mass $\mrem{}$ of the cold neutron star. 
Qualitatively, the results are very similar across all equations of state: the $f$-mode frequencies at early times ($\tpb \lesssim 0.4$ seconds) are tightly clustered for all models using the same EOS, whereas at later times ($\tpb \gtrsim 0.4$ seconds) there is a spread in the frequencies which correlates with the remnant mass. Additionally, we observe that the $f$-mode frequency at early times and at late times depends on the EOS. 
We also note that, for all models, the $f$-mode monotonically increases with time for much of its evolution, as seen in similar studies \citep{torres-forne_I_2018, torres-forne_II_2019, morozova_eigenfrequencies_2018}, allowing us to identify the early-time period with low $f$-mode frequencies and the late-time period with higher frequencies.

\begin{figure*}
    \centering
    \includegraphics[width=\linewidth]{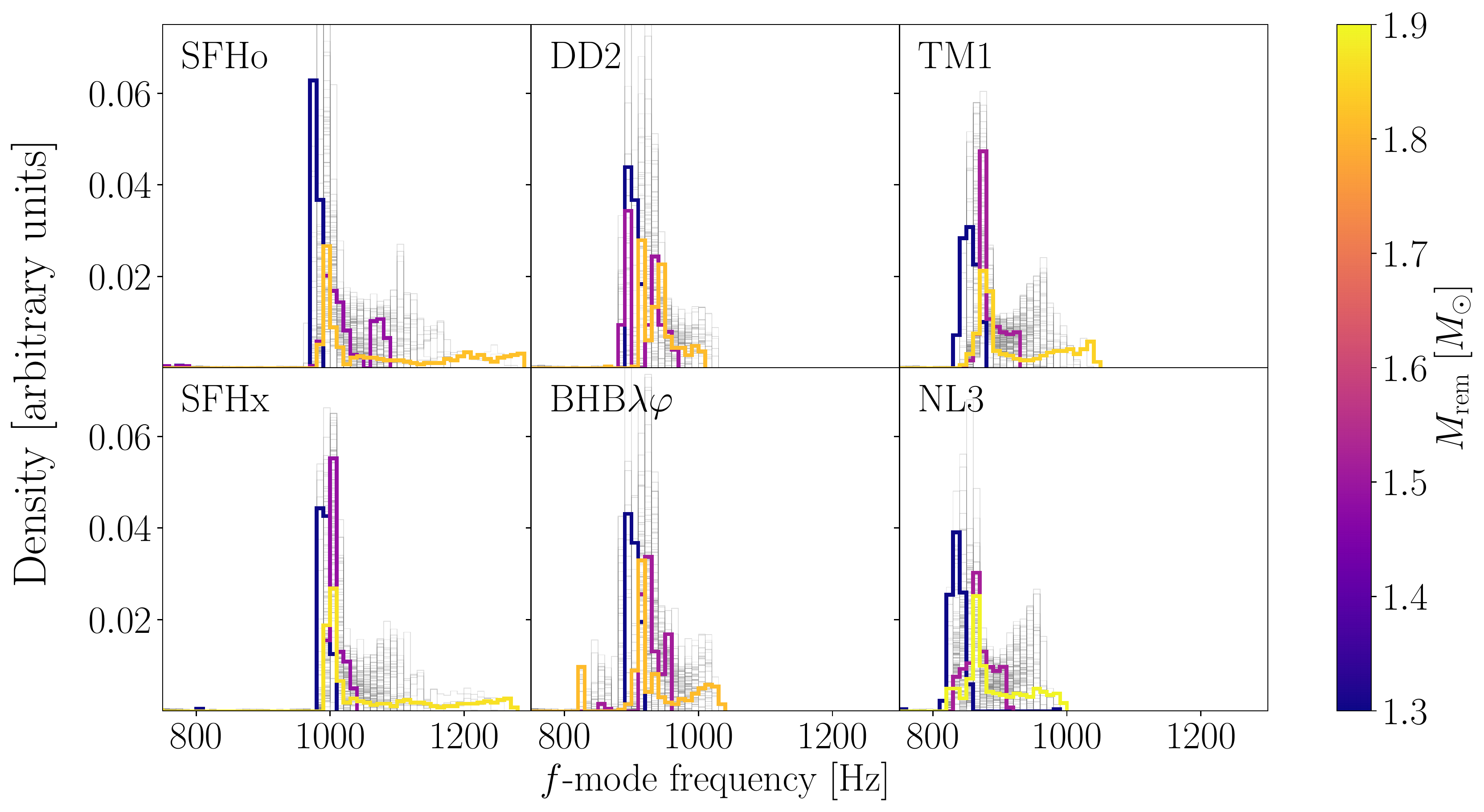}
    \caption{
        Histograms of the $f$-mode frequencies from the time interval from 0.2s to 0.7s post-bounce for each model, organized by EOS   (same data as in Figure \ref{fig:fmode-mrem-lines}). 
        Each histogram (grey or colored) corresponds to one of our simulations. 
        For each EOS, the colored histograms correspond to the models with the minimum, median, and maximum cold neutron star mass.
        The histograms use equally-spaced bins of 10 Hz in width and are normalized to have unit area. 
        Note that a low-frequency peak appears in all models. However, only models resulting in higher masses exhibit a double-peak structure in the histogram, with a second, high-frequency peak or at least a high-frequency tail.
    }
    \label{fig:fmode-mrem-hists}
\end{figure*}

In Figure \ref{fig:fmode-mrem-hists} we provide a different view of the same data as in Figure~\ref{fig:fmode-mrem-lines}, collecting the $f$-mode frequencies from 0.2s up to 0.7s post-bounce into a histogram for each model, sorted by nuclear EOS. The histograms shown in color are from the models yielding the minimum, median, and maximum remnant mass per EOS. 
This further supports the patterns we identified with Figure \ref{fig:fmode-mrem-lines}, emphasizing a low-frequency (early-time) peak whose location is dependent on equation of state and a high-frequency (late-time) tail whose location and extent correlates with the remnant mass. 
From inspecting the evolution of the $f$-mode over time in our exploding models, we conclude that there is a relationship between the structure of the proto-neutron star, as characterized by equation of state and remnant mass, and the frequency structure in the first 0.7 seconds post-bounce.

\subsection{Fit to Finite Gaussian Mixture} \label{sec:results-gmm-fits}

In the spirit of modern search pipelines for gravitational-wave signals from core-collapse supernovae such as coherent WaveBurst \citep{klimenko_cWB_2016}, we quantify the early- versus late-time features of the fundamental mode using the minimal assumptions required.
We fit the histogram of $f$-mode frequencies until 0.7 seconds post-bounce for each model to a two-component Gaussian mixture; the only input to this statistical model is the normalized density of frequency values.
A Gaussian mixture is the weighted sum of $k$ Gaussian distributions, with a probability density function of
\begin{equation} \label{eq:gmm}
    P(x | \theta) = \sum_{i = 1}^{k} \lambda_i \mathcal{N}(x | \mu_i, \sigma_i)
\end{equation}
where $\theta = \{ \mu_1, ..., \mu_k, \sigma_1, ... \sigma_k, \lambda_1, ..., \lambda_k \}$ is the set of component means $\mu_i$, standard deviations $\sigma_i$, and associated weights $\lambda_i$.
We use $\mathcal{N}$ to denote the standard Gaussian probability density function.
In a Gaussian mixture, the weights $\lambda_i$ follow a distribution conditioned on an unobserved latent variable that controls which component Gaussian an observed sample from the mixture distribution falls most closely within.
This latent variable conditioning makes Gaussian mixtures a popular choice in machine learning applications, e.g. for classification tasks \citep{viroli_deep-gmm_2019}.

Thus, a Gaussian mixture is a relevant model for quantifying the time-dependent bifurcation in behavior we identified in Figures \ref{fig:fmode-mrem-lines} and \ref{fig:fmode-mrem-hists}.
We choose to fit our frequency histograms with a two-component Gaussian mixture ($k=2$), as we observe that models with low remnant mass display one peak (captured by a Gaussian mixture where one weight $\lambda_i \sim 1$ and the other $\lambda_j \sim 0$) while models with higher remnant mass display a low-frequency peak and a high-frequency peak or tail.
In our application of the Gaussian mixture, we suspect but cannot explicitly enforce that the latent variable identifying these peaks is the post-bounce time and by extension some evolving physical parameter(s) of the proto-neutron star.
Further speculation as to the physical processes driving this temporal bifurcation will be important future work but is beyond the scope of this work.

To fit the frequency histograms, we use the expectation-maximization procedure as implemented by \code{scikit-learn} \citep{scikit-learn}.
Expectation-maximization works by repeatedly randomly or algorithmically assigning each observation to one of the two component Gaussian distributions and maximizing the likelihood of this observation over the model parameters to fit these assignments. 
This algorithm has been shown to always converge to a local optimum in $\theta$ \citep{em_dempster_1977}.
Additional details on the theory behind expectation-maximization can be found in \citet{em_dempster_1977} and implementation details can be found in the \code{scikit-learn} documentation\footnote{\url{https://scikit-learn.org/stable/modules/generated/sklearn.mixture.GaussianMixture.html}}. 

\begin{figure}
    \centering

    \begin{tabular}{c}
         \includegraphics[width=\linewidth]{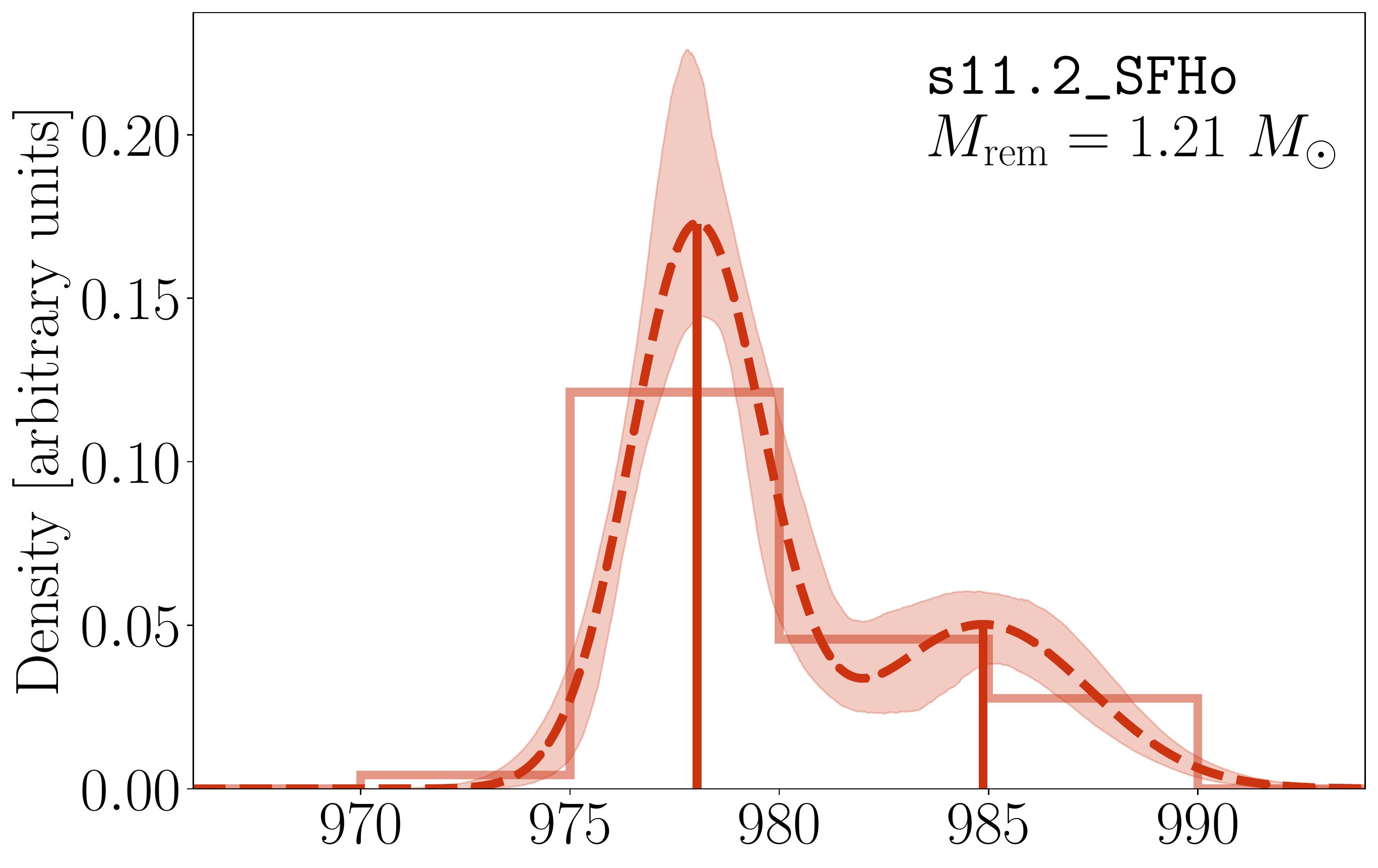}\\
         \includegraphics[width=\linewidth]{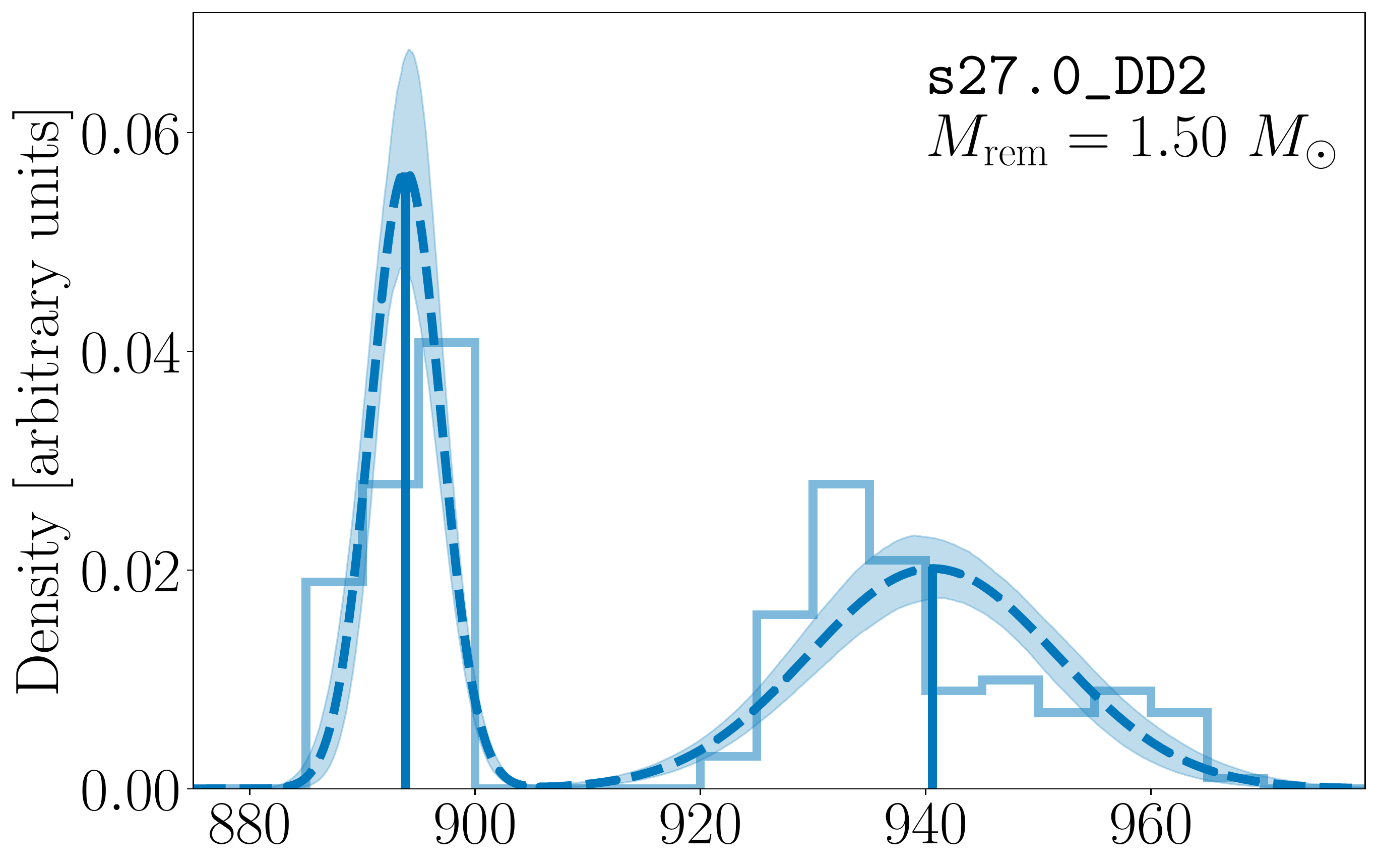}\\
         \includegraphics[width=\linewidth]{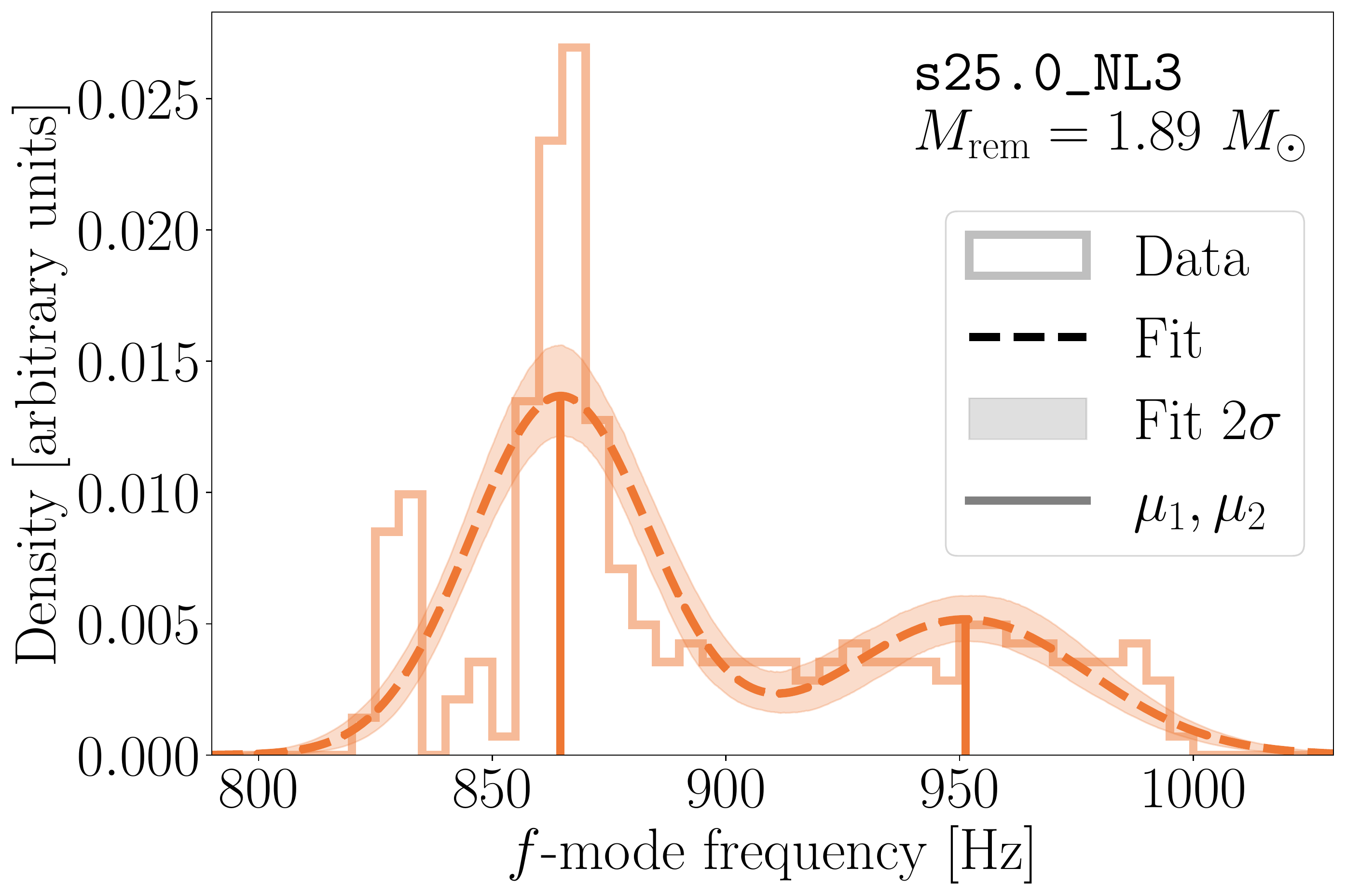}
          
    \end{tabular}
    \caption{
        Results from fitting frequency histograms to two-component Gaussian mixtures for models with the minimum (top), median (middle), and maximum (bottom) remnant mass across all our simulations. 
        We plot the histogram of frequencies from 0.2 s until 0.7 s post-bounce (solid line), with bins 5 Hz in width and normalized to have an area of one. 
        The Gaussian mixture fit to each frequency histogram via expectation-maximization (dashed line) is shown with a 2$\sigma$ confidence interval (shaded region around the dashed line). A vertical line under the lower and higher frequency peak indicate the position of $\mu_1$ and $\mu_2$, respectively.
    }
    \label{fig:gmm-fits}
\end{figure}

In Figure \ref{fig:gmm-fits}, we show frequency histograms (solid line) and fits of these probability density functions to a two-component Gaussian mixture (dashed curve) for three select models (corresponding to the minimum, median, and maximum remnant mass among all our simulations), covering the entire range of observed $f$-mode frequency structures.
The shaded area indicates the $2\sigma$ confidence region, calculated via the bootstrap algorithm described in Appendix \ref{app:bootstrap}.
We observe that while the Gaussian mixture does not perfectly replicate the probability density function for the $f$-mode from any of these models, it successfully identifies the low- and high-frequency features we observed in Figure \ref{fig:fmode-mrem-hists} through the location of the component Gaussian peaks.
We use the frequencies of the two component peaks ($\mu_1$ for the lower frequency and $\mu_2$ for the higher frequency) to characterize the time-frequency evolution of the fundamental mode. 
For a few of our models, the Gaussian mixture struggles to confidently identify these features; these difficulties are documented in Appendix~\ref{app-poor-fits}.

A natural question is whether one can define characteristic frequencies of the early-time and late-time regimes directly from the frequency evolution, without performing any fit to a Gaussian mixture model.
In an observational context, such an approach requires an absolute calibration of the time axis, e.g., by knowing the exact time of bounce, which poses an observational challenge.
Bounce may be difficult to identify due to the stochastic nature of core-collapse gravitational waveforms generated by proto-neutron star oscillations \citep{abdikamalov_gw-ccsne-review_2022}; while simultaneous neutrino signals from core-collapse may identify bounce time with a precision of ${\sim}10$~{ms}~\citep{Pagliaroli:2009,Halzen:2009}, they can also introduce additional uncertainty.
Moreover, we may not hear the full time-frequency evolution of the gravitational-wave signal unless it is especially loud \citep{szczepanczy_cWB-ccsne-detect_2021, powell_ccsne-inference_2022}.
Additionally, such a method would require a model for the temporal evolution of the amplitude, which has to be derived from two- or three-dimensional simulations. 
Given the current broad landscape of core-collapse simulations it may be difficult to find a universal model for identifying bounce.

Fitting the collection of observed frequencies from a core-collapse signal to a Gaussian mixture does not suffer the difficulty of identifying an absolute reference time (e.g., time of bounce); instead, an educated guess as to whether the observed segment of data corresponds to the low- or high-frequency peak is sufficient to conduct inference with the relevant relations.
Thus, the Gaussian mixture model is advantageous as it requires only information on the collection of frequencies observed and not the time-frequency evolution. 
However, in identifying characteristic frequencies from simulation data, we must impose a cut on the timespan of frequencies considered.
The late-time characteristic frequency $\mu_2$ may be particularly sensitive to this choice of timespan, as
the $f$-mode frequency can continue to increase beyond the first 0.7~s post-bounce (c.f. Figure~\ref{fig:fmode-mrem-lines}) and hence a longer timespan might increase $\mu_2$.
While we do not expect there to be significant power in these frequencies beyond 0.7 seconds post-bounce \citep{powell_gravitational_2019}, we cannot say this conclusively without knowledge of the gravitational-wave strain.
The time-dependence from our identification of $\mu_1$ and $\mu_2$ in simulation data could be fully relaxed by, for example, weighting the frequency histograms by the gravitational-wave power in each frequency bin, which would require future work with multi-dimensional models. 
We conclude that our approach is a step towards a time-agnostic way to characterize the GW signal from core-collapse for the purposes of parameter estimation.

\section{From Frequencies to Neutron Star Structure}
\label{sec:results3}

We applied the Gaussian mixture model to all \nns{} core-collapse simulations resulting in a successful explosion and the formation of a neutron star, and identified for each model a low ($\mu_1$) and a high ($\mu_2$) characteristic frequency using the procedure detailed in Section \ref{sec:results-gmm-fits}. 
Next, we investigate the implications for interpreting future GW signals and the constraints they might provide on the nuclear EOS and the structure of the neutron star.

\subsection{Simultaneous Structure Relations} \label{sec:sim-relations}

\begin{figure*} \label{fig:mu_results}
\includegraphics[width=0.49\linewidth]{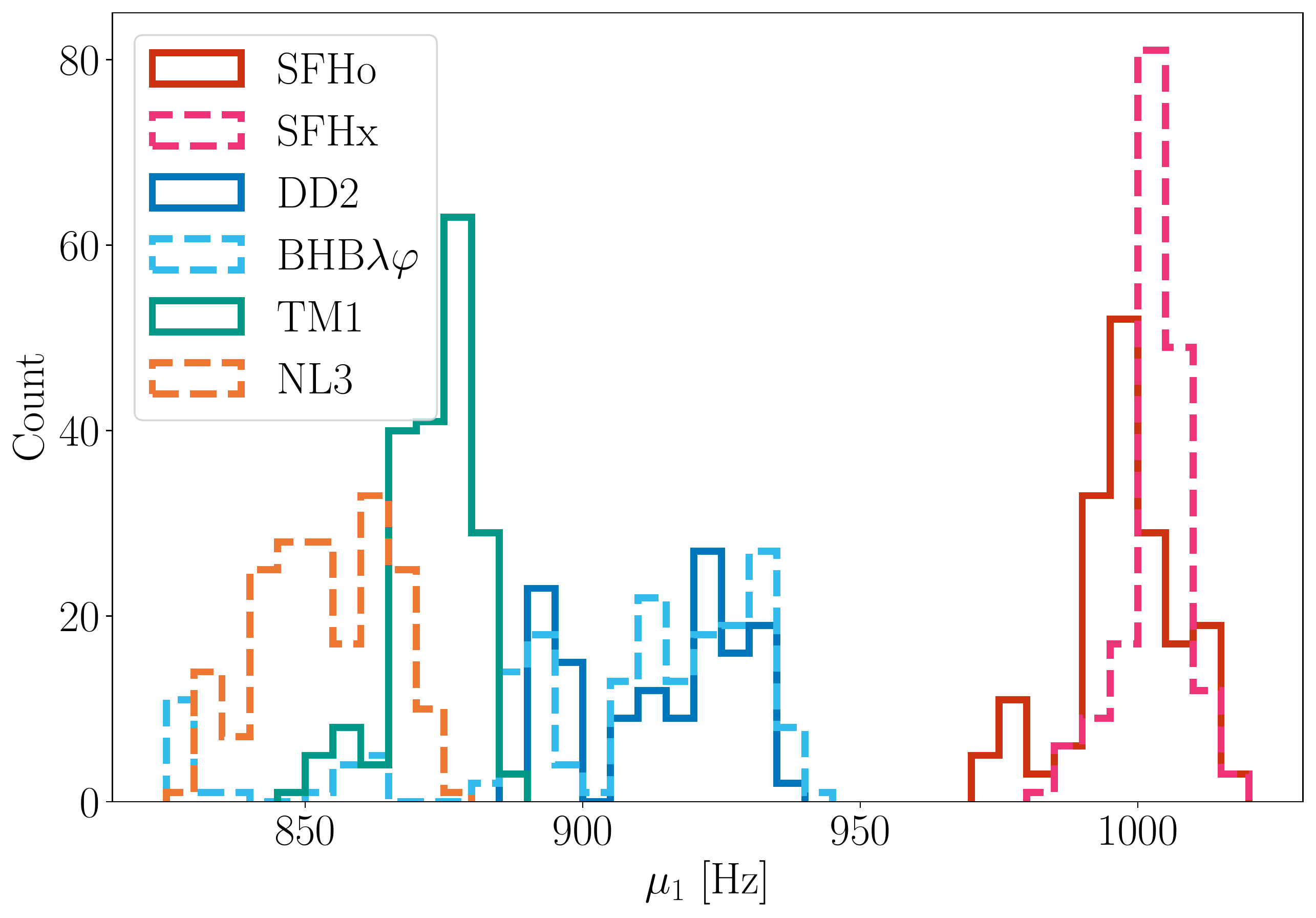}
\includegraphics[width=0.49\linewidth]{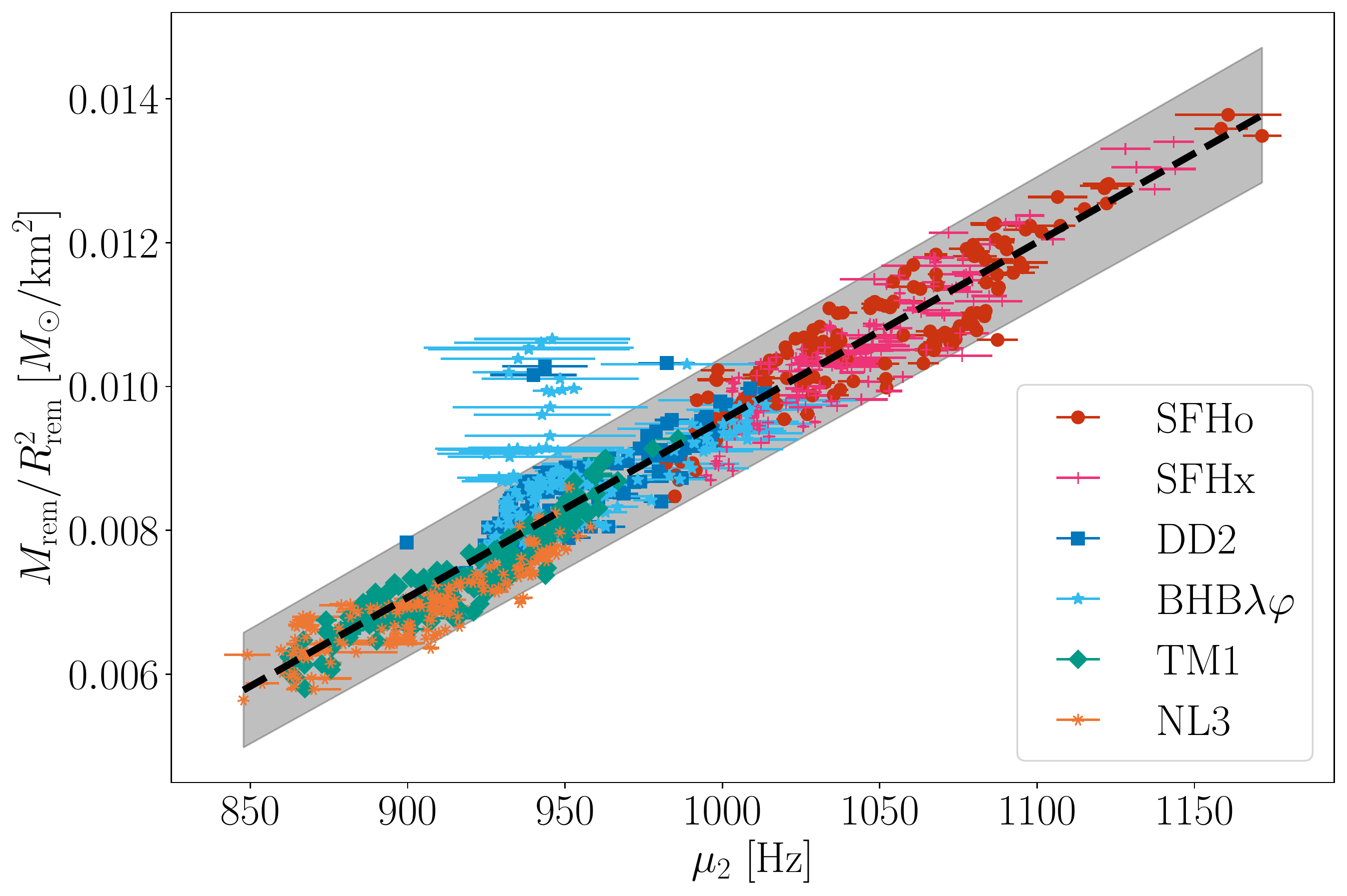}
\caption{
    Left: Histogram of the low characteristic frequencies $\mu_1$ for each EOS. Each histogram is plotted with bins 5 Hz in width. 
    Right: Surface gravity of the remnant $\mrem{} / \rrem{}^2$ (in \msun{} / km${}^2$) plotted as a function of the high characteristic frequency $\mu_2$ (points).
    The error bars on each $\mu_2$ point come from bootstrapping procedure.
    The dashed black line is a linear fit of $\mrem{} / \rrem{}^2$ as a function of $\mu_2$, with a 2$\sigma$ standard error region shaded in grey; the coefficient of determination is $r^2 = 0.961$, indicating a good linear fit.
}
\end{figure*}

For the early-time regime (until ${\sim}0.4$ seconds post-bounce, characterized by $\mu_1$), during which the frequencies are within a narrow range for any given EOS, we can identify three classes of EOS based on the $\mu_1$-frequency: 
(i) TM1 \& NL3 with frequencies below ${\sim}900$ Hz, 
(ii) DD2 \& BHB$\lambda \varphi$ with frequencies between ${\sim}875 - 950$ Hz, 
and 
(iii) SFHo \& SFHx with frequencies between ${\sim}975 - 1025$ Hz. The full histogram of all values of $\mu_1$ for each EOS is shown in the left panel of Figure~\ref{fig:mu_results}.
Thus, when we hear a gravitational-wave signal from a core-collapse supernova, we may be able to rule out some equations of state based on the characteristic frequency $\mu_1$, particularly if it falls towards the extremes of the frequency range found in this work.

For the late-time regime (after ${\sim}0.4$ seconds post-bounce, characterized by $\mu_2$), Figure~\ref{fig:fmode-mrem-lines} suggests that the remnant mass plays a role in tuning the emission of the proto-neutron star during core-collapse.
Prior work \citep{muller_fpeak_2013,torres-forne_universal_2019} also found that the surface gravity $M / R^2$ of the PNS can be directly correlated with the frequencies of this gravitational-wave emission.
In the right panel of Figure~\ref{fig:mu_results}, we plot the surface gravity of the (cold) remnant neutron star against $\mu_2$. We calculate $\rrem{}$ from $\mrem{}$ using the mass-radius relations for each equation of state; we note that, for the remnant masses obtained in our models, there is no degeneracy in the mass-radius relation (cf. Figure~\ref{fig:ns-coverage}).
While there is some substructure within this result, there is a clear linear correlation between $\mrem{} / \rrem{}^2$ and our characteristic late-time frequency that is independent of the nuclear equation of state.
We fit a linear relation for $\mrem{} / \rrem{}^2$ as a function of $\mu_2$ using the \code{scipy.stats.linregress} routine, yielding a slope of $\muslope{} \pm \muslopeerr{} \ M_{\odot} \mathrm{km}^{-2} \mathrm{Hz}^{-1}$ and an intercept of $\muintercept{} \pm \muintercepterr{} \ M_{\odot} \mathrm{km}^{-2}$.
The error in our fit is dominated by the standard error of the intercept.
Thus, given only $\mu_2$, we can recover the remnant's surface gravity $\mrem{} / \rrem{}^2$ at the $2\sigma$-level to within $\pm \mrerr{}$ $M_{\odot}$ km$^{-2}$ or within ${\sim}10 \%$ even at the smallest values of the surface gravity.
Of course, since our method for characterizing the GW frequencies is not yet fully agnostic to their temporal evolution, $\mu_2$ alone is not a measurable quantity; further effort is necessary to find a single, physically-relevant frequency like $\mu_2$ that can be equally identified in simulations like ours and in an observed time series of GW frequencies.

Finally, we note that there are a collection of models which lie outside of the 2$\sigma$ confidence region for this fit, however, these are driven by non-physical features in the frequencies, as detailed in Appendix~\ref{app-poor-fits}.
We also repeat this analysis using a different approach to characterizing the early- and late-time frequencies.
We replace $\mu_1$ and $\mu_2$ with $f$-mode frequencies at particular times, to check whether the Gaussian mixture fits introduced unexpected correlations into our data (see Appendix \ref{app:single-times}).
We find that the results obtained with both methods are consistent with each other.

\subsection{Dominant Frequency of Emission} \label{sec:avoided-crossing}

We have so far identified two characteristic frequencies of the nodeless resonant mode of emission, i.e. of the $f$-mode.
However, we must ask whether the $f$-mode corresponds to the \textit{dominant} mode of emission, in the sense that it traces the highest amplitude of emission at each time. 
If so, then $\mu_{1,2}$, which characterize the $f$-mode, could be measurable in practice.
In this section, we discuss the implications for our analysis if the $f$-mode is not the dominant contribution to the observed GW signal.

Similar studies with 2D/3D models (and thus can also access the GW amplitude evolution) have established phenomenology that we can use to address this question.

In particular, TF19 and \citet{sotani_avoided_crossing_2020} observed that the mode labels provided by the Cowling scheme are at times inconsistent with the behavior of the eigenfrequencies and their associated radial eigenfunctions.
Both studies observed an avoided crossing between the $g_1$-mode and the $f$-mode frequencies, at which point their radial eigenfunctions either converge or swap behavior entirely (in terms of the regions of the star where their eigenfunctions peak; see, for example, Figure~5 of \citet{torres-forne_II_2019} or Figure~3 of \citet{sotani_avoided_crossing_2020}). 
Correspondingly, these studies have found that the dominant mode of emission typically follows the $g_1$-mode immediately after bounce, and then follows the $f$-mode after the avoided crossing. 
Thus, the avoided crossing between the $f$- and $g_1$-modes serves as an approximate marker for how to identify the dominant mode of gravitational-wave emission. 
We note that \citet{torres-forne_universal_2019} reclassifies the modes with a custom scheme developed in TF19, and identifies the dominant gravitational-wave mode as a $g$-mode (named there $^2g_2$) and the mode avoiding the crossing at $~0.4$~s a different g-mode ($^2g_3$).
Their scheme is designed so that the radial eigenfunction associated with a mode label has a consistent shape throughout the course of a core-collapse simulation.
However, we want to stress that the difference between our work and the work of TF19 and \citet{torres-forne_universal_2019} is just the labeling of the modes, in particular at the avoided crossing. While the names differ, the phenomenology of the avoided crossing is ultimately consistent, as their $^2g_2$ mode corresponds to our $g_1$-mode before the avoided crossing, and our $f$-mode after.

In all of our models, we observe this avoided crossing at ${\sim}0.4$ seconds post-bounce; see, for example, Figure \ref{fig:eigenfrequencies}, where the $f$-mode (blue) and $g_1$-mode (purple) nearly meet, and after which the $f$-mode suddenly begins increasing.
So, guided by multi-D phenomenology, the frequencies that we call the $f$-mode after ${\sim}0.4$ seconds post-bounce are likely the dominant frequencies of emission, while prior to the avoided crossing, they are not. 

Since $\mu_2$ and $\mu_1$ appear to characterize these two distinct temporal regimes, we expect $\mu_2$ to characterize the mode of emission that dominates the GW amplitude, while $\mu_1$ characterizes a mode that may not have a large enough amplitude to be observable in practice.
Nevertheless, some numerical simulations show the presence of this mode in the gravitational wave signature \citep[e.g.][]{torres-forne_universal_2019} or a gap in the spectrograms related to the avoided frequency location \citep{morozova_eigenfrequencies_2018,Bruel2023} that could be used as a proxy for $\mu_1$.

\begin{figure}
    \centering
    \includegraphics[width=\linewidth]{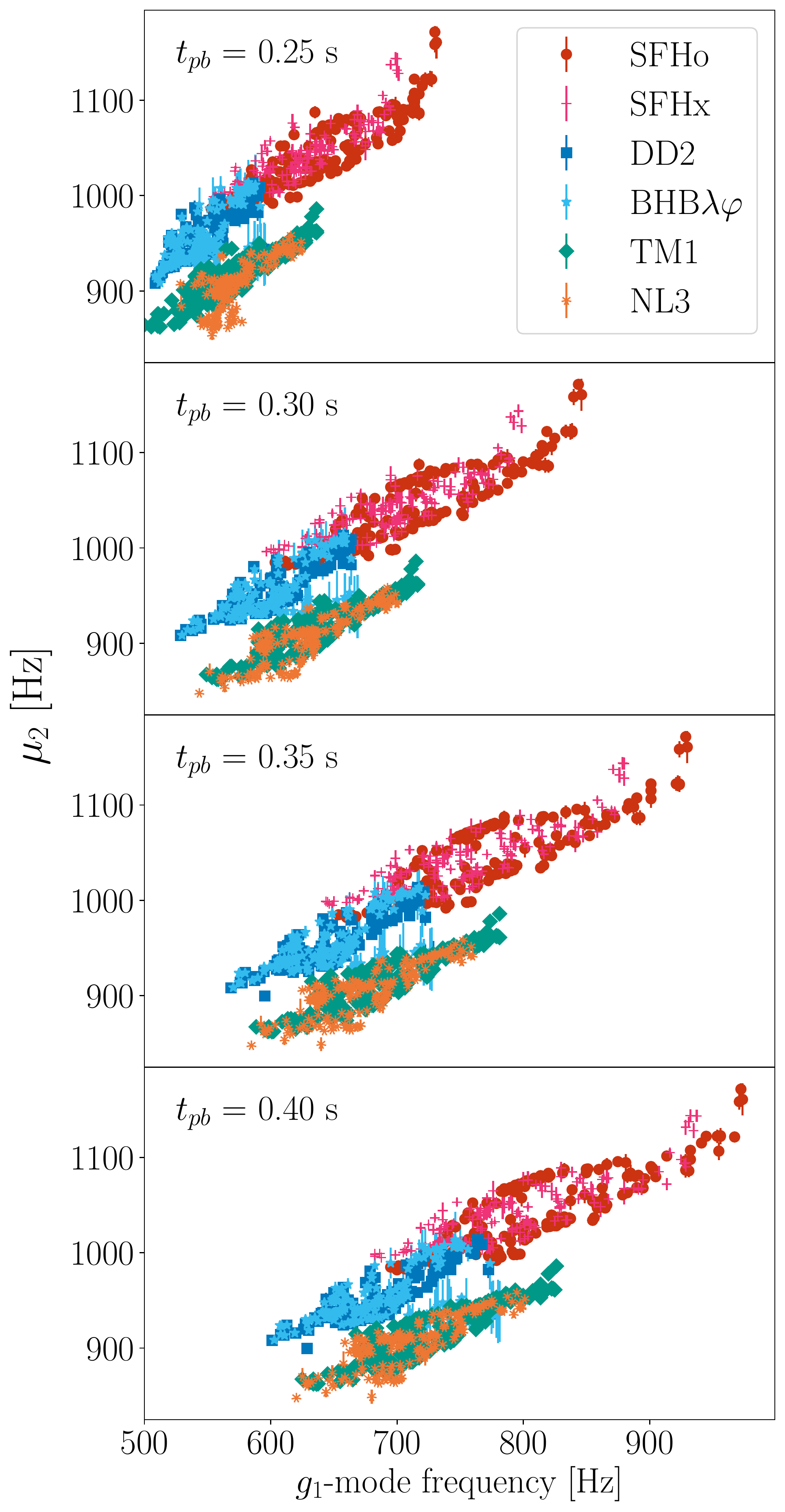}
    \caption{
        Late-time characteristic frequency $\mu_2$ versus the $g_1$-mode frequency at four times post-bounce ($t_{pb}$) approximately before the avoided crossing between the $f-$ and $g_1$-mode frequencies. At each time, there are clear regions of this frequency-frequency space that exclude a majority of the equations of state studied in this work. 
    }
    \label{fig:g1mr2-eos-4panel}
\end{figure}

In light of this phenomenology, we now seek an alternative frequency to $\mu_1$ to characterize the early-time GW emission.
Possibly, we could stitch together the $g_1$-mode frequencies prior to the avoided crossing time with the $f$-mode frequencies after this time, however, this requires manual identification of the avoided crossing time in each model which is infeasible for our large suite of models. 
For now, we can simply select the $g_1$-mode at a few times before the avoided crossing to characterize the early-time emission. 
We reanalyze our data with this different choice of early-time characteristic frequency. Specifically, we select the $g$-mode frequency at four different post-bounce times (0.25, 0.3, 0.35, and 0.4 s), some or all of which are before the avoided crossing time for each model. 
We pair these frequencies with the late-time characteristic frequency $\mu_2$ (which should characterize an observable portion of the GW signal) to create a frequency-frequency space in which an observation of gravitational waves from a core-collapse event could be placed. 
In Figure \ref{fig:g1mr2-eos-4panel}, we plot $\mu_2$ versus these four alternate choices for the early-time frequency ($g_1$-mode at 0.25, 0.3, 0.35, and 0.4 s post-bounce).
We see that combining this early-time and late-time frequency information uniquely identifies different classes of the nuclear equation of state.
When combined with an EOS-independent measurement of $\mrem{} / \rrem{}^2$ with $\mu_2$ alone, one could simultaneously constrain the mass, radius, and equation of state of the proto-neutron star.
For that analysis one would need to accurately identify the time of bounce. This could be done following the approach of \cite{Bizouard2021,Bruel2023}.

\subsection{Comparison to Other Work} \label{sec:relation-comparison}

\begin{figure}
    \centering
    \includegraphics[width=\linewidth]{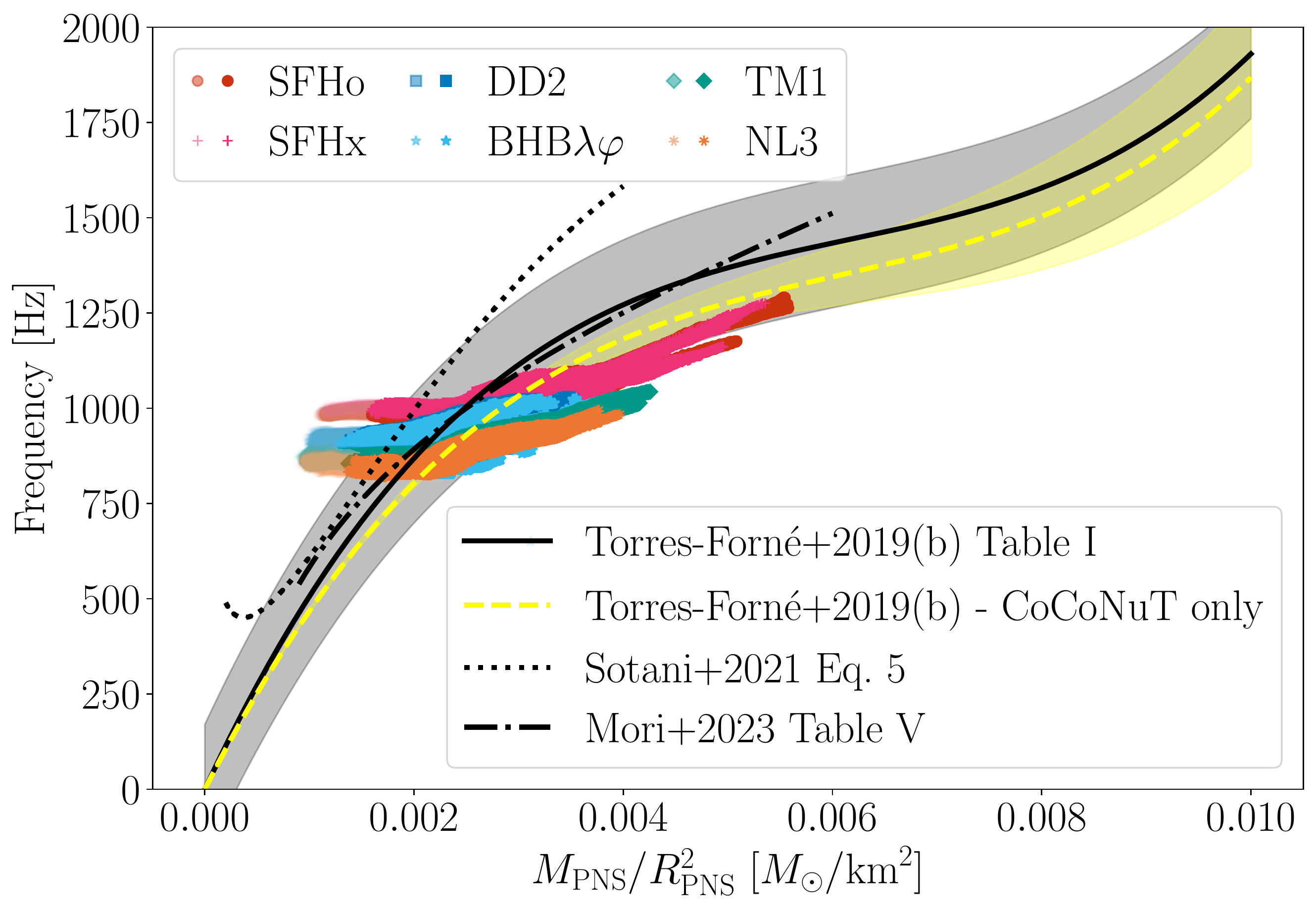}
    \caption{
    The $f$-mode frequency at each time against the hot proto-neutron star surface gravity $\mpns{} / \rpns{}^2$ at the same times from the suite of exploding models considered in this work (colored points).
    We overplot fits of the dominant frequency of gravitational-wave emission as a function of $\mpns{} / \rpns{}^2$ from the erratum \citep{torres-forne_universal_erratum} to \citet{torres-forne_universal_2019} (solid black line fit, shaded region for $2\sigma$ confidence interval), and a similar fit only considering the ``CoCoNuT'' general-relativistic models used in this same work (dashed yellow line fit and shaded confidence interval).
    This fit takes the same form as the fit to all of the models (see Table~1 of \citet{torres-forne_universal_2019}), with $a = 0$~Hz, $b = 5.88 \times 10^5$~Hz \msun{}$^{-1}$ km$^{2}$, $c = -86.2 \times 10^6$~Hz \msun{}$^{-2}$ km$^{4}$, and $c = 4.67\times 10^9$~Hz \msun{}$^{-3}$ km$^{6}$. 
    We also include a similar fit from \citet{Sotani.2021} in the form of their Equation~5 (dotted black line), and the fit from Table~V of \citet{Mori2023} labeled as ``0.2-20'' that extends to 20 seconds post-bounce (dash-dotted black line).
    As we expect the $f$-mode after the avoided crossing at ${\sim}0.4$ seconds to correspond to the dominant mode of emission, points from times after the avoided crossing are comparable to the displayed fits from the literature.
    Points prior to the avoided crossing at ${\sim}0.4$ seconds post-bounce are partly transparent and included for completeness.
    \label{fig:universal-relation-comparison}
    }
\end{figure}

So far, we have identified a linear relationship between the surface gravity of the cold neutron star, $\mrem{} / \rrem{}^2$, and the characteristic frequency of the late-time gravitational-wave signal, $\mu_2$, that is independent of the equation of state.
Similar relations have been previously presented for the hot proto-neutron star, in \citet{torres-forne_universal_2019}, \cite{Sotani.2021} and \cite{Mori2023} 

In \citet{torres-forne_universal_2019}, the authors conducted a linear perturbation analysis of the region inside the shock in 25 1D simulations from two different supernova simulation codes.
None of the models they analyze successfully explode.
They use a subset of the same \code{KEPLER} progenitors employed in this work, with solar-metallicity progenitors spanning 11 - 75~\msun{} (and one $10^{-4}$ metallicity progenitor). 
The majority of their models used an approximate treatment of gravity and a single nuclear equation of state, LS220 \citep{ls220}; they included two additional models with the BHB$\lambda$ equation of state (a variant of BHB$\lambda\varphi$, without $\varphi$-mesons), as well as three additional models with other equations of state.
While the linear perturbation analysis was also done using \code{GREAT}, they employ a different outer boundary condition. They impose that $\eta_r = 0$ at the edge of the shock (as opposed to setting $\Delta P = 0$ as in this work, c.f.\ Section~\ref{sec:eigenfreqs}).

\citet{Sotani.2021} take a different approach, modeling a single 20~\msun{} progenitor in 2D with an approximate treatment of gravity, and four different equations of state including DD2 and SFHo.
All of their models explode (H.~Sotani 2023, private communication).
Their linear perturbation analysis adopts the Cowling approximation (only allowing the lapse to vary), but employs the same boundary condition of $\Delta P = 0$ at the proto-neutron star surface as in this work.
Generally, their methodology appears to be consistent with \code{GREAT} at the ${\sim}10\%$ level \citep{Sotani.2020}.

The recent work by \citet{Mori2023} presents results for one exploding low-mass progenitor (9.6 \msun{}).
They also use the \code{GREAT} code to compute the eigenfrequencies, and also impose boundary conditions at the PNS surface. However,  they focus on the long term evolution of the eigenfrequencies

In Figure~\ref{fig:universal-relation-comparison} we show the relations identified in each of these works together with the $f$-mode frequency vs. hot proto-neutron star surface gravity from our suite of models. Frequencies from the time interval 0.4 to 0.7 s post-bounce (i.e.\ after the avoided crossing, see also Section \ref{sec:avoided-crossing}) are shown in fully opaque symbols; frequencies from times before 0.4 s post-bounce are shown with semi-transparent symbols.
Our models occupy a (relatively small) subset of the frequency -- surface gravity space indicated by the various universal fit relations. The points corresponding to times before the avoided crossing (semi-transparent symbols) agree the least with the universal relations from the literature, both in their location in the frequency -- surface gravity space and in the shape of their evolution with time (and hence increasing surface gravity). 

We do not expect a perfect match between the universal relations from the literature and our models, as the underlying simulations of core collapse differ in their underlying physical and numerical assumptions.
For example, our simulations are based on fully general relativistic hydrodynamics. Only a subset of models in \citet{torres-forne_universal_2019}, specifically those from the ``CoCoNuT'' code \citep{CoCoNut}, also employ general-relativistic hydrodynamics.
We also include the fit to these models only (yellow line and shaded confidence interval) in the figure, which shows slightly lower frequencies for a given surface gravity than the fit to all models (black solid line).
This fit and our models match somewhat better, pointing to the importance of how gravity is treated in the simulations of core collapse for the resulting GW frequencies.
There are other differences between CoCoNuT and our simulation code: CoCoNut uses a leakage scheme for neutrino transport, while our code employs the IDSA scheme for electron-flavor neutrinos and spectral leakage for heavy-flavor neutrinos (see Section \ref{sec:models}). 

Yet another aspect to consider is that in our work we only include simulations of core collapse that resulted in a successful explosion.
This is also the case for \citet{Sotani.2021} and \citet{Mori2023}, but not for \citet{torres-forne_universal_2019}.
The range of hot PNS surface gravity values covered by our models is comparable to that of \citet{Sotani.2021} and \citet{Mori2023}, all of which are smaller than the range covered by \citet{torres-forne_universal_2019}. 
Generally, the surface gravity of the proto-neutron star will increase over time as the PNS continues to accrete material (especially for non-exploding simulations), and at later times shrinks in radius from neutrino emission. 
Thus, the relations identified in \citet{torres-forne_universal_2019} cover a larger range of $\mpns{} / \rpns{}^2$ as their models do not explode, in contrast to the other works and our work.

Finally, our analysis exhibits some dependence on the equation of state in the differential evolution of the GW frequency versus the surface gravity (cf.\ the vertical offsets between groups of points for different EOS choices in Figure~\ref{fig:universal-relation-comparison}). This points to a further difference between this work and the literature. Our work spans a larger range of initial stellar masses and metallicities and more different nuclear equations of state. However, without a systematic exploration of all these confounding factors we cannot identify how they individually contribute to overall differences we find.

\section{Summary} \label{sec:summary}

In this work, we constructed a self-consistent suite of predictions for the time-frequency evolution of gravitational waves generated by perturbations of dense nuclear material during stellar collapse.
Using the \code{PUSH} method, we modeled \nns{} successful neutrino-driven core-collapse supernovae in fully-general relativistic hydrodynamics spanning six nuclear equations of state using progenitors at three different metallicities and ZAMS masses from \minmass{} \msun{} to \maxmass{} \msun{}.
Then, we performed a general-relativistic linear perturbation analysis of the proto-neutron star in each model using the \code{GREAT} code to compute eigenfrequencies of the proto-neutron star over time.
We saw that the fundamental ($f$) mode of emission had two distinct regimes in time, and it appeared that the frequency evolution in each regime was controlled by some combination of the equation of state and remnant mass.

We developed a method for identifying characteristic frequencies of the gravitational-wave signal by fitting the histogram of $f$-mode frequencies to a two-component Gaussian mixture.
We extracted two separate frequencies, $\mu_1$ and $\mu_2$, to characterize the early- and late-time frequency evolution, respectively. 
We found that $\mu_1$ distinguishes between different classes of the equation of state. 
Simultaneously, $\mu_2$ correlates linearly with the surface gravity of the remnant neutron star, allowing for an equation of state-independent estimation of $\mrem{} / \rrem{}^2$ to within ${\sim}10\%$ of its true value.

Due to an avoided crossing, we expect the $f$-mode to dominate the gravitational-wave emission only after ${\sim}0.4$ seconds post-bounce, while the $g_1$-mode should dominate the emission prior to this time.
Thus, $\mu_2$ characterizes potentially observable frequencies, while $\mu_1$ may not characterize an observable mode of emission.
However, re-interpreting our data by replacing $\mu_1$ with the $g_1$-mode at select times prior to $0.4$ seconds, still allows us to recover information on the equation of state in the early-time signal.
In particular, by placing each modeled supernova in the space of $\mu_2$ versus the early-time $g_1$-mode, we can still clearly distinguish different classes of the equation of state.
By combining frequency information from the early- and late-time gravitational-wave emission of the proto-neutron star, one could simultaneously measure the surface gravity of the remnant neutron star and the nuclear equation of state.

Would these characteristic frequencies be observable in practice?
Without access to amplitude information, we cannot answer this question definitively.
However, recent work has studied the detectability of features in simulated gravitational wave signals from 2D and 3D supernova models using modern gravitational-wave search pipelines. For example, \citet{gw_ccsne_current_gossan2016} and later \citet{szczepanczy_cWB-ccsne-detect_2021} found that non-rotating core-collapse signals could be detectable from galactic events, within the Milky Way.
Using the BayesWave algorithm \citep{bayeswave}, which reconstructs the signal and detector noise with minimal assumptions about signal phenomenology, \citet{raza_bayeswave_ccsne_reconstruct_2022} investigated the ability to confidently reconstruct the signal waveform from 2D and 3D core-collapse models.
For a galactic, non-rotating supernova whose location on the sky is identified by coincident electromagnetic and/or neutrino observations, \citet{raza_bayeswave_ccsne_reconstruct_2022} finds that it is difficult to reconstruct most of the signal but \textit{can} confidently identify ${\sim} 2$ temporally- and frequentially-distinct\footnote{Since the dominant frequency monotonically increases, the signal at two different times will also be dominated by two different frequencies.} features (see Figure 5 of that work).
The data reduction method that we developed in Section \ref{sec:results-gmm-fits} would confidently identify these features of the signal as an early- and late-time characteristic frequency.
When combined with the results from our suite of core-collapse models, our work yields a self-consistent framework for constraining the hot equation of state, and the cold mass and radius of the remnant neutron star from its gravitational-wave emission during stellar core collapse.

Parameter studies of proto-neutron star asteroseismology like this work are a necessary step towards conducting parameter estimation on an eventual core-collapse gravitational-wave signal, as has been been studied in \cite{Bizouard2021} and \cite{Bruel2023}. 
In particular, understanding how the results from PNS seismology depend on the assumptions made in the modeling of core collapse is necessary for contextualizing the conclusions drawn from a future GW signal. While the current work is focused on exploding models, an equivalent analysis of non-exploding models can performed in the future. In parallel, multi-dimensional studies of core collapse are needed to disentangle how model assumptions affect the resulting GW signal.

\begin{acknowledgements}
We acknowledge fruitful discussions with Somdutta Ghosh, Sanjana Curtis, Charles Stapleford, Andrew Connelly, Grant Sherrill, Thomas Steckmann, and Celine Wang.
The work at NC State was supported by United States Department of Energy, Office of Science, Office of Nuclear Physics (award number DE-FG02-02ER41216).
This material is based upon work supported by NSF's LIGO Laboratory which is a major facility fully funded by the National Science Foundation.
Throughout their time at NC State, NEW was supported by a Park Scholarship; at MIT, NEW is supported by the La Gattuta Physics Fund and the Henry W. Kendall (1955) Fellowship Fund.
The work at Los Alamos National Laboratory (LANL) was supported through the Laboratory Directed Research and Development program under project number 20220564ECR. LANL is operated by Triad National Security, LLC, for the National Nuclear Security Administration of U.S. Department of Energy (Contract No. 89233218CNA000001).This work is approved for unlimited release with LA-UR-23-21532.
The work of ATF and PCD was funded by the Spanish Agencia Estatal de Investigaci\'on (Grants No. PGC2018-095984-B-I00 and PID2021-125485NB-C21) funded by MCIN/AEI/10.13039/501100011033 and ERDF A way of making Europe and by the Generalitat Valenciana (PROMETEO/2019/071).

\software{\code{Agile} \citep{Liebendoerfer.Agile},
          \code{matplotlib} \citep{matplotlib}, \code{numpy} \citep{numpy}, \code{scipy} \citep{2020SciPy-NMeth}, \code{pandas}  \citep{reback2020pandas, mckinney-proc-scipy-2010}, \code{scikit-learn} \citep{scikit-learn}, ``vibrant'' color scheme from \citet{paul-tol-colors}
}
          
\end{acknowledgements}

\clearpage
\bibliography{refs-main,refs-multi-d,refs-pns-asteroseismology,refs-stats,refs-ccsne-gw-astronomy,refs-CF-added,refs-software}

\begin{thebibliography}{}
\expandafter\ifx\csname natexlab\endcsname\relax\def\natexlab#1{#1}\fi
\providecommand{\url}[1]{\href{#1}{#1}}
\providecommand{\dodoi}[1]{doi:~\href{http://doi.org/#1}{\nolinkurl{#1}}}
\providecommand{\doeprint}[1]{\href{http://ascl.net/#1}{\nolinkurl{http://ascl.net/#1}}}
\providecommand{\doarXiv}[1]{\href{https://arxiv.org/abs/#1}{\nolinkurl{https://arxiv.org/abs/#1}}}

\bibitem[{{Abbott} {et~al.}(2018){Abbott}, {Abbott}, {Abbott}, {Acernese},
  {Ackley}, {Adams}, {Adams}, {Addesso}, {Adhikari}, {Adya}, {Affeldt},
  {Agarwal}, {Agathos}, {Agatsuma}, {Aggarwal}, {Aguiar}, {Aiello}, {Ain},
  {Ajith}, {Allen}, {Allen}, {Allocca}, {Aloy}, {Altin}, {Amato}, {Ananyeva},
  {Anderson}, {Anderson}, {Angelova}, {Antier}, {Appert}, {Arai}, {Araya},
  {Areeda}, {Ar{\`e}ne}, {Arnaud}, {Arun}, {Ascenzi}, {Ashton}, {Ast}, {Aston},
  {Astone}, {Atallah}, {Aubin}, {Aufmuth}, {Aulbert}, {AultONeal}, {Austin},
  {Avila-Alvarez}, {Babak}, {Bacon}, {Badaracco}, {Bader}, {Bae}, {Baker},
  {Baldaccini}, {Ballardin}, {Ballmer}, {Banagiri}, {Barayoga}, {Barclay},
  {Barish}, {Barker}, {Barkett}, {Barnum}, {Barone}, {Barr}, {Barsotti},
  {Barsuglia}, {Barta}, {Bartlett}, {Bartos}, {Bassiri}, {Basti}, {Batch},
  {Bawaj}, {Bayley}, {Bazzan}, {B{\'e}csy}, {Beer}, {Bejger}, {Belahcene},
  {Bell}, {Beniwal}, {Bensch}, {Berger}, {Bergmann}, {Bernuzzi}, {Bero},
  {Berry}, {Bersanetti}, {Bertolini}, {Betzwieser}, {Bhandare}, {Bilenko},
  {Bilgili}, {Billingsley}, {Billman}, {Birch}, {Birney}, {Birnholtz},
  {Biscans}, {Biscoveanu}, {Bisht}, {Bitossi}, {Bizouard}, {Blackburn},
  {Blackman}, {Blair}, {Blair}, {Blair}, {Bloemen}, {Bock}, {Bode}, {Boer},
  {Boetzel}, {Bogaert}, {Bohe}, {Bondu}, {Bonilla}, {Bonnand}, {Booker},
  {Boom}, {Booth}, {Bork}, {Boschi}, {Bose}, {Bossie}, {Bossilkov}, {Bosveld},
  {Bouffanais}, {Bozzi}, {Bradaschia}, {Brady}, {Bramley}, {Branchesi}, {Brau},
  {Briant}, {Brighenti}, {Brillet}, {Brinkmann}, {Brisson}, {Brockill},
  {Brooks}, {Brown}, {Brunett}, {Buchanan}, {Buikema}, {Bulik}, {Bulten},
  {Buonanno}, {Buskulic}, {Buy}, {Byer}, {Cabero}, {Cadonati}, {Cagnoli},
  {Cahillane}, {Calder{\'o}n Bustillo}, {Callister}, {Calloni}, {Camp},
  {Canepa}, {Canizares}, {Cannon}, {Cao}, {Cao}, {Capano}, {Capocasa},
  {Carbognani}, {Caride}, {Carney}, {Carullo}, {Casanueva Diaz}, {Casentini},
  {Caudill}, {Cavagli{\`a}}, {Cavalier}, {Cavalieri}, {Cella}, {Cepeda},
  {Cerd{\'a}-Dur{\'a}n}, {Cerretani}, {Cesarini}, {Chaibi}, {Chamberlin},
  {Chan}, {Chao}, {Charlton}, {Chase}, {Chassande-Mottin}, {Chatterjee},
  {Chatziioannou}, {Cheeseboro}, {Chen}, {Chen}, {Chen}, {Cheng}, {Chia},
  {Chincarini}, {Chiummo}, {Chmiel}, {Cho}, {Cho}, {Chow}, {Christensen},
  {Chu}, {Chua}, {Chua}, {Chung}, {Chung}, {Ciani}, {Ciobanu}, {Ciolfi},
  {Cipriano}, {Cirelli}, {Cirone}, {Clara}, {Clark}, {Clearwater}, {Cleva},
  {Cocchieri}, {Coccia}, {Cohadon}, {Cohen}, {Colla}, {Collette}, {Collins},
  {Cominsky}, {Constancio}, {Conti}, {Cooper}, {Corban}, {Corbitt},
  {Cordero-Carri{\'o}n}, {Corley}, {Cornish}, {Corsi}, {Cortese}, {Costa},
  {Cotesta}, {Coughlin}, {Coughlin}, {Coulon}, {Countryman}, {Couvares},
  {Covas}, {Cowan}, {Coward}, {Cowart}, {Coyne}, {Coyne}, {Creighton},
  {Creighton}, {Cripe}, {Crowder}, {Cullen}, {Cumming}, {Cunningham}, {Cuoco},
  {Canton}, {D{\'a}lya}, {Danilishin}, {D'Antonio}, {Danzmann}, {Dasgupta}, {Da
  Silva Costa}, {Dattilo}, {Dave}, {Davier}, {Davis}, {Daw}, {Day}, {DeBra},
  {Deenadayalan}, {Degallaix}, {De Laurentis}, {Del{\'e}glise}, {Del Pozzo},
  {Demos}, {Denker}, {Dent}, {De Pietri}, {Derby}, {Dergachev}, {De Rosa}, {De
  Rossi}, {DeSalvo}, {de Varona}, {Dhurandhar}, {D{\'\i}az}, {Dietrich}, {Di
  Fiore}, {Di Giovanni}, {Di Girolamo}, {Di Lieto}, {Ding}, {Di Pace}, {Di
  Palma}, {Di Renzo}, {Dmitriev}, {Doctor}, {Dolique}, {Donovan}, {Dooley},
  {Doravari}, {Dorrington}, {Dovale {\'A}lvarez}, {Downes}, {Drago},
  {Dreissigacker}, {Driggers}, {Du}, {Dupej}, {Dwyer}, {Easter}, {Edo},
  {Edwards}, {Effler}, {Eggenstein}, {Ehrens}, {Eichholz}, {Eikenberry},
  {Eisenmann}, {Eisenstein}, {Essick}, {Estelles}, {Estevez}, {Etienne},
  {Etzel}, {Evans}, {Evans}, {Fafone}, {Fair}, {Fairhurst}, {Fan}, {Farinon},
  {Farr}, {Farr}, {Fauchon-Jones}, {Favata}, {Fays}, {Fee}, {Fehrmann},
  {Feicht}, {Fejer}, {Feng}, {Fernandez-Galiana}, {Ferrante}, {Ferreira},
  {Ferrini}, {Fidecaro}, {Fiori}, {Fiorucci}, {Fishbach}, {Fisher}, {Fishner},
  {Fitz-Axen}, {Flaminio}, {Fletcher}, {Fong}, {Font}, {Forsyth}, {Forsyth},
  {Fournier}, {Frasca}, {Frasconi}, {Frei}, {Freise}, {Frey}, {Frey},
  {Fritschel}, {Frolov}, {Fulda}, {Fyffe}, {Gabbard}, {Gadre}, {Gaebel},
  {Gair}, {Gammaitoni}, {Ganija}, {Gaonkar}, {Garcia},
  {Garc{\'\i}a-Quir{\'o}s}, {Garufi}, {Gateley}, {Gaudio}, {Gaur}, {Gayathri},
  {Gemme}, {Genin}, {Gennai}, {George}, {George}, {Gergely}, {Germain},
  {Ghonge}, {Ghosh}, {Ghosh}, {Ghosh}, {Giacomazzo}, {Giaime}, {Giardina},
  {Giazotto}, {Gill}, {Giordano}, {Glover}, {Goetz}, {Goetz}, {Goncharov},
  {Gonz{\'a}lez}, {Gonzalez Castro}, {Gopakumar}, {Gorodetsky}, {Gossan},
  {Gosselin}, {Gouaty}, {Grado}, {Graef}, {Granata}, {Grant}, {Gras}, {Gray},
  {Greco}, {Green}, {Green}, {Gretarsson}, {Groot}, {Grote}, {Grunewald},
  {Gruning}, {Guidi}, {Gulati}, {Guo}, {Gupta}, {Gupta}, {Gushwa}, {Gustafson},
  {Gustafson}, {Halim}, {Hall}, {Hall}, {Hamilton}, {Hamilton}, {Hammond},
  {Haney}, {Hanke}, {Hanks}, {Hanna}, {Hannam}, {Hannuksela}, {Hanson},
  {Hardwick}, {Harms}, {Harry}, {Harry}, {Hart}, {Haster}, {Haughian}, {Healy},
  {Heidmann}, {Heintze}, {Heitmann}, {Hello}, {Hemming}, {Hendry}, {Heng},
  {Hennig}, {Heptonstall}, {Hernandez}, {Heurs}, {Hild}, {Hinderer}, {Ho},
  {Hoak}, {Hochheim}, {Hofman}, {Holland}, {Holt}, {Holz}, {Hopkins}, {Horst},
  {Hough}, {Houston}, {Howell}, {Hreibi}, {Huerta}, {Huet}, {Hughey}, {Hulko},
  {Husa}, {Huttner}, {Huynh-Dinh}, {Iess}, {Indik}, {Ingram}, {Inta}, {Intini},
  {Irwin}, {Isa}, {Isac}, {Isi}, {Iyer}, {Izumi}, {Jacqmin}, {Jani},
  {Jaranowski}, {Johnson}, {Johnson}, {Jones}, {Jones}, {Jonker}, {Ju},
  {Junker}, {Kalaghatgi}, {Kalogera}, {Kamai}, {Kandhasamy}, {Kang}, {Kanner},
  {Kapadia}, {Karki}, {Karvinen}, {Kasprzack}, {Katolik}, {Katsanevas},
  {Katsavounidis}, {Katzman}, {Kaufer}, {Kawabe}, {Keerthana},
  {K{\'e}f{\'e}lian}, {Keitel}, {Kemball}, {Kennedy}, {Key}, {Khalili},
  {Khamesra}, {Khan}, {Khan}, {Khan}, {Khan}, {Khazanov}, {Kijbunchoo}, {Kim},
  {Kim}, {Kim}, {Kim}, {Kim}, {Kim}, {King}, {King}, {Kinley-Hanlon},
  {Kirchhoff}, {Kissel}, {Kleybolte}, {Klimenko}, {Knowles}, {Koch},
  {Koehlenbeck}, {Koley}, {Kondrashov}, {Kontos}, {Korobko}, {Korth},
  {Kowalska}, {Kozak}, {Kr{\"a}mer}, {Kringel}, {Krishnan}, {Kr{\'o}lak},
  {Kuehn}, {Kumar}, {Kumar}, {Kumar}, {Kuo}, {Kutynia}, {Kwang}, {Lackey},
  {Lai}, {Landry}, {Landry}, {Lang}, {Lange}, {Lantz}, {Lanza},
  {Lartaux-Vollard}, {Lasky}, {Laxen}, {Lazzarini}, {Lazzaro}, {Leaci},
  {Leavey}, {Lee}, {Lee}, {Lee}, {Lee}, {Lee}, {Lehmann}, {Lenon}, {Leonardi},
  {Leroy}, {Letendre}, {Levin}, {Li}, {Li}, {Li}, {Linker}, {Littenberg},
  {Liu}, {Liu}, {Lo}, {Lockerbie}, {London}, {Longo}, {Lorenzini}, {Loriette},
  {Lormand}, {Losurdo}, {Lough}, {Lousto}, {Lovelace}, {L{\"u}ck}, {Lumaca},
  {Lundgren}, {Lynch}, {Ma}, {Macas}, {Macfoy}, {Machenschalk}, {MacInnis},
  {Macleod}, {Maga{\~n}a Hernandez}, {Maga{\~n}a-Sandoval}, {Maga{\~n}a
  Zertuche}, {Magee}, {Majorana}, {Maksimovic}, {Man}, {Mandic}, {Mangano},
  {Mansell}, {Manske}, {Mantovani}, {Marchesoni}, {Marion}, {M{\'a}rka},
  {M{\'a}rka}, {Markakis}, {Markosyan}, {Markowitz}, {Maros}, {Marquina},
  {Martelli}, {Martellini}, {Martin}, {Martin}, {Martynov}, {Mason}, {Massera},
  {Masserot}, {Massinger}, {Masso-Reid}, {Mastrogiovanni}, {Matas},
  {Matichard}, {Matone}, {Mavalvala}, {Mazumder}, {McCann}, {McCarthy},
  {McClelland}, {McCormick}, {McCuller}, {McGuire}, {McIver}, {McManus},
  {McRae}, {McWilliams}, {Meacher}, {Meadors}, {Mehmet}, {Meidam},
  {Mejuto-Villa}, {Melatos}, {Mendell}, {Mendoza-Gandara}, {Mercer}, {Mereni},
  {Merilh}, {Merzougui}, {Meshkov}, {Messenger}, {Messick}, {Metzdorff},
  {Meyers}, {Miao}, {Michel}, {Middleton}, {Mikhailov}, {Milano}, {Miller},
  {Miller}, {Miller}, {Miller}, {Millhouse}, {Mills}, {Milovich-Goff},
  {Minazzoli}, {Minenkov}, {Ming}, {Mishra}, {Mitra}, {Mitrofanov},
  {Mitselmakher}, {Mittleman}, {Moffa}, {Mogushi}, {Mohan}, {Mohapatra},
  {Montani}, {Moore}, {Moraru}, {Moreno}, {Morisaki}, {Mours}, {Mow-Lowry},
  {Mueller}, {Muir}, {Mukherjee}, {Mukherjee}, {Mukherjee}, {Mukund},
  {Mullavey}, {Munch}, {Mu{\~n}iz}, {Muratore}, {Murray}, {Nagar}, {Napier},
  {Nardecchia}, {Naticchioni}, {Nayak}, {Neilson}, {Nelemans}, {Nelson},
  {Nery}, {Neunzert}, {Nevin}, {Newport}, {Ng}, {Ng}, {Nguyen}, {Nguyen},
  {Nichols}, {Nielsen}, {Nissanke}, {Nitz}, {Nocera}, {Nolting}, {North},
  {Nuttall}, {Obergaulinger}, {Oberling}, {O'Brien}, {O'Dea}, {Ogin}, {Oh},
  {Oh}, {Ohme}, {Ohta}, {Okada}, {Oliver}, {Oppermann}, {Oram}, {O'Reilly},
  {Ormiston}, {Ortega}, {O'Shaughnessy}, {Ossokine}, {Ottaway}, {Overmier},
  {Owen}, {Pace}, {Pagano}, {Page}, {Page}, {Pai}, {Pai}, {Palamos},
  {Palashov}, {Palomba}, {Pal-Singh}, {Pan}, {Pan}, {Pang}, {Pang}, {Pankow},
  {Pannarale}, {Pant}, {Paoletti}, {Paoli}, {Papa}, {Parida}, {Parker},
  {Pascucci}, {Pasqualetti}, {Passaquieti}, {Passuello}, {Patil}, {Patricelli},
  {Pearlstone}, {Pedersen}, {Pedraza}, {Pedurand}, {Pekowsky}, {Pele}, {Penn},
  {Perego}, {Perez}, {Perreca}, {Perri}, {Pfeiffer}, {Phelps}, {Phukon},
  {Piccinni}, {Pichot}, {Piergiovanni}, {Pierro}, {Pillant}, {Pinard}, {Pinto},
  {Pirello}, {Pitkin}, {Poggiani}, {Popolizio}, {Porter}, {Possenti}, {Post},
  {Powell}, {Prasad}, {Pratt}, {Pratten}, {Predoi}, {Prestegard}, {Principe},
  {Privitera}, {Prodi}, {Prokhorov}, {Puncken}, {Punturo}, {Puppo},
  {P{\"u}rrer}, {Qi}, {Quetschke}, {Quintero}, {Quitzow-James}, {Raab},
  {Rabeling}, {Radkins}, {Raffai}, {Raja}, {Rajan}, {Rajbhandari}, {Rakhmanov},
  {Ramirez}, {Ramos-Buades}, {Rana}, {Rapagnani}, {Raymond}, {Razzano}, {Read},
  {Regimbau}, {Rei}, {Reid}, {Reitze}, {Ren}, {Ricci}, {Ricker},
  {Riemenschneider}, {Riles}, {Rizzo}, {Robertson}, {Robie}, {Robinet},
  {Robson}, {Rocchi}, {Rolland}, {Rollins}, {Roma}, {Romano}, {Romel}, {Romie},
  {Rosi{\'n}ska}, {Ross}, {Rowan}, {R{\"u}diger}, {Ruggi}, {Rutins}, {Ryan},
  {Sachdev}, {Sadecki}, {Sakellariadou}, {Salconi}, {Saleem}, {Salemi},
  {Samajdar}, {Sammut}, {Sampson}, {Sanchez}, {Sanchez}, {Sanchis-Gual},
  {Sandberg}, {Sanders}, {Sarin}, {Sassolas}, {Sathyaprakash}, {Saulson},
  {Sauter}, {Savage}, {Sawadsky}, {Schale}, {Scheel}, {Scheuer}, {Schmidt},
  {Schnabel}, {Schofield}, {Sch{\"o}nbeck}, {Schreiber}, {Schuette}, {Schulte},
  {Schutz}, {Schwalbe}, {Scott}, {Scott}, {Seidel}, {Sellers}, {Sengupta},
  {Sentenac}, {Sequino}, {Sergeev}, {Setyawati}, {Shaddock}, {Shaffer}, {Shah},
  {Shahriar}, {Shaner}, {Shao}, {Shapiro}, {Shawhan}, {Shen}, {Shoemaker},
  {Shoemaker}, {Siellez}, {Siemens}, {Sieniawska}, {Sigg}, {Silva}, {Singer},
  {Singh}, {Singhal}, {Sintes}, {Slagmolen}, {Slaven-Blair}, {Smith}, {Smith},
  {Smith}, {Somala}, {Son}, {Sorazu}, {Sorrentino}, {Souradeep}, {Spencer},
  {Srivastava}, {Staats}, {Steinke}, {Steinlechner}, {Steinlechner},
  {Steinmeyer}, {Steltner}, {Stevenson}, {Stocks}, {Stone}, {Stops}, {Strain},
  {Stratta}, {Strigin}, {Strunk}, {Sturani}, {Stuver}, {Summerscales}, {Sun},
  {Sunil}, {Suresh}, {Sutton}, {Swinkels}, {Szczepa{\'n}czyk}, {Tacca}, {Tait},
  {Talbot}, {Talukder}, {Tanner}, {T{\'a}pai}, {Taracchini}, {Tasson},
  {Taylor}, {Taylor}, {Tewari}, {Theeg}, {Thies}, {Thomas}, {Thomas}, {Thomas},
  {Thorne}, {Thrane}, {Tiwari}, {Tiwari}, {Tokmakov}, {Toland}, {Tonelli},
  {Tornasi}, {Torres-Forn{\'e}}, {Torrie}, {T{\"o}yr{\"a}}, {Travasso},
  {Traylor}, {Trinastic}, {Tringali}, {Trovato}, {Trozzo}, {Tsang}, {Tse},
  {Tso}, {Tsuna}, {Tsukada}, {Tuyenbayev}, {Ueno}, {Ugolini}, {Urban}, {Usman},
  {Vahlbruch}, {Vajente}, {Valdes}, {van Bakel}, {van Beuzekom}, {van den
  Brand}, {Van Den Broeck}, {Vander-Hyde}, {van der Schaaf}, {van Heijningen},
  {van Veggel}, {Vardaro}, {Varma}, {Vass}, {Vas{\'u}th}, {Vecchio},
  {Vedovato}, {Veitch}, {Veitch}, {Venkateswara}, {Venugopalan}, {Verkindt},
  {Vetrano}, {Vicer{\'e}}, {Viets}, {Vinciguerra}, {Vine}, {Vinet}, {Vitale},
  {Vo}, {Vocca}, {Vorvick}, {Vyatchanin}, {Wade}, {Wade}, {Wade}, {Walet},
  {Walker}, {Wallace}, {Walsh}, {Wang}, {Wang}, {Wang}, {Wang}, {Wang}, {Ward},
  {Warner}, {Was}, {Watchi}, {Weaver}, {Wei}, {Weinert}, {Weinstein}, {Weiss},
  {Wellmann}, {Wen}, {Wessel}, {We{\ss}els}, {Westerweck}, {Wette}, {Whelan},
  {Whiting}, {Whittle}, {Wilken}, {Williams}, {Williams}, {Williamson},
  {Willis}, {Willke}, {Wimmer}, {Winkler}, {Wipf}, {Wittel}, {Woan}, {Woehler},
  {Wofford}, {Wong}, {Worden}, {Wright}, {Wu}, {Wysocki}, {Xiao}, {Yam},
  {Yamamoto}, {Yancey}, {Yang}, {Yap}, {Yazback}, {Yu}, {Yu}, {Yvert},
  {Zadro{\.Z}ny}, {Zanolin}, {Zelenova}, {Zendri}, {Zevin}, {Zhang}, {Zhang},
  {Zhang}, {Zhang}, {Zhang}, {Zhao}, {Zhou}, {Zhou}, {Zhu}, {Zhu}, {Zimmerman},
  {Zlochower}, {Zucker}, {Zweizig}, {LIGO Scientific Collaboration}, \& {Virgo
  Collaboration}}]{lvk-ns-radii-2018}
{Abbott}, B.~P., {Abbott}, R., {Abbott}, T.~D., {et~al.} 2018, \prl, 121,
  161101, \dodoi{10.1103/PhysRevLett.121.161101}

\bibitem[{Abdikamalov {et~al.}(2022)Abdikamalov, Pagliaroli, \&
  Radice}]{abdikamalov_gw-ccsne-review_2022}
Abdikamalov, E., Pagliaroli, G., \& Radice, D. 2022, in Handbook of
  Gravitational Wave Astronomy (Springer), 1--37

\bibitem[{{Acernese} {et~al.}(2015)}]{Virgo}
{Acernese}, F., {et~al.} 2015, Classical and Quantum Gravity, 32, 024001,
  \dodoi{10.1088/0264-9381/32/2/024001}

\bibitem[{{Ackley} {et~al.}(2020){Ackley}, {Adya}, {Agrawal}, {Altin},
  {Ashton}, {Bailes}, {Baltinas}, {Barbuio}, {Beniwal}, {Blair}, {Blair},
  {Bolingbroke}, {Bossilkov}, {Shachar Boublil}, {Brown}, {Burridge}, {Calderon
  Bustillo}, {Cameron}, {Tuong Cao}, {Carlin}, {Chang}, {Charlton},
  {Chatterjee}, {Chattopadhyay}, {Chen}, {Chi}, {Chow}, {Chu}, {Ciobanu},
  {Clarke}, {Clearwater}, {Cooke}, {Coward}, {Crisp}, {Dattatri}, {Deller},
  {Dobie}, {Dunn}, {Easter}, {Eichholz}, {Evans}, {Flynn}, {Foran}, {Forsyth},
  {Gai}, {Galaudage}, {Galloway}, {Gendre}, {Goncharov}, {Goode}, {Gozzard},
  {Grace}, {Graham}, {Heger}, {Hernandez Vivanco}, {Hirai}, {Holland},
  {Holmes}, {Howard}, {Howell}, {Howitt}, {H{\"u}bner}, {Hurley}, {Ingram},
  {Jaberian Hamedan}, {Jenner}, {Ju}, {Kapasi}, {Kaur}, {Kijbunchoo},
  {Kovalam}, {Kumar Choudhary}, {Lasky}, {Lau}, {Leung}, {Liu}, {Loh},
  {Mailvagan}, {Mandel}, {McCann}, {McClelland}, {McKenzie}, {McManus},
  {McRae}, {Melatos}, {Meyers}, {Middleton}, {Miles}, {Millhouse}, {Lun Mong},
  {Mueller}, {Munch}, {Musiov}, {Muusse}, {Nathan}, {Naveh}, {Neijssel},
  {Neil}, {Ng}, {Oloworaran}, {Ottaway}, {Page}, {Pan}, {Pathak}, {Payne},
  {Powell}, {Pritchard}, {Puckridge}, {Raidani}, {Rallabhandi}, {Reardon},
  {Riley}, {Roberts}, {Romero-Shaw}, {Roocke}, {Rowell}, {Sahu}, {Sarin},
  {Sarre}, {Sattari}, {Schiworski}, {Scott}, {Sengar}, {Shaddock}, {Shannon},
  {SHI}, {Sibley}, {Slagmolen}, {Slaven-Blair}, {Smith}, {Spollard}, {Steed},
  {Strang}, {Sun}, {Sunderland}, {Suvorova}, {Talbot}, {Thrane},
  {T{\"o}yr{\"a}}, {Trahanas}, {Vajpeyi}, {van Heijningen}, {Vargas}, {Veitch},
  {Vigna-Gomez}, {Wade}, {Walker}, {Wang}, {Ward}, {Ward}, {Webb}, {Wen},
  {Wette}, {Wilcox}, {Winterflood}, {Wolf}, {Wu}, {Jet Yap}, {You}, {Yu},
  {Zhang}, {Zhang}, {Zhao}, \& {Zhu}}]{nemo}
{Ackley}, K., {Adya}, V.~B., {Agrawal}, P., {et~al.} 2020, \pasa, 37, e047,
  \dodoi{10.1017/pasa.2020.39}

\bibitem[{{Adams} {et~al.}(2013){Adams}, {Kochanek}, {Beacom}, {Vagins}, \&
  {Stanek}}]{Adam2013}
{Adams}, S.~M., {Kochanek}, C.~S., {Beacom}, J.~F., {Vagins}, M.~R., \&
  {Stanek}, K.~Z. 2013, \apj, 778, 164, \dodoi{10.1088/0004-637X/778/2/164}

\bibitem[{Alcubierre(2008)}]{alcubierre}
Alcubierre, M. 2008, Introduction to 3+1 {Numerical} {Relativity},
  International {Series} of {Monographs} on {Physics} (Oxford: Oxford
  University Press), \dodoi{10.1093/acprof:oso/9780199205677.001.0001}

\bibitem[{{Andresen} {et~al.}(2021){Andresen}, {Glas}, \&
  {Janka}}]{andresen_2021}
{Andresen}, H., {Glas}, R., \& {Janka}, H.~T. 2021, \mnras, 503, 3552,
  \dodoi{10.1093/mnras/stab675}

\bibitem[{Andresen {et~al.}(2017)Andresen, Müller, Müller, \&
  Janka}]{andresen_gravitational_2017}
Andresen, H., Müller, B., Müller, E., \& Janka, H.-T. 2017, Monthly Notices
  of the Royal Astronomical Society, 468, 2032, \dodoi{10.1093/mnras/stx618}

\bibitem[{{Arnaud} {et~al.}(1999){Arnaud}, {Cavalier}, {Davier}, \&
  {Hello}}]{arnaud_gw-bursts_1999}
{Arnaud}, N., {Cavalier}, F., {Davier}, M., \& {Hello}, P. 1999, \prd, 59,
  082002, \dodoi{10.1103/PhysRevD.59.082002}

\bibitem[{{Banik} {et~al.}(2014){Banik}, {Hempel}, \& {Bandyopadhyay}}]{bhblp}
{Banik}, S., {Hempel}, M., \& {Bandyopadhyay}, D. 2014, \apjs, 214, 22,
  \dodoi{10.1088/0067-0049/214/2/22}

\bibitem[{{Bizouard} {et~al.}(2021){Bizouard}, {Maturana-Russel},
  {Torres-Forn{\'e}}, {Obergaulinger}, {Cerd{\'a}-Dur{\'a}n}, {Christensen},
  {Font}, \& {Meyer}}]{Bizouard2021}
{Bizouard}, M.-A., {Maturana-Russel}, P., {Torres-Forn{\'e}}, A., {et~al.}
  2021, \prd, 103, 063006, \dodoi{10.1103/PhysRevD.103.063006}

\bibitem[{{Bruel} {et~al.}(2023){Bruel}, {Bizouard}, {Obergaulinger},
  {Maturana-Russel}, {Torres-Forn{\'e}}, {Cerd{\'a}-Dur{\'a}n}, {Christensen},
  {Font}, \& {Meyer}}]{Bruel2023}
{Bruel}, T., {Bizouard}, M.-A., {Obergaulinger}, M., {et~al.} 2023, arXiv
  e-prints, arXiv:2301.10019, \dodoi{10.48550/arXiv.2301.10019}

\bibitem[{{Bugli} {et~al.}(2022){Bugli}, {Guilet}, {Foglizzo}, \&
  {Obergaulinger}}]{bugli_2022}
{Bugli}, M., {Guilet}, J., {Foglizzo}, T., \& {Obergaulinger}, M. 2022, arXiv
  e-prints, arXiv:2210.05012, \dodoi{10.48550/arXiv.2210.05012}

\bibitem[{Burrows \& Hayes(1996)}]{gw_pns-convection_burrows1996}
Burrows, A., \& Hayes, J. 1996, Physical Review Letters, 76, 352

\bibitem[{{Cornish} \& {Littenberg}(2015)}]{bayeswave}
{Cornish}, N.~J., \& {Littenberg}, T.~B. 2015, Classical and Quantum Gravity,
  32, 135012, \dodoi{10.1088/0264-9381/32/13/135012}

\bibitem[{{Cowling}(1941)}]{cowling_1941}
{Cowling}, T.~G. 1941, \mnras, 101, 367, \dodoi{10.1093/mnras/101.8.367}

\bibitem[{Dempster {et~al.}(1977)Dempster, Laird, \& Rubin}]{em_dempster_1977}
Dempster, A.~P., Laird, N.~M., \& Rubin, D.~B. 1977, Journal of the Royal
  Statistical Society: Series B (Methodological), 39, 1

\bibitem[{{Dimmelmeier} {et~al.}(2001){Dimmelmeier}, {Font}, \&
  {M{\"u}ller}}]{dimmelmeier_relativistic-ccsne-gw_2001}
{Dimmelmeier}, H., {Font}, J.~A., \& {M{\"u}ller}, E. 2001, \apjl, 560, L163,
  \dodoi{10.1086/324406}

\bibitem[{{Dimmelmeier} {et~al.}(2005){Dimmelmeier}, {Novak}, {Font},
  {Ib{\'a}{\~n}ez}, \& {M{\"u}ller}}]{CoCoNut}
{Dimmelmeier}, H., {Novak}, J., {Font}, J.~A., {Ib{\'a}{\~n}ez}, J.~M., \&
  {M{\"u}ller}, E. 2005, \prd, 71, 064023, \dodoi{10.1103/PhysRevD.71.064023}

\bibitem[{{Ebinger} {et~al.}(2019){Ebinger}, {Curtis}, {Fr{\"o}hlich},
  {Hempel}, {Perego}, {Liebend{\"o}rfer}, \& {Thielemann}}]{push2}
{Ebinger}, K., {Curtis}, S., {Fr{\"o}hlich}, C., {et~al.} 2019, \apj, 870, 1,
  \dodoi{10.3847/1538-4357/aae7c9}

\bibitem[{{Ebinger} {et~al.}(2020){Ebinger}, {Curtis}, {Ghosh}, {Fr{\"o}hlich},
  {Hempel}, {Perego}, {Liebend{\"o}rfer}, \& {Thielemann}}]{push4}
{Ebinger}, K., {Curtis}, S., {Ghosh}, S., {et~al.} 2020, \apj, 888, 91,
  \dodoi{10.3847/1538-4357/ab5dcb}

\bibitem[{{Evans} {et~al.}(2021){Evans}, {Adhikari}, {Afle}, {Ballmer},
  {Biscoveanu}, {Borhanian}, {Brown}, {Chen}, {Eisenstein}, {Gruson}, {Gupta},
  {Hall}, {Huxford}, {Kamai}, {Kashyap}, {Kissel}, {Kuns}, {Landry}, {Lenon},
  {Lovelace}, {McCuller}, {Ng}, {Nitz}, {Read}, {Sathyaprakash}, {Shoemaker},
  {Slagmolen}, {Smith}, {Srivastava}, {Sun}, {Vitale}, \&
  {Weiss}}]{ce_horizon_study_evans2021}
{Evans}, M., {Adhikari}, R.~X., {Afle}, C., {et~al.} 2021, arXiv e-prints,
  arXiv:2109.09882.
\newblock \doarXiv{2109.09882}

\bibitem[{{Fryer} \& {New}(2011)}]{fryer_ccsne-gw-review_2010s}
{Fryer}, C.~L., \& {New}, K. C.~B. 2011, Living Reviews in Relativity, 14, 1,
  \dodoi{10.12942/lrr-2011-1}

\bibitem[{{Gossan} {et~al.}(2016){Gossan}, {Sutton}, {Stuver}, {Zanolin},
  {Gill}, \& {Ott}}]{gw_ccsne_current_gossan2016}
{Gossan}, S.~E., {Sutton}, P., {Stuver}, A., {et~al.} 2016, \prd, 93, 042002,
  \dodoi{10.1103/PhysRevD.93.042002}

\bibitem[{Halzen \& Raffelt(2009)}]{Halzen:2009}
Halzen, F., \& Raffelt, G.~G. 2009, Phys. Rev. D, 80, 087301,
  \dodoi{10.1103/PhysRevD.80.087301}

\bibitem[{Harris {et~al.}(2020)Harris, Millman, van~der Walt, Gommers,
  Virtanen, Cournapeau, Wieser, Taylor, Berg, Smith, Kern, Picus, Hoyer, van
  Kerkwijk, Brett, Haldane, del R{\'{i}}o, Wiebe, Peterson,
  G{\'{e}}rard-Marchant, Sheppard, Reddy, Weckesser, Abbasi, Gohlke, \&
  Oliphant}]{numpy}
Harris, C.~R., Millman, K.~J., van~der Walt, S.~J., {et~al.} 2020, Nature, 585,
  357, \dodoi{10.1038/s41586-020-2649-2}

\bibitem[{{Hayama} {et~al.}(2015){Hayama}, {Kuroda}, {Kotake}, \&
  {Takiwaki}}]{takiwaki_ccsne-network-analysis_2015}
{Hayama}, K., {Kuroda}, T., {Kotake}, K., \& {Takiwaki}, T. 2015, \prd, 92,
  122001, \dodoi{10.1103/PhysRevD.92.122001}

\bibitem[{{Hempel} \& {Schaffner-Bielich}(2010)}]{dd2_1}
{Hempel}, M., \& {Schaffner-Bielich}, J. 2010, Nuclear Physics A, 837, 210,
  \dodoi{10.1016/j.nuclphysa.2010.02.010}

\bibitem[{{Hild} {et~al.}(2011){Hild}, {Abernathy}, {Acernese}, {Amaro-Seoane},
  {Andersson}, {Arun}, {Barone}, {Barr}, {Barsuglia}, {Beker}, {Beveridge},
  {Birindelli}, {Bose}, {Bosi}, {Braccini}, {Bradaschia}, {Bulik}, {Calloni},
  {Cella}, {Chassande Mottin}, {Chelkowski}, {Chincarini}, {Clark}, {Coccia},
  {Colacino}, {Colas}, {Cumming}, {Cunningham}, {Cuoco}, {Danilishin},
  {Danzmann}, {De Salvo}, {Dent}, {De Rosa}, {Di Fiore}, {Di Virgilio},
  {Doets}, {Fafone}, {Falferi}, {Flaminio}, {Franc}, {Frasconi}, {Freise},
  {Friedrich}, {Fulda}, {Gair}, {Gemme}, {Genin}, {Gennai}, {Giazotto},
  {Glampedakis}, {Gr{\"a}f}, {Granata}, {Grote}, {Guidi}, {Gurkovsky},
  {Hammond}, {Hannam}, {Harms}, {Heinert}, {Hendry}, {Heng}, {Hennes}, {Hough},
  {Husa}, {Huttner}, {Jones}, {Khalili}, {Kokeyama}, {Kokkotas}, {Krishnan},
  {Li}, {Lorenzini}, {L{\"u}ck}, {Majorana}, {Mandel}, {Mandic}, {Mantovani},
  {Martin}, {Michel}, {Minenkov}, {Morgado}, {Mosca}, {Mours},
  {M{\"u}ller{\textendash}Ebhardt}, {Murray}, {Nawrodt}, {Nelson},
  {Oshaughnessy}, {Ott}, {Palomba}, {Paoli}, {Parguez}, {Pasqualetti},
  {Passaquieti}, {Passuello}, {Pinard}, {Plastino}, {Poggiani}, {Popolizio},
  {Prato}, {Punturo}, {Puppo}, {Rabeling}, {Rapagnani}, {Read}, {Regimbau},
  {Rehbein}, {Reid}, {Ricci}, {Richard}, {Rocchi}, {Rowan}, {R{\"u}diger},
  {Santamar{\'\i}a}, {Sassolas}, {Sathyaprakash}, {Schnabel}, {Schwarz},
  {Seidel}, {Sintes}, {Somiya}, {Speirits}, {Strain}, {Strigin}, {Sutton},
  {Tarabrin}, {Th{\"u}ring}, {van den Brand}, {van Veggel}, {van den Broeck},
  {Vecchio}, {Veitch}, {Vetrano}, {Vicere}, {Vyatchanin}, {Willke}, {Woan}, \&
  {Yamamoto}}]{Hild2011}
{Hild}, S., {Abernathy}, M., {Acernese}, F., {et~al.} 2011, Classical and
  Quantum Gravity, 28, 094013, \dodoi{10.1088/0264-9381/28/9/094013}

\bibitem[{Hunter(2007)}]{matplotlib}
Hunter, J.~D. 2007, Computing In Science \& Engineering, 9, 90

\bibitem[{{Jakobus} {et~al.}(2023){Jakobus}, {M{\"u}ller}, {Heger}, {Zha},
  {Powell}, {Motornenko}, {Steinheimer}, \& {Stoecker}}]{jakobus-eos-2023}
{Jakobus}, P., {M{\"u}ller}, B., {Heger}, A., {et~al.} 2023, arXiv e-prints,
  arXiv:2301.06515, \dodoi{10.48550/arXiv.2301.06515}

\bibitem[{Klimenko {et~al.}(2016)Klimenko, Vedovato, Drago, Salemi, Tiwari,
  Prodi, Lazzaro, Ackley, Tiwari, Da~Silva, {et~al.}}]{klimenko_cWB_2016}
Klimenko, S., Vedovato, G., Drago, M., {et~al.} 2016, Physical Review D, 93,
  042004

\bibitem[{Kuroda {et~al.}(2016)Kuroda, Kotake, \& Takiwaki}]{kuroda_new_2016}
Kuroda, T., Kotake, K., \& Takiwaki, T. 2016, The Astrophysical Journal, 829,
  L14, \dodoi{10.3847/2041-8205/829/1/L14}

\bibitem[{Lattimer \& {Douglas Swesty}(1991)}]{ls220}
Lattimer, J.~M., \& {Douglas Swesty}, F. 1991, Nuclear Physics A, 535, 331,
  \dodoi{10.1016/0375-9474(91)90452-C}

\bibitem[{{Liebend{\"o}rfer}(2000)}]{liebendoerfer_thesis_2000}
{Liebend{\"o}rfer}, M. 2000, PhD thesis, University of Basel, Switzerland

\bibitem[{{Liebend{\"o}rfer} {et~al.}(2001){Liebend{\"o}rfer}, {Mezzacappa}, \&
  {Thielemann}}]{Liebendoerfer.Agile}
{Liebend{\"o}rfer}, M., {Mezzacappa}, A., \& {Thielemann}, F.-K. 2001, \prd,
  63, 104003, \dodoi{10.1103/PhysRevD.63.104003}

\bibitem[{{Liebend{\"o}rfer} {et~al.}(2009){Liebend{\"o}rfer}, {Whitehouse}, \&
  {Fischer}}]{Liebendoerfer.IDSA:2009}
{Liebend{\"o}rfer}, M., {Whitehouse}, S.~C., \& {Fischer}, T. 2009, \apj, 698,
  1174, \dodoi{10.1088/0004-637X/698/2/1174}

\bibitem[{{LIGO Scientific Collaboration} {et~al.}(2015)}]{LIGO}
{LIGO Scientific Collaboration}, {et~al.} 2015, Classical and Quantum Gravity,
  32, 074001, \dodoi{10.1088/0264-9381/32/7/074001}

\bibitem[{Marek {et~al.}(2009)Marek, Janka, \&
  Müller}]{marek_equation-of-state_2009}
Marek, A., Janka, H.-T., \& Müller, E. 2009, Astronomy \& Astrophysics, 496,
  475, \dodoi{10.1051/0004-6361/200810883}

\bibitem[{{Meskhi} {et~al.}(2022){Meskhi}, {Wolfe}, {Dai}, {Fr{\"o}hlich},
  {Miller}, {Wong}, \& {Vilalta}}]{meskhi_eos-remnant-distribution_2022}
{Meskhi}, M.~M., {Wolfe}, N.~E., {Dai}, Z., {et~al.} 2022, \apjl, 932, L3,
  \dodoi{10.3847/2041-8213/ac7054}

\bibitem[{Mezzacappa {et~al.}(2020)Mezzacappa, Marronetti, Landfield, Lentz,
  Yakunin, Bruenn, Hix, Messer, Endeve, Blondin, \&
  Harris}]{mezzacappa_gravitational-wave_2020}
Mezzacappa, A., Marronetti, P., Landfield, R.~E., {et~al.} 2020, Physical
  Review D, 102, 023027, \dodoi{10.1103/PhysRevD.102.023027}

\bibitem[{{Mezzacappa} {et~al.}(2023){Mezzacappa}, {Marronetti}, {Landfield},
  {Lentz}, {Murphy}, {Raphael Hix}, {Harris}, {Bruenn}, {Blondin}, {Bronson
  Messer}, {Casanova}, \& {Kronzer}}]{mezzacappa_gravitational-wave_2023}
{Mezzacappa}, A., {Marronetti}, P., {Landfield}, R.~E., {et~al.} 2023, \prd,
  107, 043008, \dodoi{10.1103/PhysRevD.107.043008}

\bibitem[{{Moenchmeyer} {et~al.}(1991){Moenchmeyer}, {Schaefer}, {Mueller}, \&
  {Kates}}]{moenchmeyer_rotating-core-collapse-gw_1991}
{Moenchmeyer}, R., {Schaefer}, G., {Mueller}, E., \& {Kates}, R.~E. 1991, \aap,
  246, 417

\bibitem[{{Mori} {et~al.}(2023){Mori}, {Suwa}, \& {Takiwaki}}]{Mori2023}
{Mori}, M., {Suwa}, Y., \& {Takiwaki}, T. 2023, arXiv e-prints,
  arXiv:2302.00292, \dodoi{10.48550/arXiv.2302.00292}

\bibitem[{Morozova {et~al.}(2018)Morozova, Radice, Burrows, \&
  Vartanyan}]{morozova_eigenfrequencies_2018}
Morozova, V., Radice, D., Burrows, A., \& Vartanyan, D. 2018, The Astrophysical
  Journal, 861, 10, \dodoi{10.3847/1538-4357/aac5f1}

\bibitem[{M\"uller {et~al.}(2013)M\"uller, Janka, \& Marek}]{muller_fpeak_2013}
M\"uller, B., Janka, H.-T., \& Marek, A. 2013, The Astrophysical Journal, 766,
  43, \dodoi{10.1088/0004-637X/766/1/43}

\bibitem[{M{\"u}ller \& Janka(1997)}]{gw_pns-convection_muller1997}
M{\"u}ller, E., \& Janka, H.-T. 1997, Astronomy and Astrophysics, 317, 140

\bibitem[{M{\"u}ller {et~al.}(2004)M{\"u}ller, Rampp, Buras, Janka, \&
  Shoemaker}]{gw_pns-convection_muller2004}
M{\"u}ller, E., Rampp, M., Buras, R., Janka, H.-T., \& Shoemaker, D.~H. 2004,
  The Astrophysical Journal, 603, 221

\bibitem[{{Murphy} {et~al.}(2009){Murphy}, {Ott}, \& {Burrows}}]{murphy_2009}
{Murphy}, J.~W., {Ott}, C.~D., \& {Burrows}, A. 2009, \apj, 707, 1173,
  \dodoi{10.1088/0004-637X/707/2/1173}

\bibitem[{Nakamura {et~al.}(2016)Nakamura, Horiuchi, Tanaka, Hayama, Takiwaki,
  \& Kotake}]{nakamura_multimessenger_2016}
Nakamura, K., Horiuchi, S., Tanaka, M., {et~al.} 2016, Monthly Notices of the
  Royal Astronomical Society, 461, 3296, \dodoi{10.1093/mnras/stw1453}

\bibitem[{O'Connor \& Ott(2011)}]{oconnor_compactness_2011}
O'Connor, E., \& Ott, C.~D. 2011, The Astrophysical Journal, 730, 70,
  \dodoi{10.1088/0004-637X/730/2/70}

\bibitem[{O'Connor \& Couch(2018)}]{oconnor_exploring_2018}
O'Connor, E.~P., \& Couch, S.~M. 2018, The Astrophysical Journal, 865, 81,
  \dodoi{10.3847/1538-4357/aadcf7}

\bibitem[{Pagliaroli {et~al.}(2009)Pagliaroli, Vissani, Coccia, \&
  Fulgione}]{Pagliaroli:2009}
Pagliaroli, G., Vissani, F., Coccia, E., \& Fulgione, W. 2009, Phys. Rev.
  Lett., 103, 031102, \dodoi{10.1103/PhysRevLett.103.031102}

\bibitem[{pandas~development team(2020)}]{reback2020pandas}
pandas~development team, T. 2020, pandas-dev/pandas: Pandas, latest,  Zenodo,
  \dodoi{10.5281/zenodo.3509134}

\bibitem[{Pedregosa {et~al.}(2011)Pedregosa, Varoquaux, Gramfort, Michel,
  Thirion, Grisel, Blondel, Prettenhofer, Weiss, Dubourg, Vanderplas, Passos,
  Cournapeau, Brucher, Perrot, \& Duchesnay}]{scikit-learn}
Pedregosa, F., Varoquaux, G., Gramfort, A., {et~al.} 2011, Journal of Machine
  Learning Research, 12, 2825

\bibitem[{{Perego} {et~al.}(2016){Perego}, {Cabez{\'o}n}, \&
  {K{\"a}ppeli}}]{perego16.asl}
{Perego}, A., {Cabez{\'o}n}, R.~M., \& {K{\"a}ppeli}, R. 2016, \apjs, 223, 22,
  \dodoi{10.3847/0067-0049/223/2/22}

\bibitem[{{Perego} {et~al.}(2015){Perego}, {Hempel}, {Fr{\"o}hlich}, {Ebinger},
  {Eichler}, {Casanova}, {Liebend{\"o}rfer}, \& {Thielemann}}]{push1}
{Perego}, A., {Hempel}, M., {Fr{\"o}hlich}, C., {et~al.} 2015, Astrophys.\ J.,
  806, 275, \dodoi{10.1088/0004-637X/806/2/275}

\bibitem[{Powell \& M\"uller(2019)}]{powell_gravitational_2019}
Powell, J., \& M\"uller, B. 2019, Monthly Notices of the Royal Astronomical
  Society, 487, 1178, \dodoi{10.1093/mnras/stz1304}

\bibitem[{{Powell} \& {M{\"u}ller}(2022)}]{powell_ccsne-inference_2022}
{Powell}, J., \& {M{\"u}ller}, B. 2022, \prd, 105, 063018,
  \dodoi{10.1103/PhysRevD.105.063018}

\bibitem[{{Raaijmakers} {et~al.}(2021){Raaijmakers}, {Greif}, {Hebeler},
  {Hinderer}, {Nissanke}, {Schwenk}, {Riley}, {Watts}, {Lattimer}, \&
  {Ho}}]{raaijmakers_2021}
{Raaijmakers}, G., {Greif}, S.~K., {Hebeler}, K., {et~al.} 2021, \apjl, 918,
  L29, \dodoi{10.3847/2041-8213/ac089a}

\bibitem[{Radice {et~al.}(2019)Radice, Morozova, Burrows, Vartanyan, \&
  Nagakura}]{radice_characterizing_2019}
Radice, D., Morozova, V., Burrows, A., Vartanyan, D., \& Nagakura, H. 2019, The
  Astrophysical Journal, 876, L9, \dodoi{10.3847/2041-8213/ab191a}

\bibitem[{{Raza} {et~al.}(2022){Raza}, {McIver}, {D{\'a}lya}, \&
  {Raffai}}]{raza_bayeswave_ccsne_reconstruct_2022}
{Raza}, N., {McIver}, J., {D{\'a}lya}, G., \& {Raffai}, P. 2022, \prd, 106,
  063014, \dodoi{10.1103/PhysRevD.106.063014}

\bibitem[{{Richers} {et~al.}(2017){Richers}, {Ott}, {Abdikamalov}, {O'Connor},
  \& {Sullivan}}]{richers_2017}
{Richers}, S., {Ott}, C.~D., {Abdikamalov}, E., {O'Connor}, E., \& {Sullivan},
  C. 2017, \prd, 95, 063019, \dodoi{10.1103/PhysRevD.95.063019}

\bibitem[{{Rozwadowska} {et~al.}(2021){Rozwadowska}, {Vissani}, \&
  {Cappellaro}}]{Rozwadowska2021}
{Rozwadowska}, K., {Vissani}, F., \& {Cappellaro}, E. 2021, \na, 83, 101498,
  \dodoi{10.1016/j.newast.2020.101498}

\bibitem[{{Ruffini} \& {Wheeler}(1971)}]{ruffini_1971}
{Ruffini}, R., \& {Wheeler}, J.~A. 1971, ESRO, 52, 45

\bibitem[{{Sieniawska} \& {Bejger}(2019)}]{ns-continuous-gw-review-2019}
{Sieniawska}, M., \& {Bejger}, M. 2019, Universe, 5, 217,
  \dodoi{10.3390/universe5110217}

\bibitem[{{Sotani} {et~al.}(2019){Sotani}, {Kuroda}, {Takiwaki}, \&
  {Kotake}}]{Sotani2019}
{Sotani}, H., {Kuroda}, T., {Takiwaki}, T., \& {Kotake}, K. 2019, \prd, 99,
  123024, \dodoi{10.1103/PhysRevD.99.123024}

\bibitem[{{Sotani} \& {Takiwaki}(2016)}]{sotani_simple_2016}
{Sotani}, H., \& {Takiwaki}, T. 2016, \prd, 94, 044043,
  \dodoi{10.1103/PhysRevD.94.044043}

\bibitem[{{Sotani} \& {Takiwaki}(2020{\natexlab{a}})}]{Sotani.2020}
---. 2020{\natexlab{a}}, \prd, 102, 063025, \dodoi{10.1103/PhysRevD.102.063025}

\bibitem[{{Sotani} \&
  {Takiwaki}(2020{\natexlab{b}})}]{sotani_avoided_crossing_2020}
---. 2020{\natexlab{b}}, \mnras, 498, 3503, \dodoi{10.1093/mnras/staa2597}

\bibitem[{{Sotani} {et~al.}(2021){Sotani}, {Takiwaki}, \&
  {Togashi}}]{Sotani.2021}
{Sotani}, H., {Takiwaki}, T., \& {Togashi}, H. 2021, \prd, 104, 123009,
  \dodoi{10.1103/PhysRevD.104.123009}

\bibitem[{{Steiner} {et~al.}(2013){Steiner}, {Hempel}, \& {Fischer}}]{sfho}
{Steiner}, A.~W., {Hempel}, M., \& {Fischer}, T. 2013, \apj, 774, 17,
  \dodoi{10.1088/0004-637X/774/1/17}

\bibitem[{{Szczepa{\'n}czyk} {et~al.}(2021){Szczepa{\'n}czyk}, {Antelis},
  {Benjamin}, {Cavagli{\`a}}, {Gondek-Rosi{\'n}ska}, {Hansen}, {Klimenko},
  {Morales}, {Moreno}, {Mukherjee}, {Nurbek}, {Powell}, {Singh},
  {Sitmukhambetov}, {Szewczyk}, {Valdez}, {Vedovato}, {Westhouse}, {Zanolin},
  \& {Zheng}}]{szczepanczy_cWB-ccsne-detect_2021}
{Szczepa{\'n}czyk}, M.~J., {Antelis}, J.~M., {Benjamin}, M., {et~al.} 2021,
  \prd, 104, 102002, \dodoi{10.1103/PhysRevD.104.102002}

\bibitem[{{Takiwaki} {et~al.}(2021){Takiwaki}, {Kotake}, \&
  {Foglizzo}}]{takiwaki_2021}
{Takiwaki}, T., {Kotake}, K., \& {Foglizzo}, T. 2021, \mnras, 508, 966,
  \dodoi{10.1093/mnras/stab2607}

\bibitem[{{The LIGO Scientific Collaboration} {et~al.}(2021){The LIGO
  Scientific Collaboration}, {the Virgo Collaboration}, {the KAGRA
  Collaboration}, {Abbott}, {Abe}, {Acernese}, {Ackley}, {Adhikari},
  {Adhikari}, {Adkins}, {Adya}, {Affeldt}, {Agarwal}, {Agathos}, {Agatsuma},
  {Aggarwal}, {Aguiar}, {Aiello}, {Ain}, {Ajith}, {Akutsu}, {de Alarc{\'o}n},
  {Albanesi}, {Alfaidi}, {Allocca}, {Altin}, {Amato}, {Anand}, {Anand},
  {Ananyeva}, {Anderson}, {Anderson}, {Ando}, {Andrade}, {Andres},
  {Andr{\'e}s-Carcasona}, {Andri{\'c}}, {Angelova}, {Ansoldi}, {Antelis},
  {Antier}, {Apostolatos}, {Appavuravther}, {Appert}, {Apple}, {Arai}, {Araya},
  {Araya}, {Areeda}, {Ar{\`e}ne}, {Aritomi}, {Arnaud}, {Arogeti}, {Aronson},
  {Arun}, {Asada}, {Asali}, {Ashton}, {Aso}, {Assiduo}, {Assis de Souza Melo},
  {Aston}, {Astone}, {Aubin}, {AultONeal}, {Austin}, {Babak}, {Badaracco},
  {Bader}, {Badger}, {Bae}, {Bae}, {Baer}, {Bagnasco}, {Bai}, {Baird},
  {Bajpai}, {Baka}, {Ball}, {Ballardin}, {Ballmer}, {Balsamo}, {Baltus},
  {Banagiri}, {Banerjee}, {Bankar}, {Barayoga}, {Barbieri}, {Barish}, {Barker},
  {Barneo}, {Barone}, {Barr}, {Barsotti}, {Barsuglia}, {Barta}, {Bartlett},
  {Barton}, {Bartos}, {Basak}, {Bassiri}, {Basti}, {Bawaj}, {Bayley}, {Bazzan},
  {Becher}, {B{\'e}csy}, {Bedakihale}, {Beirnaert}, {Bejger}, {Belahcene},
  {Benedetto}, {Beniwal}, {Benjamin}, {Bennett}, {Bentley}, {BenYaala}, {Bera},
  {Berbel}, {Bergamin}, {Berger}, {Bernuzzi}, {Berry}, {Bersanetti},
  {Bertolini}, {Betzwieser}, {Beveridge}, {Bhandare}, {Bhandari}, {Bhardwaj},
  {Bhatt}, {Bhattacharjee}, {Bhaumik}, {Bianchi}, {Bilenko}, {Billingsley},
  {Bini}, {Birney}, {Birnholtz}, {Biscans}, {Bischi}, {Biscoveanu}, {Bisht},
  {Biswas}, {Bitossi}, {Bizouard}, {Blackburn}, {Blair}, {Blair}, {Blair},
  {Bobba}, {Bode}, {Bo{\"e}r}, {Bogaert}, {Boldrini}, {Bolingbroke},
  {Bonavena}, {Bondu}, {Bonilla}, {Bonnand}, {Booker}, {Boom}, {Bork},
  {Boschi}, {Bose}, {Bose}, {Bossilkov}, {Boudart}, {Bouffanais}, {Bozzi},
  {Bradaschia}, {Brady}, {Bramley}, {Branch}, {Branchesi}, {Brau}, {Breschi},
  {Briant}, {Briggs}, {Brillet}, {Brinkmann}, {Brockill}, {Brooks}, {Brooks},
  {Brown}, {Brunett}, {Bruno}, {Bruntz}, {Bryant}, {Bucci}, {Bulik}, {Bulten},
  {Buonanno}, {Burtnyk}, {Buscicchio}, {Buskulic}, {Buy}, {Byer}, {Cabourn
  Davies}, {Cabras}, {Cabrita}, {Cadonati}, {Caesar}, {Cagnoli}, {Cahillane},
  {Calder{\'o}n Bustillo}, {Callaghan}, {Callister}, {Calloni}, {Cameron},
  {Camp}, {Canepa}, {Canevarolo}, {Cannavacciuolo}, {Cannon}, {Cao}, {Cao},
  {Capocasa}, {Capote}, {Carapella}, {Carbognani}, {Carlassara}, {Carlin},
  {Carney}, {Carpinelli}, {Carrillo}, {Carullo}, {Carver}, {Casanueva Diaz},
  {Casentini}, {Castaldi}, {Caudill}, {Cavagli{\`a}}, {Cavalier}, {Cavalieri},
  {Cella}, {Cerd{\'a}-Dur{\'a}n}, {Cesarini}, {Chaibi}, {Chalathadka
  Subrahmanya}, {Champion}, {Chan}, {Chan}, {Chan}, {Chan}, {Chan}, {Chandra},
  {Chang}, {Chanial}, {Chao}, {Chapman-Bird}, {Charlton}, {Chase},
  {Chassande-Mottin}, {Chatterjee}, {Chatterjee}, {Chatterjee}, {Chaturvedi},
  {Chaty}, {Chatziioannou}, {Chen}, {Chen}, {Chen}, {Chen}, {Chen}, {Chen},
  {Chen}, {Chen}, {Chen}, {Cheng}, {Cheong}, {Cheung}, {Chia}, {Chiadini},
  {Chiang}, {Chiarini}, {Chierici}, {Chincarini}, {Chiofalo}, {Chiummo},
  {Choudhary}, {Choudhary}, {Christensen}, {Chu}, {Chu}, {Chua}, {Chung},
  {Ciani}, {Ciecielag}, {Cie{\'s}lar}, {Cifaldi}, {Ciobanu}, {Ciolfi},
  {Cipriano}, {Clara}, {Clark}, {Clearwater}, {Clesse}, {Cleva}, {Coccia},
  {Codazzo}, {Cohadon}, {Cohen}, {Colleoni}, {Collette}, {Colombo}, {Colpi},
  {Compton}, {Constancio}, {Conti}, {Cooper}, {Corban}, {Corbitt},
  {Cordero-Carri{\'o}n}, {Corezzi}, {Corley}, {Cornish}, {Corre}, {Corsi},
  {Cortese}, {Costa}, {Cotesta}, {Cottingham}, {Coughlin}, {Coulon},
  {Countryman}, {Cousins}, {Couvares}, {Coward}, {Cowart}, {Coyne}, {Coyne},
  {Creighton}, {Creighton}, {Criswell}, {Croquette}, {Crowder}, {Cudell},
  {Cullen}, {Cumming}, {Cummings}, {Cunningham}, {Cuoco}, {Cury{\l}o},
  {Dabadie}, {Dal Canton}, {Dall'Osso}, {D{\'a}lya}, {Dana}, {D'Angelo},
  {Danilishin}, {D'Antonio}, {Danzmann}, {Darsow-Fromm}, {Dasgupta}, {Datrier},
  {Datta}, {Datta}, {Dattilo}, {Dave}, {Davier}, {Davis}, {Davis}, {Daw},
  {Dean}, {DeBra}, {Deenadayalan}, {Degallaix}, {De Laurentis},
  {Del{\'e}glise}, {Del Favero}, {De Lillo}, {De Lillo}, {Dell'Aquila}, {Del
  Pozzo}, {DeMarchi}, {De Matteis}, {D'Emilio}, {Demos}, {Dent}, {Depasse}, {De
  Pietri}, {De Rosa}, {De Rossi}, {DeSalvo}, {De Simone}, {Dhurandhar},
  {D{\'\i}az}, {Didio}, {Dietrich}, {Di Fiore}, {Di Fronzo}, {Di Giorgio}, {Di
  Giovanni}, {Di Giovanni}, {Di Girolamo}, {Di Lieto}, {Di Michele}, {Ding},
  {Di Pace}, {Di Palma}, {Di Renzo}, {Divakarla}, {Divyajyoti}, {Dmitriev},
  {Doctor}, {Donahue}, {D'Onofrio}, {Donovan}, {Dooley}, {Doravari}, {Drago},
  {Driggers}, {Drori}, {Ducoin}, {Dupej}, {Dupletsa}, {Durante}, {D'Urso},
  {Duverne}, {Dwyer}, {Eassa}, {Easter}, {Ebersold}, {Eckhardt}, {Eddolls},
  {Edelman}, {Edo}, {Edy}, {Effler}, {Eguchi}, {Eichholz}, {Eikenberry},
  {Eisenmann}, {Eisenstein}, {Ejlli}, {Engelby}, {Enomoto}, {Errico}, {Essick},
  {Estell{\'e}s}, {Estevez}, {Etienne}, {Etzel}, {Evans}, {Evans},
  {Evstafyeva}, {Ewing}, {Fabrizi}, {Faedi}, {Fafone}, {Fair}, {Fairhurst},
  {Fan}, {Farah}, {Farinon}, {Farr}, {Farr}, {Fauchon-Jones}, {Favaro},
  {Favata}, {Fays}, {Fazio}, {Feicht}, {Fejer}, {Fenyvesi}, {Ferguson},
  {Fernandez-Galiana}, {Ferrante}, {Ferreira}, {Fidecaro}, {Figura}, {Fiori},
  {Fiori}, {Fishbach}, {Fisher}, {Fittipaldi}, {Fiumara}, {Flaminio}, {Floden},
  {Fong}, {Font}, {Fornal}, {Forsyth}, {Franke}, {Frasca}, {Frasconi}, {Freed},
  {Frei}, {Freise}, {Freitas}, {Frey}, {Fritschel}, {Frolov}, {Fronz{\'e}},
  {Fujii}, {Fujikawa}, {Fujimoto}, {Fulda}, {Fyffe}, {Gabbard}, {Gabella},
  {Gadre}, {Gair}, {Gais}, {Galaudage}, {Gamba}, {Ganapathy}, {Ganguly}, {Gao},
  {Gaonkar}, {Garaventa}, {Garc{\'\i}a N{\'u}{\~n}ez},
  {Garc{\'\i}a-Quir{\'o}s}, {Garufi}, {Gateley}, {Gayathri}, {Ge}, {Gemme},
  {Gennai}, {George}, {Gerberding}, {Gergely}, {Gewecke}, {Ghonge}, {Ghosh},
  {Ghosh}, {Ghosh}, {Ghosh}, {Ghosh}, {Giacomazzo}, {Giacoppo}, {Giaime},
  {Giardina}, {Gibson}, {Gier}, {Giesler}, {Giri}, {Gissi}, {Gkaitatzis},
  {Glanzer}, {Gleckl}, {Godwin}, {Goetz}, {Goetz}, {Gohlke}, {Golomb},
  {Goncharov}, {Gonz{\'a}lez}, {Gosselin}, {Gouaty}, {Gould}, {Goyal}, {Grace},
  {Grado}, {Graham}, {Granata}, {Granata}, {Grant}, {Gras}, {Grassia}, {Gray},
  {Gray}, {Greco}, {Green}, {Green}, {Gretarsson}, {Gretarsson}, {Griffith},
  {Griffiths}, {Griggs}, {Grignani}, {Grimaldi}, {Grimes}, {Grimm}, {Grote},
  {Grunewald}, {Gruning}, {Gruson}, {Guerra}, {Guidi}, {Guimaraes},
  {Guix{\'e}}, {Gulati}, {Gunny}, {Guo}, {Guo}, {Gupta}, {Gupta}, {Gupta},
  {Gupta}, {Gupta}, {Gustafson}, {Guzman}, {Ha}, {Hadiputrawan}, {Haegel},
  {Haino}, {Halim}, {Hall}, {Hamilton}, {Hammond}, {Han}, {Haney}, {Hanks},
  {Hanna}, {Hannam}, {Hannuksela}, {Hansen}, {Hansen}, {Hanson}, {Harder},
  {Haris}, {Harms}, {Harry}, {Harry}, {Hartwig}, {Hasegawa}, {Haskell},
  {Haster}, {Hathaway}, {Hattori}, {Haughian}, {Hayakawa}, {Hayama}, {Hayes},
  {Healy}, {Heidmann}, {Heidt}, {Heintze}, {Heinze}, {Heinzel}, {Heitmann},
  {Hellman}, {Hello}, {Helmling-Cornell}, {Hemming}, {Hendry}, {Heng},
  {Hennes}, {Hennig}, {Hennig}, {Henshaw}, {Hernandez}, {Hernandez Vivanco},
  {Heurs}, {Hewitt}, {Higginbotham}, {Hild}, {Hill}, {Himemoto}, {Hines},
  {Hirata}, {Hirose}, {Ho}, {Hochheim}, {Hofman}, {Hohmann}, {Holcomb},
  {Holland}, {Hollows}, {Holmes}, {Holt}, {Holz}, {Hong}, {Hough}, {Hourihane},
  {Howell}, {Hoy}, {Hoyland}, {Hreibi}, {Hsieh}, {Hsieh}, {Hsiung}, {Hsu},
  {Huang}, {Huang}, {Huang}, {Huang}, {Huang}, {Huang}, {H{\"u}bner},
  {Huddart}, {Hughey}, {Hui}, {Hui}, {Husa}, {Huttner}, {Huxford},
  {Huynh-Dinh}, {Ide}, {Idzkowski}, {Iess}, {Inayoshi}, {Inoue}, {Iosif},
  {Isi}, {Isleif}, {Ito}, {Itoh}, {Iyer}, {JaberianHamedan}, {Jacqmin},
  {Jacquet}, {Jadhav}, {Jadhav}, {Jain}, {James}, {Jan}, {Jani}, {Janquart},
  {Janssens}, {Janthalur}, {Jaranowski}, {Jariwala}, {Jaume}, {Jenkins},
  {Jenner}, {Jeon}, {Jia}, {Jiang}, {Jin}, {Johns}, {Johnson-McDaniel},
  {Johnston}, {Jones}, {Jones}, {Jones}, {Jones}, {Joshi}, {Ju}, {Jue}, {Jung},
  {Jung}, {Junker}, {Juste}, {Kaihotsu}, {Kajita}, {Kakizaki}, {Kalaghatgi},
  {Kalogera}, {Kamai}, {Kamiizumi}, {Kanda}, {Kandhasamy}, {Kang}, {Kanner},
  {Kao}, {Kapadia}, {Kapasi}, {Karathanasis}, {Karki}, {Kashyap}, {Kasprzack},
  {Kastaun}, {Kato}, {Katsanevas}, {Katsavounidis}, {Katzman}, {Kaur},
  {Kawabe}, {Kawaguchi}, {K{\'e}f{\'e}lian}, {Keitel}, {Key}, {Khadka},
  {Khalili}, {Khan}, {Khanam}, {Khazanov}, {Khetan}, {Khursheed}, {Kijbunchoo},
  {Kim}, {Kim}, {Kim}, {Kim}, {Kim}, {Kim}, {Kim}, {Kimball}, {Kimura},
  {Kinley-Hanlon}, {Kirchhoff}, {Kissel}, {Klimenko}, {Klinger}, {Knee},
  {Knowles}, {Knust}, {Knyazev}, {Kobayashi}, {Koch}, {Koekoek}, {Kohri},
  {Kokeyama}, {Koley}, {Kolitsidou}, {Kolstein}, {Komori}, {Kondrashov},
  {Kong}, {Kontos}, {Koper}, {Korobko}, {Kovalam}, {Koyama}, {Kozak},
  {Kozakai}, {Kringel}, {Krishnendu}, {Kr{\'o}lak}, {Kuehn}, {Kuei}, {Kuijer},
  {Kulkarni}, {Kumar}, {Kumar}, {Kumar}, {Kumar}, {Kume}, {Kuns}, {Kuromiya},
  {Kuroyanagi}, {Kwak}, {Lacaille}, {Lagabbe}, {Laghi}, {Lalande}, {Lalleman},
  {Lam}, {Lamberts}, {Landry}, {Lane}, {Lang}, {Lange}, {Lantz}, {La Rosa},
  {Lartaux-Vollard}, {Lasky}, {Laxen}, {Lazzarini}, {Lazzaro}, {Leaci},
  {Leavey}, {LeBohec}, {Lecoeuche}, {Lee}, {Lee}, {Lee}, {Lee}, {Lee},
  {Legred}, {Lehmann}, {Lema{\^\i}tre}, {Lenti}, {Leonardi}, {Leonova},
  {Leroy}, {Letendre}, {Levesque}, {Levin}, {Leviton}, {Leyde}, {Li}, {Li},
  {Li}, {Li}, {Li}, {Li}, {Li}, {Lin}, {Lin}, {Lin}, {Lin}, {Lin}, {Lin},
  {Linde}, {Linker}, {Linley}, {Littenberg}, {Liu}, {Liu}, {Liu}, {Liu},
  {Llamas}, {Lo}, {Lo}, {London}, {Longo}, {Lopez}, {Lopez Portilla},
  {Lorenzini}, {Loriette}, {Lormand}, {Losurdo}, {Lott}, {Lough}, {Lousto},
  {Lovelace}, {Lucaccioni}, {L{\"u}ck}, {Lumaca}, {Lundgren}, {Luo}, {Lynam},
  {Ma'arif}, {Macas}, {Machtinger}, {MacInnis}, {Macleod}, {MacMillan},
  {Macquet}, {Maga{\~n}a Hernandez}, {Magazz{\`u}}, {Magee}, {Maggiore},
  {Magnozzi}, {Mahesh}, {Majorana}, {Maksimovic}, {Maliakal}, {Malik}, {Man},
  {Mandic}, {Mangano}, {Mansell}, {Manske}, {Mantovani}, {Mapelli},
  {Marchesoni}, {Mar{\'\i}n Pina}, {Marion}, {Mark}, {M{\'a}rka}, {M{\'a}rka},
  {Markakis}, {Markosyan}, {Markowitz}, {Maros}, {Marquina}, {Marsat},
  {Martelli}, {Martin}, {Martin}, {Martinez}, {Martinez}, {Martinez},
  {Martinovic}, {Martynov}, {Marx}, {Masalehdan}, {Mason}, {Massera},
  {Masserot}, {Masso-Reid}, {Mastrogiovanni}, {Matas}, {Mateu-Lucena},
  {Matichard}, {Matiushechkina}, {Mavalvala}, {McCann}, {McCarthy},
  {McClelland}, {McClincy}, {McCormick}, {McCuller}, {McGhee}, {McGuire},
  {McIsaac}, {McIver}, {McRae}, {McWilliams}, {Meacher}, {Mehmet}, {Mehta},
  {Meijer}, {Melatos}, {Melchor}, {Mendell}, {Menendez-Vazquez}, {Menoni},
  {Mercer}, {Mereni}, {Merfeld}, {Merilh}, {Merritt}, {Merzougui}, {Meshkov},
  {Messenger}, {Messick}, {Meyers}, {Meylahn}, {Mhaske}, {Miani}, {Miao},
  {Michaloliakos}, {Michel}, {Michimura}, {Middleton}, {Mihaylov}, {Milano},
  {Miller}, {Miller}, {Miller}, {Millhouse}, {Mills}, {Milotti}, {Minenkov},
  {Mio}, {Mir}, {Miravet-Ten{\'e}s}, {Mishkin}, {Mishra}, {Mishra}, {Mistry},
  {Mitra}, {Mitrofanov}, {Mitselmakher}, {Mittleman}, {Miyakawa}, {Miyo},
  {Miyoki}, {Mo}, {Modafferi}, {Moguel}, {Mogushi}, {Mohapatra}, {Mohite},
  {Molina}, {Molina-Ruiz}, {Mondin}, {Montani}, {Moore}, {Moragues}, {Moraru},
  {Morawski}, {More}, {Moreno}, {Moreno}, {Mori}, {Morisaki}, {Morisue},
  {Moriwaki}, {Mours}, {Mow-Lowry}, {Mozzon}, {Muciaccia}, {Mukherjee},
  {Mukherjee}, {Mukherjee}, {Mukherjee}, {Mukherjee}, {Mukund}, {Mullavey},
  {Munch}, {Mu{\~n}iz}, {Murray}, {Musenich}, {Muusse}, {Nadji}, {Nagano},
  {Nagar}, {Nakamura}, {Nakano}, {Nakano}, {Nakayama}, {Napolano},
  {Nardecchia}, {Narikawa}, {Narola}, {Naticchioni}, {Nayak}, {Nayak}, {Neil},
  {Neilson}, {Nelson}, {Nelson}, {Nery}, {Neubauer}, {Neunzert}, {Ng}, {Ng},
  {Nguyen}, {Nguyen}, {Nguyen}, {Nguyen Quynh}, {Ni}, {Ni}, {Nichols},
  {Nishimoto}, {Nishizawa}, {Nissanke}, {Nitoglia}, {Nocera}, {Norman},
  {North}, {Nozaki}, {Nurbek}, {Nuttall}, {Obayashi}, {Oberling}, {O'Brien},
  {O'Dell}, {Oelker}, {Ogaki}, {Oganesyan}, {Oh}, {Oh}, {Oh}, {Ohashi},
  {Ohashi}, {Ohkawa}, {Ohme}, {Ohta}, {Okada}, {Okutani}, {Olivetto}, {Oohara},
  {Oram}, {O'Reilly}, {Ormiston}, {Ormsby}, {O'Shaughnessy}, {O'Shea},
  {Oshino}, {Ossokine}, {Osthelder}, {Otabe}, {Ottaway}, {Overmier}, {Pace},
  {Pagano}, {Pagano}, {Page}, {Pagliaroli}, {Pai}, {Pai}, {Pal}, {Palamos},
  {Palashov}, {Palomba}, {Pan}, {Pan}, {Panda}, {Pang}, {Pankow}, {Pannarale},
  {Pant}, {Panther}, {Paoletti}, {Paoli}, {Paolone}, {Pappas}, {Parisi},
  {Park}, {Park}, {Parker}, {Pascucci}, {Pasqualetti}, {Passaquieti},
  {Passuello}, {Patel}, {Pathak}, {Patricelli}, {Patron}, {Paul}, {Payne},
  {Pedraza}, {Pedurand}, {Pegoraro}, {Pele}, {Pe{\~n}a Arellano}, {Penano},
  {Penn}, {Perego}, {Pereira}, {Pereira}, {Perez}, {P{\'e}rigois}, {Perkins},
  {Perreca}, {Perri{\`e}s}, {Pesios}, {Petermann}, {Petterson}, {Pfeiffer},
  {Pham}, {Pham}, {Phukon}, {Phurailatpam}, {Piccinni}, {Pichot}, {Piendibene},
  {Piergiovanni}, {Pierini}, {Pierro}, {Pillant}, {Pillas}, {Pilo}, {Pinard},
  {Pineda-Bosque}, {Pinto}, {Pinto}, {Piotrzkowski}, {Piotrzkowski}, {Pirello},
  {Pitkin}, {Placidi}, {Placidi}, {Planas}, {Plastino}, {Pluchar}, {Poggiani},
  {Polini}, {Pong}, {Ponrathnam}, {Porter}, {Poulton}, {Poverman}, {Powell},
  {Pracchia}, {Pradier}, {Prajapati}, {Prasai}, {Prasanna}, {Pratten},
  {Principe}, {Prodi}, {Prokhorov}, {Prosposito}, {Prudenzi}, {Puecher},
  {Punturo}, {Puosi}, {Puppo}, {P{\"u}rrer}, {Qi}, {Quartey}, {Quetschke},
  {Quinonez}, {Quitzow-James}, {Qutob}, {Raab}, {Raaijmakers}, {Radkins},
  {Radulesco}, {Raffai}, {Rail}, {Raja}, {Rajan}, {Ramirez}, {Ramirez},
  {Ramos-Buades}, {Rana}, {Rapagnani}, {Ray}, {Raymond}, {Raza}, {Razzano},
  {Read}, {Rees}, {Regimbau}, {Rei}, {Reid}, {Reid}, {Reitze}, {Relton},
  {Renzini}, {Rettegno}, {Revenu}, {Reza}, {Rezac}, {Ricci}, {Richards},
  {Richardson}, {Richardson}, {Riemenschneider}, {Riles}, {Rinaldi}, {Rink},
  {Robertson}, {Robie}, {Robinet}, {Rocchi}, {Rodriguez}, {Rolland}, {Rollins},
  {Romanelli}, {Romano}, {Romel}, {Romero}, {Romero-Shaw}, {Romie}, {Ronchini},
  {Rosa}, {Rose}, {Rosi{\'n}ska}, {Ross}, {Rowan}, {Rowlinson}, {Roy}, {Roy},
  {Roy}, {Rozza}, {Ruggi}, {Ruiz-Rocha}, {Ryan}, {Sachdev}, {Sadecki}, {Sadiq},
  {Saha}, {Saito}, {Sakai}, {Sakellariadou}, {Sakon}, {Salafia},
  {Salces-Carcoba}, {Salconi}, {Saleem}, {Salemi}, {Samajdar}, {Sanchez},
  {Sanchez}, {Sanchez}, {Sanchis-Gual}, {Sanders}, {Sanuy}, {Saravanan},
  {Sarin}, {Sassolas}, {Satari}, {Sathyaprakash}, {Sauter}, {Savage}, {Savant},
  {Sawada}, {Sawant}, {Sayah}, {Schaetzl}, {Scheel}, {Scheuer}, {Schiworski},
  {Schmidt}, {Schmidt}, {Schnabel}, {Schneewind}, {Schofield}, {Sch{\"o}nbeck},
  {Schulte}, {Schutz}, {Schwartz}, {Scott}, {Scott}, {Seglar-Arroyo},
  {Sekiguchi}, {Sellers}, {Sengupta}, {Sentenac}, {Seo}, {Sequino}, {Sergeev},
  {Setyawati}, {Shaffer}, {Shahriar}, {Shaikh}, {Shams}, {Shao}, {Sharma},
  {Sharma}, {Shawhan}, {Shcheblanov}, {Sheela}, {Shikano}, {Shikauchi},
  {Shimizu}, {Shimode}, {Shinkai}, {Shishido}, {Shoda}, {Shoemaker},
  {Shoemaker}, {ShyamSundar}, {Sieniawska}, {Sigg}, {Silenzi}, {Singer},
  {Singh}, {Singh}, {Singh}, {Singha}, {Sintes}, {Sipala}, {Skliris},
  {Slagmolen}, {Slaven-Blair}, {Smetana}, {Smith}, {Smith}, {Smith},
  {Soldateschi}, {Somala}, {Somiya}, {Song}, {Soni}, {Soni}, {Sordini},
  {Sorrentino}, {Sorrentino}, {Soulard}, {Souradeep}, {Sowell}, {Spagnuolo},
  {Spencer}, {Spera}, {Spinicelli}, {Srivastava}, {Srivastava}, {Staats},
  {Stachie}, {Stachurski}, {Steer}, {Steinhoff}, {Steinlechner},
  {Steinlechner}, {Stergioulas}, {Stops}, {Stover}, {Strain}, {Strang},
  {Stratta}, {Strong}, {Strunk}, {Sturani}, {Stuver}, {Suchenek}, {Sudhagar},
  {Sudhir}, {Sugimoto}, {Suh}, {Sullivan}, {Sullivan}, {Summerscales}, {Sun},
  {Sunil}, {Sur}, {Suresh}, {Sutton}, {Suzuki}, {Suzuki}, {Suzuki}, {Swinkels},
  {Szczepa{\'n}czyk}, {Szewczyk}, {Tacca}, {Tagoshi}, {Tait}, {Takahashi},
  {Takahashi}, {Takano}, {Takeda}, {Takeda}, {Talbot}, {Talbot}, {Tanaka},
  {Tanaka}, {Tanaka}, {Tanasijczuk}, {Tanioka}, {Tanner}, {Tao}, {Tao},
  {Tapia}, {Tapia San Mart{\'\i}n}, {Taranto}, {Taruya}, {Tasson}, {Tenorio},
  {Terhune}, {Terkowski}, {Thirugnanasambandam}, {Thomas}, {Thomas},
  {Thompson}, {Thompson}, {Thondapu}, {Thorne}, {Thrane}, {Tiwari}, {Tiwari},
  {Tiwari}, {Toivonen}, {Tolley}, {Tomaru}, {Tomura}, {Tonelli}, {Tornasi},
  {Torres-Forn{\'e}}, {Torrie}, {Tosta e Melo}, {T{\"o}yr{\"a}}, {Trapananti},
  {Travasso}, {Traylor}, {Trevor}, {Tringali}, {Tripathee}, {Troiano},
  {Trovato}, {Trozzo}, {Trudeau}, {Tsai}, {Tsang}, {Tsang}, {Tsao}, {Tse},
  {Tso}, {Tsuchida}, {Tsukada}, {Tsuna}, {Tsutsui}, {Turbang}, {Turconi},
  {Tuyenbayev}, {Ubhi}, {Uchikata}, {Uchiyama}, {Udall}, {Ueda}, {Uehara},
  {Ueno}, {Ueshima}, {Unnikrishnan}, {Urban}, {Ushiba}, {Utina}, {Vajente},
  {Vajpeyi}, {Valdes}, {Valentini}, {Valsan}, {van Bakel}, {van Beuzekom}, {van
  Dael}, {van den Brand}, {Van Den Broeck}, {Vander-Hyde}, {van Haevermaet},
  {van Heijningen}, {van Putten}, {van Remortel}, {Vardaro}, {Vargas}, {Varma},
  {Vas{\'u}th}, {Vecchio}, {Vedovato}, {Veitch}, {Veitch}, {Venneberg},
  {Venugopalan}, {Verkindt}, {Verma}, {Verma}, {Vermeulen}, {Veske}, {Vetrano},
  {Vicer{\'e}}, {Vidyant}, {Viets}, {Vijaykumar}, {Villa-Ortega}, {Vinet},
  {Virtuoso}, {Vitale}, {Vocca}, {von Reis}, {von Wrangel}, {Vorvick},
  {Vyatchanin}, {Wade}, {Wade}, {Wagner}, {Wald}, {Walet}, {Walker}, {Wallace},
  {Wallace}, {Wang}, {Wang}, {Wang}, {Ward}, {Warner}, {Was}, {Washimi},
  {Washington}, {Watchi}, {Weaver}, {Weaving}, {Webster}, {Weinert},
  {Weinstein}, {Weiss}, {Weller}, {Weller}, {Wellmann}, {Wen}, {We{\ss}els},
  {Wette}, {Whelan}, {White}, {Whiting}, {Whittle}, {Wilken}, {Williams},
  {Williams}, {Williamson}, {Willis}, {Willke}, {Wilson}, {Wipf}, {Wlodarczyk},
  {Woan}, {Woehler}, {Wofford}, {Wong}, {Wong}, {Wright}, {Wu}, {Wu}, {Wu},
  {Wysocki}, {Xiao}, {Yamada}, {Yamamoto}, {Yamamoto}, {Yamamoto}, {Yamashita},
  {Yamazaki}, {Yang}, {Yang}, {Yang}, {Yang}, {Yang}, {Yang}, {Yap}, {Yeeles},
  {Yeh}, {Yelikar}, {Ying}, {Yokoyama}, {Yokozawa}, {Yoo}, {Yoshioka}, {Yu},
  {Yu}, {Yuzurihara}, {Zadro{\.z}ny}, {Zanolin}, {Zeidler}, {Zelenova},
  {Zendri}, {Zevin}, {Zhan}, {Zhang}, {Zhang}, {Zhang}, {Zhang}, {Zhang},
  {Zhang}, {Zhao}, {Zhao}, {Zhao}, {Zhao}, {Zhou}, {Zhou}, {Zhu}, {Zhu},
  {Zimmerman}, {Zucker}, \& {Zweizig}}]{2021arXiv211206861T}
{The LIGO Scientific Collaboration}, {the Virgo Collaboration}, {the KAGRA
  Collaboration}, {et~al.} 2021, arXiv e-prints, arXiv:2112.06861.
\newblock \doarXiv{2112.06861}

\bibitem[{{Tol}(2021)}]{paul-tol-colors}
{Tol}, P. 2021, {Colour Schemes}, Tech. rep.
\newblock \url{https://personal.sron.nl/~pault/data/colourschemes.pdf}

\bibitem[{{Torres-Forn{\'e}} {et~al.}(2019{\natexlab{a}}){Torres-Forn{\'e}},
  {Cerd{\'a}-Dur{\'a}n}, {Obergaulinger}, {M\"uller}, \&
  {Font}}]{torres-forne_universal_2019}
{Torres-Forn{\'e}}, A., {Cerd{\'a}-Dur{\'a}n}, P., {Obergaulinger}, M.,
  {M\"uller}, B., \& {Font}, J.~A. 2019{\natexlab{a}}, Physical Review Letters,
  123, 051102, \dodoi{10.1103/PhysRevLett.123.051102}

\bibitem[{{Torres-Forn{\'e}} {et~al.}(2021){Torres-Forn{\'e}},
  {Cerd{\'a}-Dur{\'a}n}, {Obergaulinger}, {M{\"u}ller}, \&
  {Font}}]{torres-forne_universal_erratum}
{Torres-Forn{\'e}}, A., {Cerd{\'a}-Dur{\'a}n}, P., {Obergaulinger}, M.,
  {M{\"u}ller}, B., \& {Font}, J.~A. 2021, \prl, 127, 239901,
  \dodoi{10.1103/PhysRevLett.127.239901}

\bibitem[{{Torres-Forn{\'e}} {et~al.}(2018){Torres-Forn{\'e}},
  {Cerd{\'a}-Dur{\'a}n}, {Passamonti}, \& {Font}}]{torres-forne_I_2018}
{Torres-Forn{\'e}}, A., {Cerd{\'a}-Dur{\'a}n}, P., {Passamonti}, A., \& {Font},
  J.~A. 2018, \mnras, 474, 5272, \dodoi{10.1093/mnras/stx3067}

\bibitem[{{Torres-Forn{\'e}} {et~al.}(2019{\natexlab{b}}){Torres-Forn{\'e}},
  {Cerd{\'a}-Dur{\'a}n}, {Passamonti}, {Obergaulinger}, \&
  {Font}}]{torres-forne_II_2019}
{Torres-Forn{\'e}}, A., {Cerd{\'a}-Dur{\'a}n}, P., {Passamonti}, A.,
  {Obergaulinger}, M., \& {Font}, J.~A. 2019{\natexlab{b}}, \mnras, 482, 3967,
  \dodoi{10.1093/mnras/sty2854}

\bibitem[{{Vartanyan} \& {Burrows}(2020)}]{vartanyan_2020}
{Vartanyan}, D., \& {Burrows}, A. 2020, \apj, 901, 108,
  \dodoi{10.3847/1538-4357/abafac}

\bibitem[{{Vartanyan} {et~al.}(2023){Vartanyan}, {Burrows}, {Wang}, {Coleman},
  \& {White}}]{vartanyan_2023}
{Vartanyan}, D., {Burrows}, A., {Wang}, T., {Coleman}, M. S.~B., \& {White},
  C.~J. 2023, arXiv e-prints, arXiv:2302.07092,
  \dodoi{10.48550/arXiv.2302.07092}

\bibitem[{Viroli \& McLachlan(2019)}]{viroli_deep-gmm_2019}
Viroli, C., \& McLachlan, G.~J. 2019, Statistics and Computing, 29, 43

\bibitem[{Virtanen {et~al.}(2020)Virtanen, Gommers, Oliphant, Haberland, Reddy,
  Cournapeau, Burovski, Peterson, Weckesser, Bright, {van der Walt}, Brett,
  Wilson, Millman, Mayorov, Nelson, Jones, Kern, Larson, Carey, Polat, Feng,
  Moore, {VanderPlas}, Laxalde, Perktold, Cimrman, Henriksen, Quintero, Harris,
  Archibald, Ribeiro, Pedregosa, {van Mulbregt}, \& {SciPy 1.0
  Contributors}}]{2020SciPy-NMeth}
Virtanen, P., Gommers, R., Oliphant, T.~E., {et~al.} 2020, Nature Methods, 17,
  261, \dodoi{10.1038/s41592-019-0686-2}

\bibitem[{{Warren} {et~al.}(2020){Warren}, {Couch}, {O'Connor}, \&
  {Morozova}}]{warren_2020}
{Warren}, M.~L., {Couch}, S.~M., {O'Connor}, E.~P., \& {Morozova}, V. 2020,
  \apj, 898, 139, \dodoi{10.3847/1538-4357/ab97b7}

\bibitem[{{W}es {M}c{K}inney(2010)}]{mckinney-proc-scipy-2010}
{W}es {M}c{K}inney. 2010, in {P}roceedings of the 9th {P}ython in {S}cience
  {C}onference, ed. {S}t\'efan van~der {W}alt \& {J}arrod {M}illman, 56 -- 61,
  \dodoi{10.25080/Majora-92bf1922-00a}

\bibitem[{{Woosley} \& {Heger}(2007)}]{Woosley.Heger:2007}
{Woosley}, S.~E., \& {Heger}, A. 2007, \physrep, 442, 269,
  \dodoi{10.1016/j.physrep.2007.02.009}

\bibitem[{{Woosley} {et~al.}(2002){Woosley}, {Heger}, \&
  {Weaver}}]{Woosley.Heger:2002}
{Woosley}, S.~E., {Heger}, A., \& {Weaver}, T.~A. 2002, Reviews of Modern
  Physics, 74, 1015, \dodoi{10.1103/RevModPhys.74.1015}

\bibitem[{Yakunin {et~al.}(2015)Yakunin, Mezzacappa, Marronetti, Yoshida,
  Bruenn, Hix, Lentz, Bronson~Messer, Harris, Endeve, Blondin, \&
  Lingerfelt}]{yakunin_gravitational_2015}
Yakunin, K.~N., Mezzacappa, A., Marronetti, P., {et~al.} 2015, Physical Review
  D, 92, 084040, \dodoi{10.1103/PhysRevD.92.084040}

\end{thebibliography}
\bibliographystyle{aasjournal}

\appendix

\section{Comparison of Spacetime Geometries} \label{app:spacetime}
The Einstein equations of general relativity $G_{\mu\nu}= 8\pi G T_{\mu\nu}$ do not uniquely constrain the metric tensor $g_{\mu\nu}$. One must also specify a \textit{coordinate system} or \textit{gauge} to fully constrain the system. In the 3+1 formulation, the gauge can be expressed as equations for the lapse $\alpha$ and shift vector $\beta$ \citep{alcubierre}. The eigenfrequency analysis of TF19 assumes a \textit{conformally-flat} metric:
\begin{equation}
    ds^2 = \left( -\alpha^2 + \beta_i \beta^i \right) dt^2 + 2 \beta_i dt dx^i  + \psi^4\left( \eta_{ij} dx^i dx^j \right)
\end{equation}
where latin indices $i,j$ denote spatial coordinates, $t$ denotes a timelike coordinate and $x^i$ denotes spacelike coordinates, $\eta_{ij}$ is the spatial part of the Minkowski metric, and $\psi$ is known as the ``conformal factor''.
The factor $\psi$ may be a constant or a function of spacetime variables.
When $\psi$ is constant in space, the metric is spatially flat. The background metric for the linearized system from TF19 that we solved to compute the gravitational-wave eigenfrequencies of the proto-neutron star is expressed in spherical coordinates that are also isotropic, i.e. shiftless, as,
\begin{equation} \label{eq:tf-metric}
    ds^2 = -\alpha^2 dt^2 + \psi^4\left[ dr^2 + r^2 \left( d\theta^2 + sin^2(\theta) d\varphi^2 \right) \right]
\end{equation}
where $(r, \theta, \varphi)$ are the usual spherical coordinates of radius, polar angle, and azimuthal angle. In spherical symmetry, this is entirely a gauge choice. The assumption of conformal flatness constrains the lapse and the vanishing shift constrains the shift.

The metric assumed by \code{Agile} is expressed under a different gauge choice, based upon a Lagrangian formulation of general-relativistic hydrodynamics, where
%
\begin{equation} \label{eq:agile-metric}
    ds^2 = -\alpha^2 dt^2 + \left( \frac{\partial r}{\partial a} \frac{1}{\Gamma} \right)^2 da^2 + r^2 \left( d\theta^2 + \sin^2(\theta) d\varphi^2 \right).
\end{equation}
Here, $r$ is the areal radius,\footnote{Such that spheres have surface area $4\pi r^2$.} $a$ is a spatial coordinate tracking the mass of the fluid, and $\Gamma$ is the Lorentz factor, defined as 
\begin{equation}
    \Gamma = \sqrt{1 + u^2 - 2m / r}
\end{equation}
with a fluid velocity $u = \left( \partial r / \partial t \right) / \alpha$ and enclosed rest mass $m$.
Additional details on the metric used by \code{Agile} can be found in Chapter 7 of \cite{liebendoerfer_thesis_2000}.

The areal radius $r = r(a,t)$ prevents by-eye identification of a conformally-flat decomposition of this metric, as $da$ will decompose into cross terms of $dr$ and $dt$.
To express the \code{Agile} metric in a conformally-flat manner, we introduce a coordinate transformation of the areal radius into a ``conformal radius" $\tilde{r}$, where
\begin{equation} \label{eqn:conf_radius_defn}
    \tilde{r} = \Gamma r \frac{\partial \tilde{r}}{\partial r}.
\end{equation}
Now, we show that this coordinate transformation yields a metric that is isotropic like Equation~\ref{eq:tf-metric}.
By definition, $\tilde{r} = \tilde{r}(r, t)$, and so
\begin{equation} \label{eq:drtilde_initial}
    d\tilde{r} = \frac{\partial \tilde{r}}{\partial r} dr + \frac{\partial \tilde{r}}{dt} dt.
\end{equation}
Since $r = r(a,t)$,
\begin{equation}
    dr = \frac{\partial r}{\partial a} da + \frac{\partial r}{\partial t} dt,
\end{equation}
which we substitute into Equation~\ref{eq:drtilde_initial} to find
\begin{equation} \label{eq:drtilde_clean}
    d\tilde{r} = \frac{\partial \tilde{r}}{\partial r} \frac{\partial r}{\partial a} da + \left( \frac{\partial \tilde{r}}{\partial r}\frac{\partial r}{\partial t} + \frac{\partial \tilde{r}}{\partial t} \right) dt.
\end{equation}
To rewrite our metric in terms of the new coordinate $\tilde{r}$, we will need to manipulate this expression for $da$ in terms of $d\tilde{r}$.
However, as it currently stands, the additional $dt$ term would induce a cross-term $d\tilde{r} dt$, and thus a shift-full metric, when substituted in for $da$ in Equation~\ref{eq:agile-metric}.
From the definition of our proposed coordinate transformation alone, there is no immediately obvious reason that the $dt$ term should vanish.
Luckily, it turns out to be \textit{approximately} zero for our models.

For the linear perturbation analysis conducted in this work, our integration region is restricted to the proto-neutron star, whose surface is defined by a density of $10^{11}$ g/cm$^3$.
In this region of our models, the fluid velocity in geometric units $u \ll 1$, as our models are spherically-symmetric.\footnote{In higher dimensions, this may not be true due to convection within the proto-neutron star.}
We have inspected a few of our models carefully to confirm this property and found that $u$ is of
$\mathcal{O}(10^{-5})$ within the proto-neutron star, rising to at most values of $\mathcal{O}(10^{-3})$ at the outermost edge of the proto-neutron star.
Since $\partial r/\partial t = u \alpha$, and $\alpha$ is typically order 1 or smaller, we have that $\partial r/\partial t \approx 0$.
In a similar manner, we can argue that $\partial \tilde{r} / \partial t$ is small.
Given the definition of $\tilde{r}$, we can compute via the chain rule that
\begin{equation}
    \frac{\partial \tilde{r}}{\partial t} = \frac{\partial \Gamma}{\partial t} r \frac{\partial \tilde{r}}{\partial t} + \Gamma \frac{\partial r}{\partial t} \frac{\partial \tilde{r}}{\partial r} + \Gamma r \frac{\partial}{\partial t}\frac{\partial \tilde{r}}{\partial r}.
\end{equation}
The derivative of $\Gamma$ is
\begin{align}
    \frac{\partial \Gamma}{\partial t} &= \frac{1}{\Gamma}\left( 2u \frac{\partial u}{\partial t} + \frac{2 m}{r^2} \frac{\partial r}{\partial t} - \frac{2}{r} \frac{\partial m}{\partial t} \right) \approx -\frac{2}{r \Gamma} \frac{\partial m}{\partial t}
\end{align}
where we drop the first term as $u$ as small, and the second term as $\partial r / \partial t$ is small.
Generally, if the fluid velocity within and near the proto-neutron star is small, we would expect the mass flux $\partial m / \partial t$ to be small.
So, $\partial \Gamma / \partial t$ is small as well.
Therefore,
\begin{equation}
    \frac{\partial \tilde{r}}{\partial t} \approx \Gamma r \frac{\partial}{\partial t}\frac{\partial \tilde{r}}{\partial r}.
\end{equation}
By equality of mixed partial derivatives, we can rearrange the second derivatives on the right-hand side, to find
\begin{equation}
    \frac{\partial \tilde{r}}{\partial t} \approx \Gamma r \frac{\partial}{\partial r}\frac{\partial \tilde{r}}{\partial t}.
\end{equation}
One solution to this equation is that $\partial \tilde{r} / \partial t = 0$.
While there may be other solutions, we have directly computed $\partial \tilde{r} / \partial t$ (as described at the end of this appendix) and confirmed that it is approximately zero within the proto-neutron star (at most $\mathcal{O}(10^{-8})$, again in geometric units).

With $\partial r / \partial t$ and $\partial \tilde{r} / \partial t$ both small, we return to Equation~\ref{eq:drtilde_clean} to conclude that for $\tilde{r}$,
\begin{equation}
    d\tilde{r} \approx \frac{\partial \tilde{r}}{\partial r}\frac{\partial r}{\partial a} da,
\end{equation}
or alternatively,
\begin{equation}
    da \approx \left( \frac{\partial \tilde{r}}{\partial r} \frac{\partial r}{\partial a} \right)^{-1} d\tilde{r},
\end{equation}
which we substitute into the \code{Agile} metric, Equation~\ref{eq:agile-metric}, to find
\begin{align}
    ds^2 &= -\alpha^2 dt^2 + \left( \frac{\partial r}{\partial a} \frac{1}{\Gamma} \right)^2 \left( \frac{\partial \tilde{r}}{\partial r} \frac{\partial r}{\partial a} \right)^{-2} d\tilde{r}^2 + r^2 \left( d\theta^2 + \sin^2(\theta) d\varphi^2 \right) \\
    &= -\alpha^2 dt^2 + \left( \Gamma \frac{\partial \tilde{r}}{\partial r}  \right)^{-2} d\tilde{r}^2 + r^2 \left( d\theta^2 + \sin^2(\theta) d\varphi^2 \right).
\end{align}
We manipulate our definition of $\tilde{r}$ for $\tilde{r} / r$, as well as insert identity $(\tilde{r} / \tilde{r})^2$ into the last term of the metric, so
\begin{align}
    ds^2 &= -\alpha^2 dt^2 + \left( \frac{\tilde{r}}{r} \right)^{-2} d\tilde{r}^2 + r^2 \left(\frac{\tilde{r}}{\tilde{r}}\right)^2 \left( d\theta^2 + \sin^2(\theta) d\varphi^2 \right) \\
    &= -\alpha^2 dt^2 + \left( \frac{r}{\tilde{r}} \right)^{2} \left[ d\tilde{r}^2 + \tilde{r}^2 \left( d\theta^2 + \sin^2(\theta) d\varphi^2 \right) \right].
\end{align}
Comparing this to the conformally flat, isotropic metric in Equation~\ref{eq:tf-metric}, we can identify the conformal factor as
\begin{equation}
    \psi^4 = \left( \frac{r}{\tilde{r}} \right)^2.
\end{equation}

At each timestep of the eigenfrequency analysis, we calculate the conformal radius $\tilde{r}$ and conformal factor $\psi$ via the Backwards Eulerian method, and provide both to \code{GREAT} (using $\tilde{r}$ as part of the hydrodynamic background instead of the areal radius $r$ from \code{Agile}).
We manipulate our definition of $\tilde{r}$, at a particular timestep, as
\begin{equation}
    \frac{\partial \tilde{r}}{\partial r} = \frac{\tilde{r}}{r} \frac{1}{\Gamma(r)} \label{eqn:conf_diffeq}.
\end{equation}
Our outer boundary condition is that, as $r \rightarrow \infty$, we expect $\psi \rightarrow 1$, as we expect space to be flat far from the star.
In theory, we would implement this boundary condition by enforcing that $\tilde{r} = r$, $\Gamma = 1$, and thus $\partial \tilde{r} / \partial r = 1$ in the outermost zone of the star, number 180.
Then, for the remaining zones, we discretize Equation~\ref{eqn:conf_diffeq} as
\begin{equation} \label{eqn:discrete_conf_diffeq}
    \tilde{r}_i = \tilde{r}_{i + 1} - \left(r_{i+1} - r_i\right) \left( \frac{\partial \tilde{r}}{\partial r} \right)_{i + 1}
\end{equation}
which we evaluate for each zone $i = 1, ..., 179$, iterating backward from zone 179, and where after evaluating each $\tilde{r}_i$ we also calculate
\begin{equation}
    \left( \frac{\partial \tilde{r}}{\partial r} \right)_{i} = \frac{\tilde{r}_i}{r_i \Gamma(r_i)}
\end{equation}
as input to the iteration for the next zone.

In practice, due to the irregular spacing of radial zones resulting from the adaptive mesh in \code{Agile}, the numerical integration of Equation~\ref{eqn:conf_diffeq} will yield solutions where $\partial \psi / \partial r$ changes sign and falls below 1.
Specifically, we found that when a few radial zones are much closer together relative to the spacing of other zones, e.g. near the reverse shock, $\psi$ would begin decreasing towards the center of the star.
However, we expect $\psi$ to continually increase towards the center of the star as the compactness of the material increases.
So, we dynamically enforce the outer boundary condition in such a way as to always find physical solutions for $\psi$.
In particular, we attempt the integration of Equation~\ref{eqn:conf_diffeq} from zone 179 to 1, in the discrete form of Equation~\ref{eqn:discrete_conf_diffeq} with the outer boundary condition enforced in zone 180.
If $\psi < 1$ in any radial zone, we re-attempt the integration, but instead enforce the outer boundary condition in zone 180 \textit{and} 179, and then integrate from zone 178 to 1.
Again, if $\psi < 1$ in any radial zone, we re-attempt the integration once more with the outer boundary condition enforced in zones 180, 179, and 178.
We continue in this manner until $\psi \geq 1$ in every zone.
We have found that, in the most extreme cases, this procedure stops when the outer boundary has been enforced down to roughly zone number 100; however, this is still well outside the proto-neutron star at any iteration of our models, and so does not change the eigenfrequencies. 
\clearpage
\pagebreak

\section{Bootstrap Expectation-Maximization Confidence Interval} \label{app:bootstrap}
To compute the 2$\sigma$ confidence interval of the Gaussian mixture fit to the $f$-mode frequencies for each of our $M = \nns{}$ models, we use a bootstrap procedure with $N = 1000$ bootstrap samples.
The distributions $p(f | \theta_{ij})$ are evaluated at $n = 1000$ test frequencies $f$.
\begin{algorithm}
\SetAlgoLined
\KwInput{$M$: number of exploding supernova models; $N$: number of bootstrap samples; $n$: number of test frequencies over which to evaluate Gaussian mixture distributions}
\KwOutput{Bootstrap 2$\sigma$ confidence intervals of the Gaussian mixture fit for each model.}
    \For{$i=1,\dots,M$}{
        \For{$j=1,\dots,N$}{
            Randomly-sample, with replacement, from the set of $f$-mode frequencies for model $i$, $\{ f(t) \}_i$ to generate a new set of frequencies, $\{ f(t) \}_{i,\rm new}$, with equal length\;
            Apply the expectation-maximization algorithm with $k = 2$ to fit a two-component Gaussian mixture to $\{ f(t) \}_{i,\rm new}$, generating the $j$-th set of component means, standard deviations, and weights for model $i$, $\theta_{ij}$\;
            Compute the distribution $p_j(f | \theta_{ij})$ according to Equation~\ref{eq:gmm} over a range of $n$ test frequencies $f$, evenly spaced from $\min{\{ f(t) \}_i}$ to $\max{\{ f(t) \}_i}$\;
        }
        Compute the 5th and 95th quantiles of 
        $\{ p_j (f | \theta_{ij}) \}_{j = 1}^{N}$ at each test frequency $f$, to compute either side of the 2$\sigma$ confidence interval for the fit of model $i$.
    }
 \caption{Bootstrap confidence interval}
\end{algorithm}
\clearpage
\pagebreak

\section{Models That Are Poorly Fit by a Gaussian Mixture} \label{app-poor-fits}
\subsection{Misplaced Gaussian Peaks}

\begin{figure}
    \centering
    \includegraphics[width=0.48\linewidth]{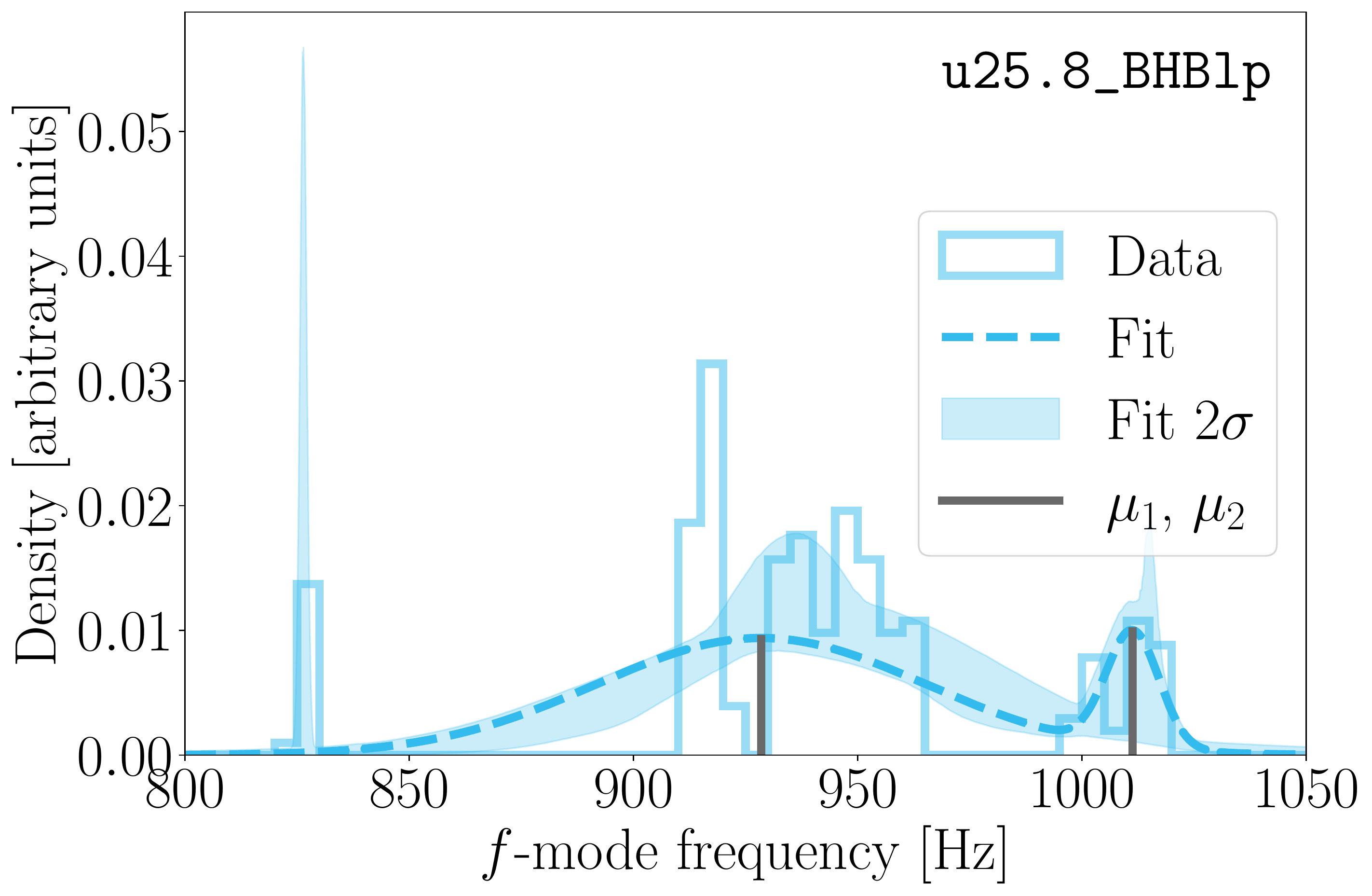}
    \includegraphics[width=0.48\linewidth]{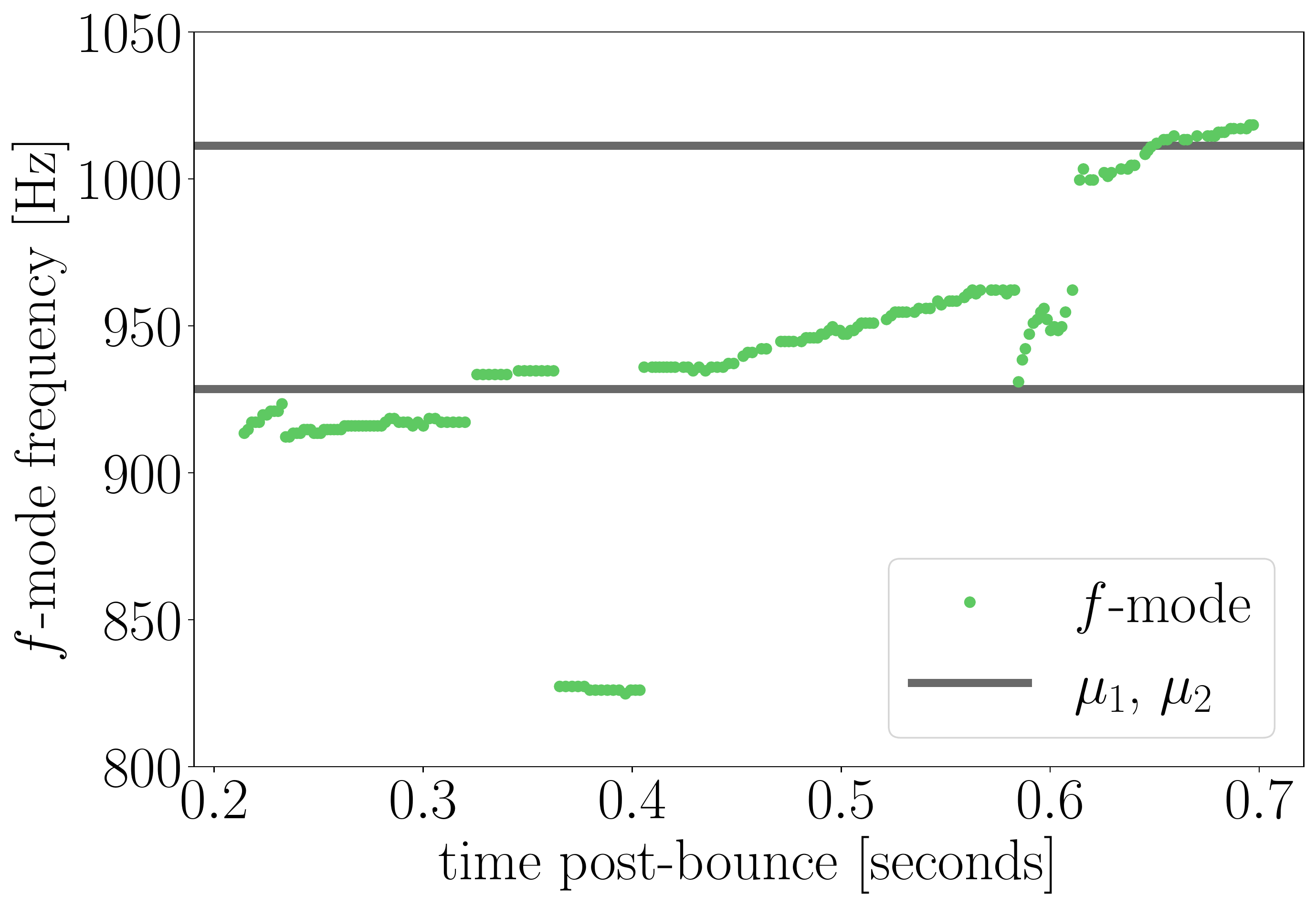} \\
    \includegraphics[width=0.48\linewidth]{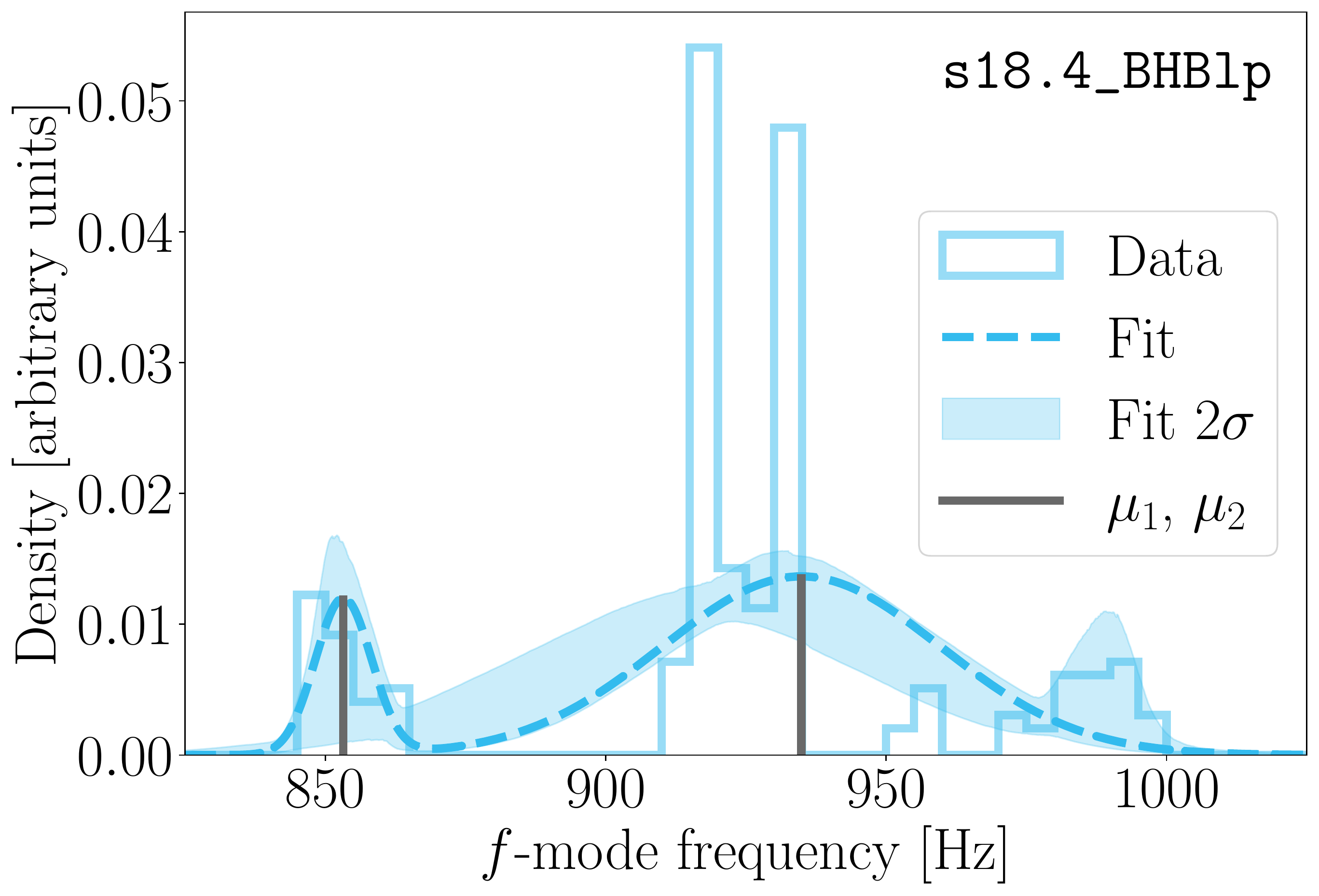}
    \includegraphics[width=0.48\linewidth]{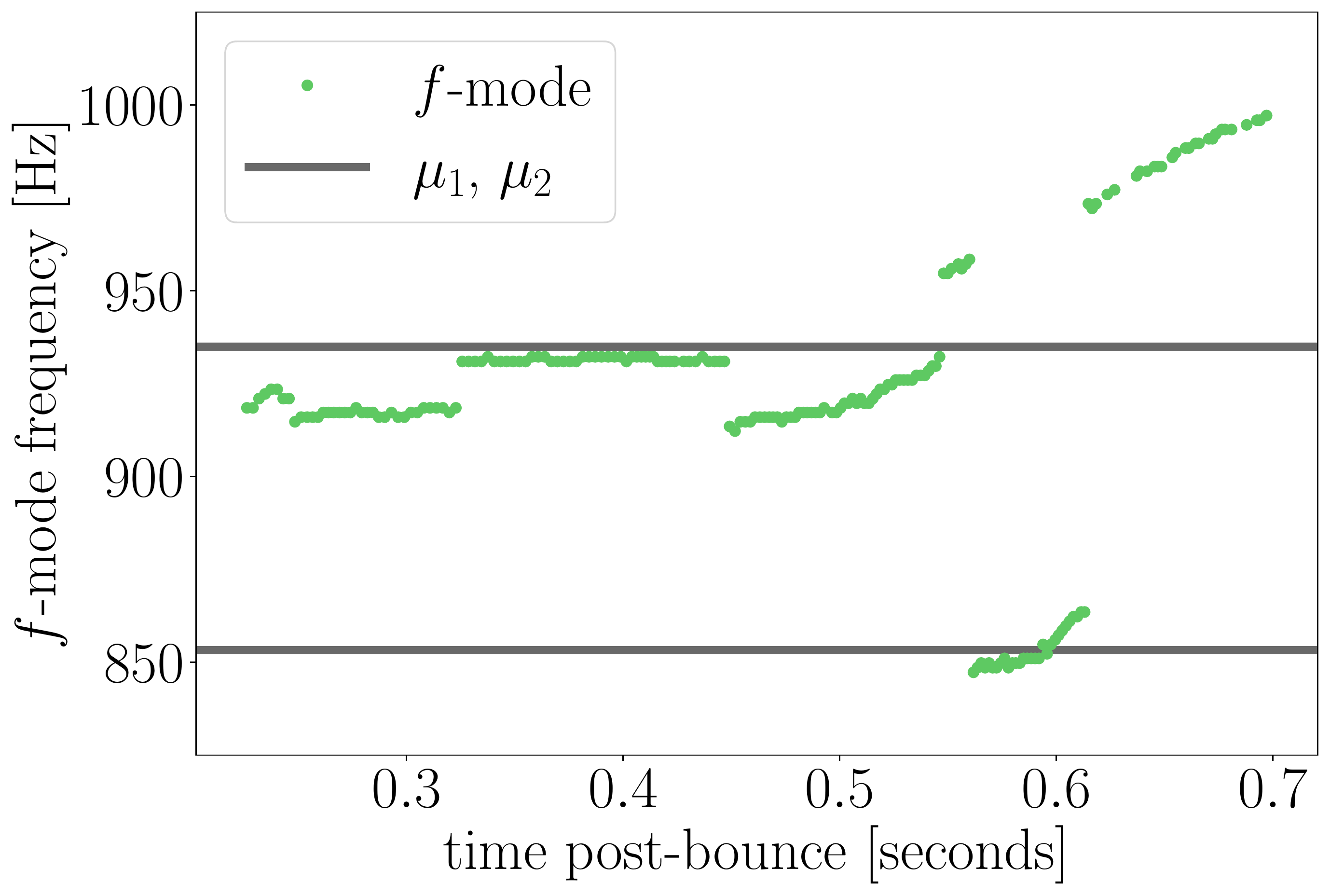} \\
    \includegraphics[width=0.48\linewidth]{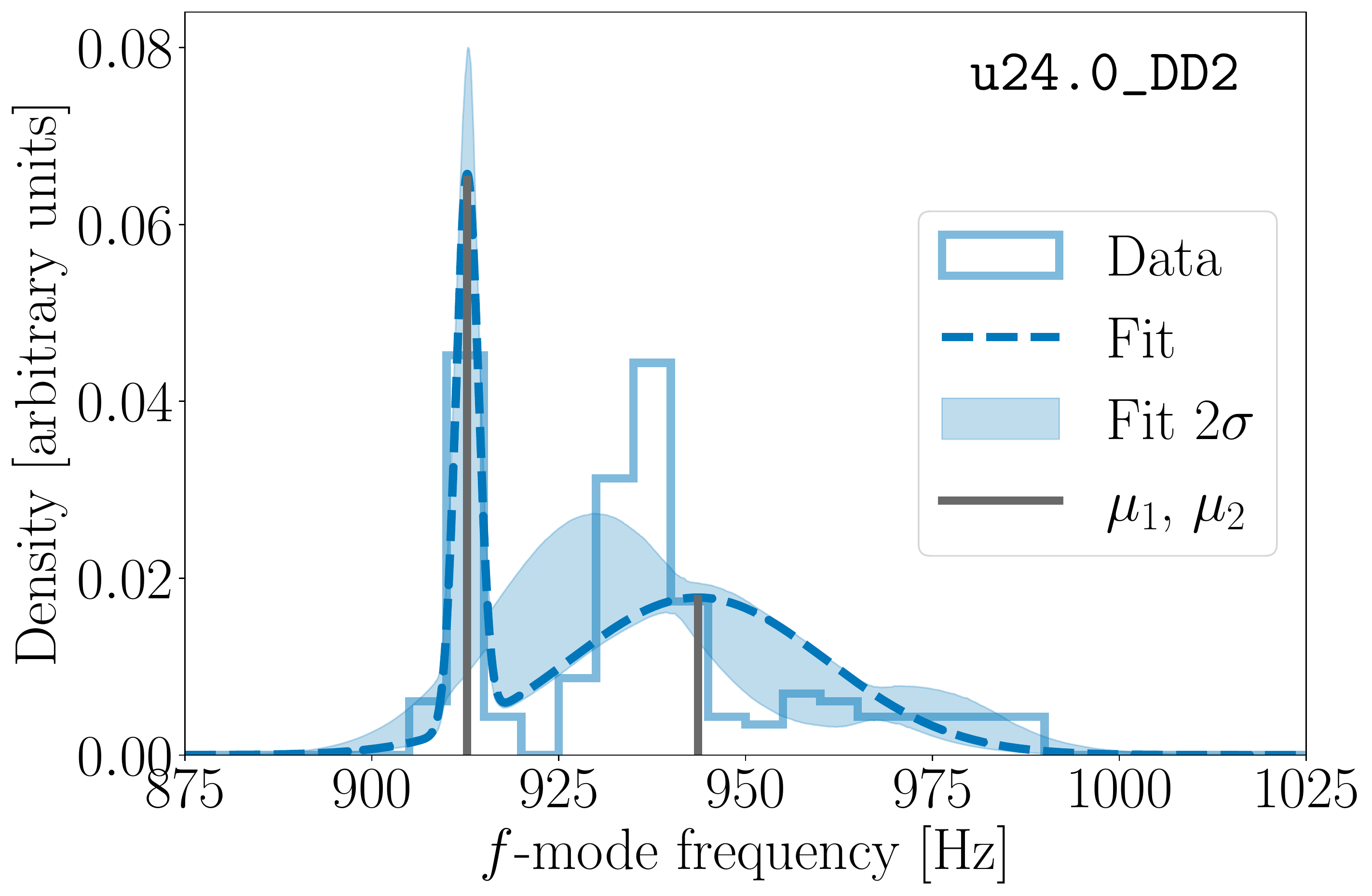}
    \includegraphics[width=0.48\linewidth]{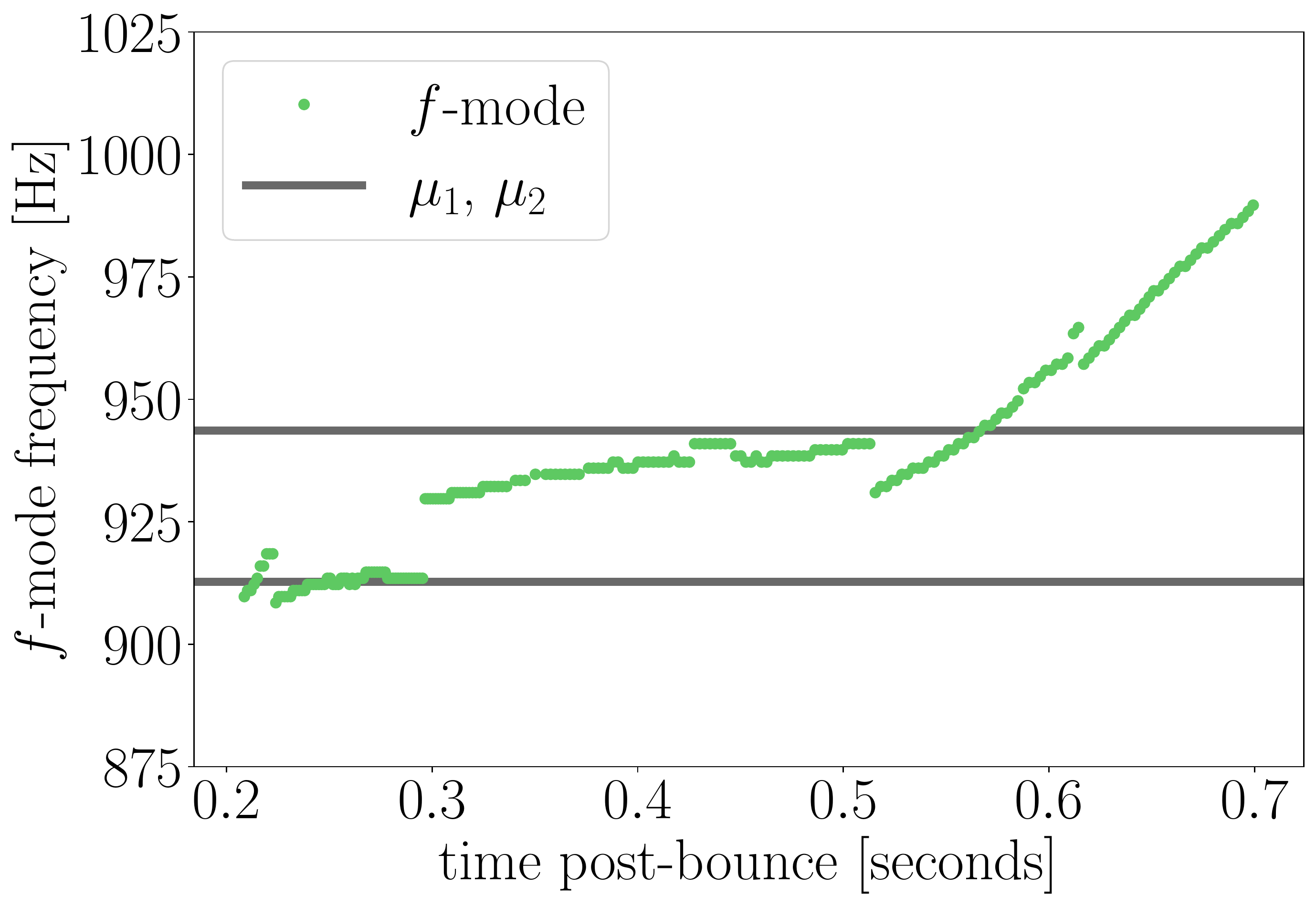}
    \caption{
    Left column: Histogram of the $f$-mode frequencies and the corresponding Gaussian mixture fit for models \code{u25.8_BHBlp} (top), \code{s18.4_BHBlp} (middle), and \code{u24.0_DD2} (bottom).
    Right column: Time evolution of the $f$-mode frequencies between 0.2 and 0.7 s post-bounce for models
    The discontinuity in $f$-mode frequencies is seen at ${\sim} 0.36$ seconds post-bounce (top), at ${\sim} 0.6$ seconds post-bounce (middle), and at ${\sim} 0.6$ seconds post-bounce (bottom). 
    Horizontal lines identify the $\mu_1$ and $\mu_2$ frequencies from the Gaussian mixture fit for each model.
    }
    \label{fig:poorly-fit}
\end{figure}

Despite the ability of a two-component Gaussian Mixture to identify characteristic $f$-mode frequencies for the majority of our models, for some models this functional form is a poor fit to the histogram of frequencies and so does not confidently identify two characteristic frequencies.
These poorly-fit models all contain some type of low-frequency discontinuity in their $f$-modes which impacts the confident identification of the Gaussian peak locations $\mu_1$ and $\mu_2$.
These discontinuities fall $\gtrsim 20$ Hz or more below the rest of the time-frequency evolution of the $f$-mode.
Since all of our models are one-dimensional, there is no physical explanation for why the resonant fundamental frequency of the proto-neutron star would change so severely and suddenly, as e.g. phenomena like downflows are not possible.
For a single model (\code{u24.0_BHBlp}) with a clear discontinuity in the $f$-mode, we have double-checked that this feature does not coincide with any abrupt changes in the evolution of the radius, density, enthalpy, and internal energy of the outermost radial zone within the PNS (defined by the $> 10^{11} g/{\rm cm}^{-3}$), nor in the central density of the PNS.
We also did not observe any coincident features in the radial profile of the rest mass, gravitational mass, fluid velocity, density, temperature, lapse, internal energy, potential energy, pressure, adiabatic index, entropy, and hyperon fraction at times near such a non-physical frequency feature. Thus, we conclude that these discontinuities are non-physical, even if we cannot precisely determine their origin.

Among the poorly-fit models, we find three distinct groups.
(i) In some models, any low-frequency discontinuities constitute only a small portion of the total histogram of frequencies, such that they do not change $\mu_1$ and $\mu_2$. However, the discontinuities noticeably increase the uncertainty in the location of $\mu_1$ (some realizations of the fit in the bootstrap procedure place the first Gaussian peak at a lower frequency to capture this discontinuity in the $f$-modes). An example of this behavior is shown in the top row of Figure~\ref{fig:poorly-fit}, which shows a sudden decrease in the $f$-mode frequencies from ${\sim} 940$ Hz down to ${\sim} 825$ Hz, at ${\sim}0.36$ seconds post-bounce. These unphysically low frequencies appear as a small peak at ${\sim} 825$ Hz in the histogram. Note that the $\mu_1$ and $\mu_2$ frequencies are close to the values one would identify by eye, ignoring the discontinuity in the $f$-mode frequencies.


(ii) The low-frequency discontinuities do constitute a large enough fraction of the total histogram of frequencies to change the identification of $\mu_1$ and $\mu_2$. For example, model \code{s18.4_BHBlp} (middle row in Figure~\ref{fig:poorly-fit}) shows a serious discontinuity in the $f$-modes at ${\sim} 0.6$ seconds post-bounce, which manifests as a peak at ${\sim} 850$ Hz in the frequency histogram. Note that this (unphysical) peak is of similar height and width as the (physical) peak at ${\sim} 985$ Hz. In these cases, the Gaussian mixture prefers to place $\mu_1$ at the lowest-frequency peak originating from the discontinuity in many realizations of the bootstrap procedure. This, in turn, lowers the position of the $\mu_2$ frequency to the peak at ${\sim}925$ Hz. These values for $\mu_1$ and $\mu_2$ are lower than what we expect if the discontinuity were not present (we would expect $\mu_1 \sim 925$ Hz and $\mu_2 \sim 985$ Hz).


(iii) The low-frequency discontinuities are a large enough fraction of the overall data to lower the value of $\mu_1$ but are still closer in value to the rest of the $f$-mode frequencies than in the other two cases. 
This increases the statistical confidence in the location of $\mu_1$, as this feature is less extreme in an absolute sense, even if still likely non-physical.
In the bottom row of Figure~\ref{fig:poorly-fit}, we can see an example of this pattern.
Here, the discontinuity at ${\sim} 0.2$ seconds post-bounce yields a low-frequency peak at ${\sim} 915$ Hz in the histogram.
Since this is within ${\sim} 20$ Hz of the next peak in the frequency histogram, the confidence interval around the location of $\mu_1$ is narrow. The resulting value for $\mu_1$ is similar (albeit slightly lower) than what we would expect without the discontinuity. As a consequence, $\mu_2$ is visibly shifted to a lower value than expected without discontinuity. 

\subsection{Impact on the Correlation between NS Surface Gravity and GW Frequencies}

\begin{figure}
    \centering
    \includegraphics[width=0.5\linewidth]{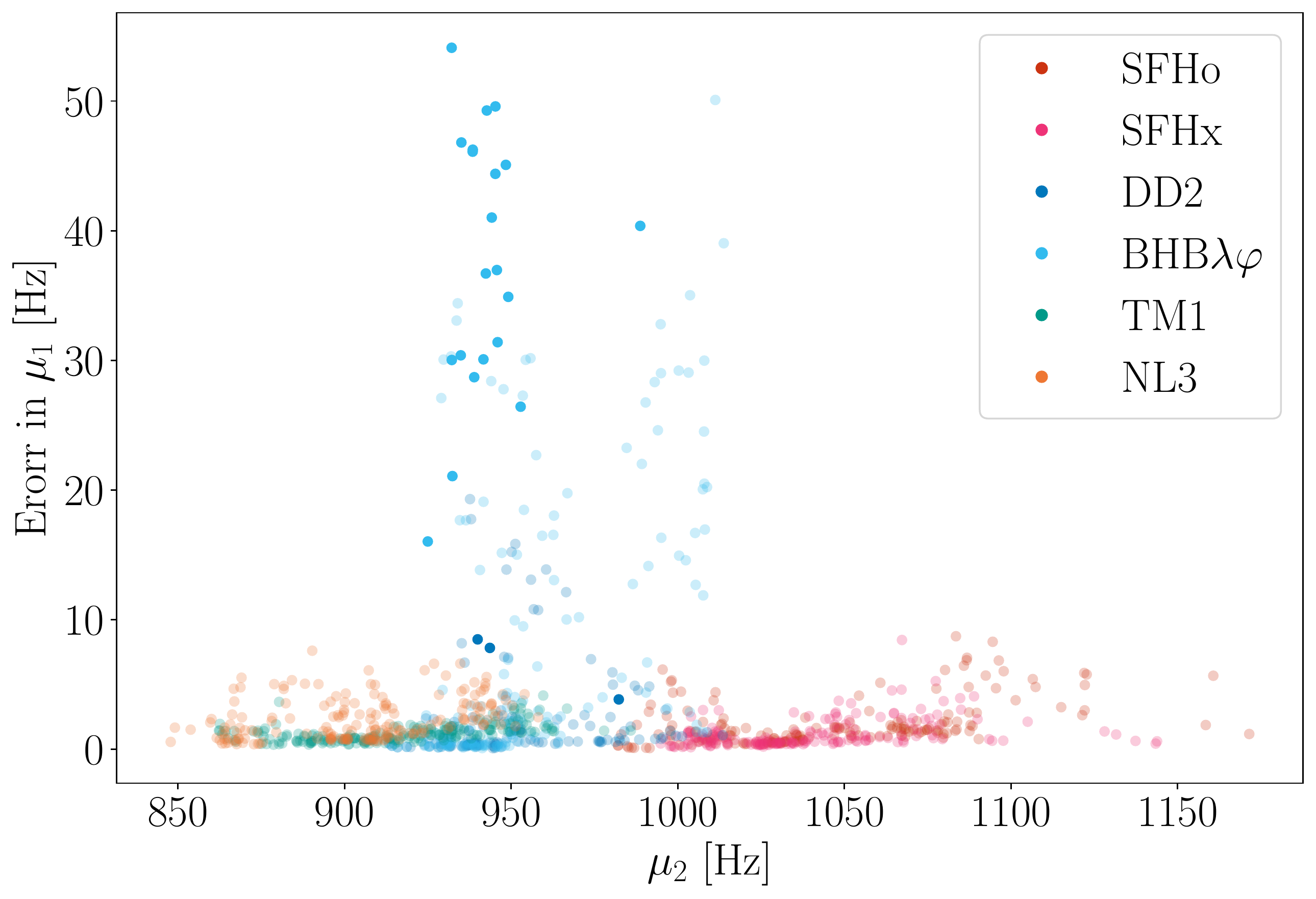}
    \caption{
    Error (1$\sigma$ confidence interval) in the location of the low-frequency Gaussian peak, $\mu_1$, versus the location of the high-frequency Gaussian peak, $\mu_2$, from the Gaussian mixture fit procedure for all models. 
    Points that lie outside the 2$\sigma$ confidence region of the linear fit in Figure~\ref{fig:mu_results} are shown with fully opaque symbols, and all other models are shown with semi-transparent symbols. All models with large $1\sigma$ confidence intervals are associated with one of the three cases of misidentified Gaussian peaks (see text for details).
    }
    \label{fig:mu1err-mu2}
\end{figure}

In the right panel of Figure~\ref{fig:mu_results}, we show a linear correlation (independent of the nuclear EOS) between the surface gravity of the cold, remnant neutron star and $\mu_2$. 
There is a clear subset of models which lie outside the 2$\sigma$ confidence region for this linear fit, consisting mostly of models with the BHB$\lambda \varphi$ EOS and three models with the DD2 EOS. 
We can fully explain these outliers with the three cases of poorly-fit Gaussian mixtures described above. 
In Figure~\ref{fig:mu1err-mu2}, we plot the 1$\sigma$ confidence interval at the location of the low-frequency Gaussian peak, $\mu_1$, against the location of the high-frequency peak, $\mu_2$, from the Gaussian mixture fit procedure for all models. 
The models that lie outside of the $2\sigma$ confidence interval in the right panel of Figure~\ref{fig:mu_results} are shown with opaque symbols. 
Of these, the BHB$\lambda\phi$ models correspond to models with a non-physical discontinuity in the frequency evolution, which ultimately shifts both $\mu_1$ and $\mu_2$ to much lower values than expected without such a discontinuity (see case (ii) above). Identifying $\mu_2$ at a lower frequency than expected shifts the models to the left in Figure~\ref{fig:mu_results}. 
The three DD2 models also shown with opaque symbols are examples of case (iii) discussed above, where $\mu_1$ is lower than would be expected without discontinuities, however, the fit is more confident in the location of $\mu_1$. 
Finally, there are models (semi-transparent points with large $\mu_1$ error) with a broad $\mu_1$ confidence interval that still lie within $2\sigma$ of the fit in Figure~\ref{fig:mu_results}. These are examples of case (i) discussed above, where the location of $\mu_1$ is not shifted much however the confidence in the location is low, resulting in a broader confidence interval.


\clearpage
\pagebreak

\section{Characteristic Frequencies Selected by Characteristic Times} \label{app:single-times}

%
%
%

\begin{figure*}
    \centering
    \includegraphics[width=\linewidth]{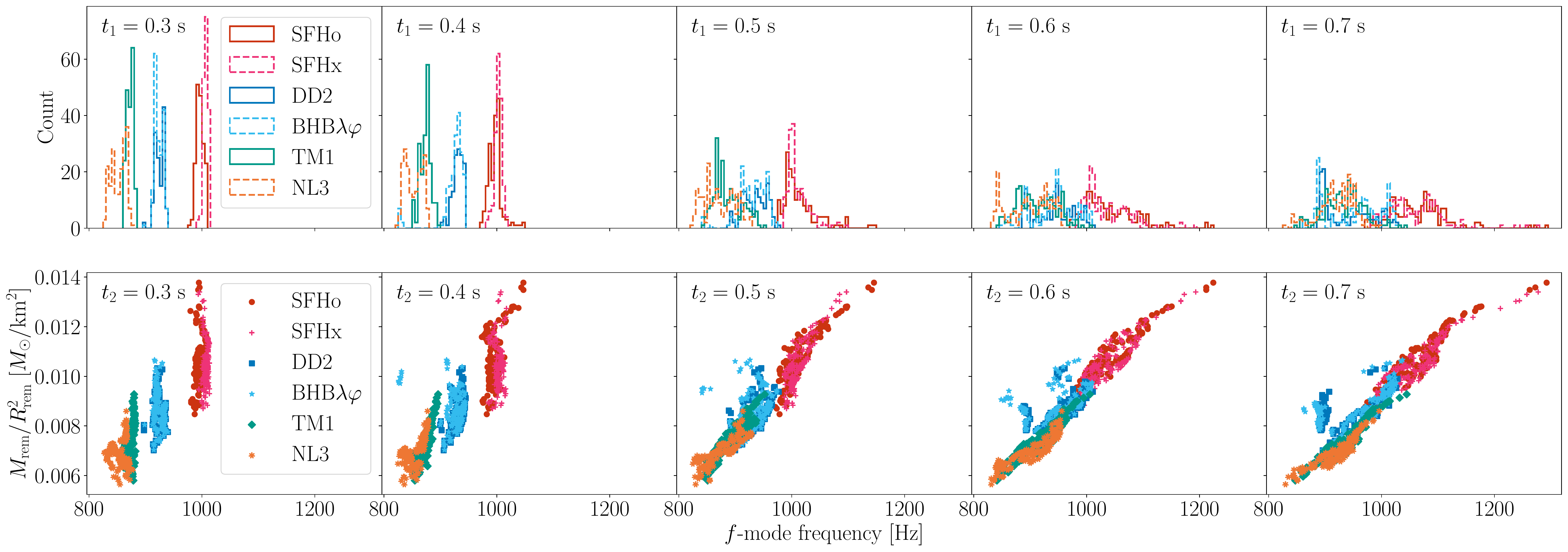}
    \caption{
    Top: Histograms of the characteristic frequency $f_1$ for each equation of state, identified at the characteristic time $t_1$ as noted in each panel.
    Each histogram is plotted with a bin width of 10 Hz and normalized to have an area of one.
    While $t_1 \lesssim 0.4$ seconds post-bounce, prior to the avoided crossing between the $f$- and $g_1$-modes, these results reflect those of the left panel of Figure~\ref{fig:mu_results}, showing a dependence of $f_1$ on the choice of nuclear equation of state.
    Bottom: We plot remnant surface gravity of each model versus characteristic frequency $f_2$ as identified at the time $t_2$ noted in each subplot. 
    For $t_2 \gtrsim 0.4$ seconds post-bounce, these results reflect those of the right panel of Figure~\ref{fig:mu_results}, showing a linear correlation between the surface gravity of the cold neutron star and $f_2$ independent of the equation of state.
    }
    \label{fig:single-times}
\end{figure*}

In Section~\ref{sec:results-gmm-fits}, we identified characteristic frequencies of the gravitational-wave signal from our models by fitting a the histogram of $f$-mode frequencies for each model to a two-component Gaussian mixture.
Then, in Section~\ref{sec:sim-relations}, we explored the correlations between these characteristic frequencies and the nuclear equation of state and surface gravity of the proto-neutron star. 
In this appendix, we check whether relying on characteristic frequencies identified by the Gaussian mixture fitting procedure have introduced spurious correlations in our data. For this, we repeat our analysis using the frequencies at a specific post-bounce time ($t_1$) instead of $\mu_1$ (cf.\ left panel of Figure~\ref{fig:mu_results}) and using the frequencies at a specific post-bounce time ($t_2$) instead of $\mu_2$ (cf.\ right panel of Figure~\ref{fig:mu_results}). We select both $t_1$ and $t_2$ to be at 0.3, 0.4, 0.5, 0.6, and 0.7 seconds post-bounce (in the language of this paper, the first two would be considered `early time' and the last three would be considered `late time'). 
The top row of Figure~\ref{fig:single-times} shows histograms for the $f$-mode frequencies at the five specific post-bounce times and is the analogous figure to the left panel of Figure~\ref{fig:mu_results}. For `early times' (i.e. for $t_1 \lesssim 0.4$~s; first two panels) these results are qualitatively similar to those obtained using the characteristic frequency $\mu_1$. At later times ($t_1 > 0.4$~s), the histograms overlap more and the distinction between different nuclear equations of state disappears. 
In the bottom row of Figure~\ref{fig:single-times}, we plot the surface gravity of the cold remnant NS star against the $f$-mode frequencies identified at the same five specific post-bounce times. Here, we find qualitatively similar results for `late times' (i.e.\ $t_2 > 0.4$~s; last three panels) as in Figure~\ref{fig:mu_results}. However, at `early times' (first two panels), the surface gravity is degenerate with the $f$-mode frequency. 
From this, we conclude that at `early times' the $f$-mode frequency can distinguish between nuclear equations of state, and at `late times' the $f$-mode frequency correlates with the surface gravity of the remnant NS star. This replicates our findings using $\mu_1$ and $\mu_2$. Hence, using time-dependent frequencies replicates the results from the time-agnostic characteristic frequency method.

\clearpage
\pagebreak

\end{document}